\RequirePackage{lineno}
\documentclass[prd,amsmath,amssymb]{revtex4}
\usepackage{rotating}
\usepackage{subfigure}
\usepackage{graphicx}% Include figure files
\usepackage{dcolumn}% Align table columns on decimal point
\usepackage{bm}% bold math
\usepackage{psfig}
\usepackage{epsfig}
\usepackage{psfrag}
\usepackage{needspace/needspace}
\usepackage{appendix}

\newcommand{\comment}[1]{} %

\def\MET{{\mbox{$E\kern-0.57em\raise0.19ex\hbox{/}_{T}$}}}
\def\met{{\mbox{$E\kern-0.57em\raise0.19ex\hbox{/}_{T}$}}}
\def\DZ{D\O\ }

\def\ifb{fb$^{-1}$}
\def\pp{$p\bar{p}$}
\def\bb{$b\bar{b}$}
\def\cc{$c\bar{c}$}
\def\ttbar{$t\bar{t}$}

\def\whl{$WH$$\rightarrow$$ \ell\nu b\bar{b}$}

\def\zhl {$ZH$$\rightarrow$$ \ell\ell b\bar{b}$}

\def\zhv {$ZH$$\rightarrow$$ \nu\bar{\nu} b\bar{b}$}

\def\vhvww{$VH \rightarrow$$ VWW$}
\def\ssem{$VH \rightarrow$$ e^\pm \mu^\pm$+$X$}
\def\lll{$VH \rightarrow$$ ee\mu/\mu\mu e$+$X$}
\def\ttm{$VH \rightarrow$$ \tau\tau\mu$+$X$}

\def\hww{$H$$\rightarrow$$ W^+ W^-$}
\def\hwwenumunu{$H$$\rightarrow$$ W^+ W^-$$\rightarrow$$e^\pm \nu \mu^\mp \nu$}
\def\hwweemm{$H$$\rightarrow$$ W^+ W^-$$\rightarrow$$ee/\mu\mu \nu\nu$}
\def\hwwlnulnu{$H$$\rightarrow$$ W^+ W^-$$\rightarrow$$(ee,\mu\mu,e\mu) \nu \nu$}
\def\hwwmvtv{$H$+$X$$\rightarrow$$ \mu^\pm \tau^{\mp}_{had}+\leq 1j$}
\def\hwwlnuqq{$H$$\rightarrow$$ W^+ W^-$$\rightarrow$$\ell \nu q \bar{q}$}
\def\lnuqqqq{$VH$$\rightarrow$$ \ell \nu q \bar{q}q \bar{q}$}
\def\hwwlvlv{$H$$\rightarrow$$ W^+ W^-$$\rightarrow$$ \ell^\pm \nu\ell^\mp \nu$}

\def\hgg{$H$$\rightarrow$$ \gamma \gamma$}
\def\hbb{$H$$\rightarrow$$ b\bar{b}$}

\def\tevE{$\sqrt{s}=1.96$~TeV}
\def\gev{~Ge\kern -0.05em V}
\def\dgev{Ge\kern -0.05em V\kern -0.1em /$c^2$}

\usepackage{etoolbox}
\newcommand\setvar[2]{\csdef{text#1}{#2}}
\newcommand\getvar[1]{\csuse{text#1}}
\setvar{lumimin}{5.4} % range of integrated luminosities
\setvar{lumimax}{9.7}
\setvar{obs115}{2.11}
\setvar{obs125}{2.94}
\setvar{obs165}{0.73}
\setvar{exp115}{1.46}
\setvar{exp125}{1.70}
\setvar{exp165}{0.72}
\setvar{exclmin}{159} %158.55 % excluded mass range
\setvar{exclmax}{170} %170.2
\setvar{exclminexp}{156} %156.2 % expected excluded mass range
\setvar{exclmaxexp}{173} %173.125
\setvar{135localpval}{2.4\%} %0.024328 %p-value of excess
\setvar{135localzval}{2.0} %1.972 %z-value of excess 
\setvar{135LEEpval}{9\%} %9.3818e-02 %p-value of excess after LEE of 4
\setvar{135LEEzval}{1.3} %1.318 %z-value of excess after LEE of 4
\setvar{120localpval}{2.8\%} %0.028888 %p-value of excess
\setvar{120localzval}{1.9} %1.897 %z-value of excess

\begin{document}

%\linenumbers

%% remove the following for publication
\begin{figure}
\leftline{\includegraphics[scale=0.5]{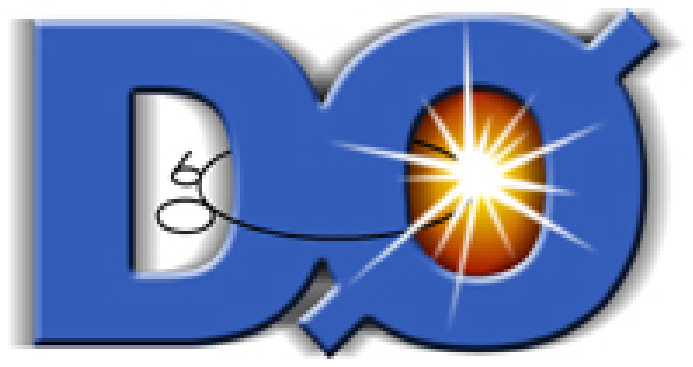}\hfill D\O~Note 6344-CONF}
\end{figure}

%% once the note is approved for conferences, remove the following line
%\centerline{\em D\O\ INTERNAL DOCUMENT -- NOT FOR PUBLIC DISTRIBUTION}
\vspace*{1.0cm}

%\leftline{Version 1.9 \hfill Send comments to d0-run2eb-004@fnal.gov, d0-run2eb-005@fnal.gov, and d0-run2eb-013@fnal.gov}
%\leftline{ \hfill by Friday, June 29th, 5PM (FNAL time)}
%\leftline{Authors: The \DZ Higgs Group}
%% remove the space for publication
%\vspace*{1cm}

\title{Updated Combination of Searches for the Standard Model Higgs Boson\\
at the D\O\ Experiment in \getvar{lumimax}~fb$^{-1}$ of Data}

% throughout the internal review, the note will be authored by individuals
% who contributed to the paper
%\author{The \DZ Higgs Group}

% once the note is approved for conferences, the following author will be used
 \author{The D\O\ Collaboration}
 \affiliation{URL http://www-d0.fnal.gov}

% use the official authorlist for publication
%\input list_of_authors_r2.tex

\date{\today}

\begin{abstract}
% remove the space for publication
\vspace*{2cm} 
Searches for standard model Higgs boson production  at the D\O\ experiment in \pp\ collisions
at \tevE~are carried out for Higgs boson masses ($m_H$) in the range
$100\leq m_H \leq 200$\gev.
Most of these searches use the full Run II data set, corresponding to an
integrated luminosity of \getvar{lumimax}~\ifb,
and are combined to maximize the sensitivity to
the standard model Higgs boson. In absence of a significant excess 
above the background expectation, 95\% confidence level upper limits
are set on the production cross section for a standard model Higgs boson.
The upper limits are found to be a factor of
\getvar{obs115}\ (\getvar{obs165}) times the predicted standard model
cross section for $m_H=115~(165)$\gev.  Under the background-only hypothesis, 
the corresponding expected limit is \getvar{exp115}\ (\getvar{exp165}) times the 
standard model prediction.  At the same confidence level, these analyses exclude a
standard model Higgs boson with a mass in the range
$\getvar{exclmin}<m_H<\getvar{exclmax}$\gev, while the {\it a priori}
expected exclusion is
$\getvar{exclminexp}<m_H<\getvar{exclmaxexp}$\gev.  
In the range $120<m_H<140$\gev, the data exhibit an excess over the
background prediction of approximately two Gaussian standard
deviations.

% remove this for publication
\vspace*{3.0cm}
\end{abstract}
% activate the following line for publication
%\pacs{Valid PACS appear here}

\maketitle
\centerline{\em Preliminary Results for Summer 2012 Conferences}

\newpage
\section{Introduction}
\label{sec:intro}

\def\citeall{\cite{dzWHl,dzZHv,dzZHl,dzHWWlnulnu,dzHWWmnutnu,dzHWWlnuqq,dzSSem,dzVVVeemu,dzVVVttmu,dzHgg,AllD0Notes}}

Despite its success as a predictive tool, the standard model (SM) of
particle physics remains incomplete without a means to explain
electroweak symmetry breaking. The simplest proposed mechanism
involves the introduction of a complex doublet of scalar fields that
generate the masses of elementary particles via their mutual
interactions.  After accounting for longitudinal polarizations for the
electroweak bosons, this so-called Higgs mechanism also gives rise to
a single scalar boson with an unpredicted mass.  Direct searches in
$e^{+} e^{-} \rightarrow Z^{*}\rightarrow ZH$ at the Large
Electron Positron (LEP) collider yielded a lower mass limit of $m_H >
114.4$\gev~\cite{leplim}, at 95\% confidence level (C.L.).  Precision
electroweak data, including the latest $W$ boson mass
measurements from CDF~\cite{CDFMW} and D0~\cite{DZMW}, constrain the
mass of a SM Higgs boson to $m_H < 152$\gev~\cite{lepewwg} at 95\%
C.L.  Direct searches at the CMS~\cite{cms} and ATLAS~\cite{atlas}
experiments limit the SM Higgs boson to have a mass between
115.5\gev\ and 127\gev\ at 95$\%$ C.L.  Additionally, both LHC
experiments report a signal-like excess around a mass of 125\gev.

In this note, we combine the results of direct searches for SM Higgs
bosons in \pp~collisions at~\tevE~recorded with the \DZ
experiment~\cite{dzero}.  The analyses combined here seek signals of
Higgs bosons produced through gluon-gluon fusion
($gg \rightarrow H$), in association with vector bosons
($q\bar{q} \rightarrow VH$, where $V=W, Z$), and through vector boson
fusion (VBF) ($q\bar{q} \rightarrow q\bar{q}H$). The Higgs boson decay 
modes studied are $H \rightarrow b{\bar{b}}$, $H \rightarrow W^+W^-$, 
$H \rightarrow \tau^+\tau^-$, and $H \rightarrow \gamma\gamma$.
Most analyses utilize data corresponding to an integrated luminosity of \getvar{lumimax}~\ifb, 
collected during the data-taking period 2002-2011 (Run II). 
In order to facilitate proper combination of signals, the analyses were constructed 
to be mutually exclusive after analysis selections. 
The searches are organized into analysis sub-channels comprising
different production, decay, and final state particle configurations
designed to maximize the sensitivity for a particular Higgs boson
production and decay mode.  These sub-channels, typically having different
sensitivity, are analyzed separately and combined at the end in order to maximize
the search sensitivity.  Details on the individual analyses and the
improvements since the last combination~\cite{M12dzcombo} are provided
in the following section.  

\section{Summary of Contributing Analyses}
\label{sec:channels}

\begin{table}[b]
\caption{\label{tab:chans}List of analysis channels ($V=W,Z$ and
  $\ell=e, \mu$) with the corresponding integrated luminosities, final
  variables used for setting limits, and mass range studied. See
  Section~\ref{sec:channels} for details. All conference notes can be found from Ref.~\cite{AllD0Notes}.}
\begin{ruledtabular}
\begin{tabular}{lcccc}
\\
Channel & Luminosity (\ifb)& Final Variable & $m_H$ Range & Reference\\\hline
\zhl,~4 (lepton) $\times$ 2 ($b$-tag) $\times$ 2 ($t\bar{t}$) categories  & 9.7      & Decision Tree Discriminant    & 100--150  & \cite{dzZHl}\\
\zhv,~2 ($b$-tag) categories                   & 9.5      & Decision Tree Discriminant    & 100--150  & \cite{dzZHv}\\
\whl,~4 ($b$-tag) $\times$ 2 (jet) categories  & 9.7      & Decision Tree Discriminant    & 100--150  & \cite{dzWHl}\\
\hwweemm,~2$\times$2+1 (jet$\times$$WW$) categories    & 9.7      & Decision Tree Discriminant    & 115--200  & \cite{dzHWWlnulnu}\\
\hwwenumunu,~3 (jet) categories    & 9.7      & Decision Tree Discriminant    & 115--200  & \cite{dzHWWlnulnu}\\
\ssem                    & 9.7      & Decision Tree Discriminant    & 115--200  & \cite{dzSSem}\\
\lll                     & 9.7      & Decision Tree Discriminant    & 100--200  & \cite{dzVVVeemu}\\
\lnuqqqq,~2 (low $b$-tag) categories   & 9.7      & Decision Tree Discriminant    & 100--200  & \cite{dzWHl}\\
\hgg                     & 9.7      & Decision Tree Discriminant    & 100--150  & \cite{dzHgg} \\
\hwwmvtv                 & 7.3      & Neural Network Discriminant   & 115--200  & \cite{dzHWWmnutnu}\\
\ttm                     & 7.0      & Summed $|p_T|$ of all objects & 115--200  & \cite{dzVVVttmu}\\
\hwwlnuqq                & 5.4      & Decision Tree Discriminant    & 155--200  & \cite{dzHWWlnuqq}\\
\end{tabular}
\end{ruledtabular}
\end{table}

A summary of the analyses used in this combination is given in
Table~\ref{tab:chans}.
The most sensitive analyses for Higgs boson masses below approximately
130\gev\ are those searching for \hbb\ in association with a
leptonically decaying weak vector boson.  To help isolate these
\hbb\ decays, the analyses use an algorithm to identify jets that are
consistent with containing the decay of a $b$ quark ($b$-tagging).
Several kinematic variables sensitive to displaced jet vertices and
jet tracks with large transverse impact parameters relative to the
hard-scatter vertices are combined in a boosted decision tree based
$b$-tagging discriminant.  This algorithm is an upgraded version of
the neural network $b$-tagger used previously~\cite{btagnim}.  By
adjusting a minimum requirement on the $b$-tagging output, a spectrum
of increasingly stringent $b$-tagging operating points is achieved,
with a range of signal efficiencies and purities.

The \zhl~($\ell=e,\mu$) analysis~\cite{dzZHl} requires two isolated
charged leptons and two or three hadronic jets, at least one of which
must pass a tight $b$-tag requirement.  The events are then divided
into ``double-tag'' and ``single-tag'' sub-channels depending on
whether or not a second jet passes a loose $b$-tag requirement.  
The typical efficiency and fake rate for taggable~\cite{btagnim} jets
for the loose (tight) $b$-tag selection is about 80\% (50\%) and 10\% (1\%), respectively.  
The analysis uses decision tree discriminants to provide the final
variables for setting limits.  For this iteration of the analysis, a
two-step process is applied.  First, the events are divided into
\ttbar-depleted or \ttbar-enriched sub-channels using decision trees
trained to discriminate signal from the \ttbar\ backgrounds in each
lepton and $b$-tag sub-channel.  This allows to isolate two regions
with different signal-to-background. Final discriminants are then
constructed to separate signal from all backgrounds.  The limit is
calculated using the output distributions of the final discriminants
for both the \ttbar-depleted and \ttbar-enriched samples. The better
signal-to-background discrimination, in addition to other optimizations
in the event selection, result in a sensitivity improvement of 
approximately 10--14\%  compared to the previous result~\cite{M12dzZHl}.

The \zhv\ analysis~\cite{dzZHv} selects events with large \met\ and
two hadronic jets.  A sizable fraction of signal comes from
\whl\ events in which the charged lepton does not pass the criteria
for the \whl\ analysis.  However, events with leptons that pass the
criteria for the \whl\ analysis are rejected to ensure orthogonality
between the two analyses.  Track-based missing transverse momentum and
\met\ significance variables are used to reduce instrumental
backgrounds with false \met.  The multijet background is further
reduced by employing a dedicated decision tree discriminant before
$b$-tagging.  This analysis defines an event-level $b$-tag quality by
summing the $b$-tag outputs from the two jets.  Two orthogonal $b$-tag
sub-channels are then defined.  The ``tight-tag'' sub-channel requires
that both jets pass rather tight $b$-tag criteria, while the
``medium-tag'' sub-channel allows for the criteria on one of the jets
to be relaxed provided that the other jet has a sufficiently high
$b$-tag output.  Decision trees classifiers trained separately for the
different $b$-tagging categories are used as the final discriminant.
Improved training of these discriminants by using larger Monte Carlo
statistics leads to a gain in sensitivity of approximately 10\% relative 
to the previous result~\cite{M12dzZHv}.

The \whl\ analysis~\cite{dzWHl} exploits topologies with a charged
lepton, missing energy, and two or three hadronic jets.
Decision trees are used to discriminate against the multijet
background.  Using the average of the two highest $b$-tag outputs from
all selected jets, six orthogonal $b$-tag categories are defined.  The
two categories with the lowest $b$ jet purity are removed to avoid
overlap with the \hwwlnuqq\ analysis~\cite{dzHWWlnuqq}, while the
remaining categories define the four $b$-tag sub-channels used in this
analysis.  A final decision tree discriminant is constructed for each
lepton flavor, jet multiplicity, and $b$-tag sub-channel.  In addition
to kinematic variables, the inputs to the final discriminants include
the output from the $b$-tagger and the output from the multijet
discriminant.  Changes with respect to the previous result~\cite{M12dzWHl}
include extending from three to four the number of $b$-tag sub-channels
considered, improving the multivariate treatment, and increasing the pseudo-rapidity
acceptance of muons, ultimately improving the sensitivity by 
approximately 10--17\%.

The \hwwlvlv\ analyses target Higgs boson decays to two $W$ bosons 
and consider the three
dominant production mechanisms: gluon-gluon fusion, associated
production, and vector-boson fusion.  The three dominant
search channels are $e^+ \nu e^-\nu$, $e^\pm \nu \mu^\mp \nu$, and
$\mu^+ \nu \mu^-\nu$~\cite{dzHWWlnulnu}.  Events are characterized by
large \met~and two oppositely-charged isolated leptons, which can have
rather low transverse momentum for $m_H<2 m_W$, where at least one of 
the $W$ bosons is off-mass shell.  The presence
of neutrinos in the final state prevents the reconstruction
of the candidate Higgs boson mass.  Each final state is further
subdivided according to the number of jets in the event: zero, one, or
two or more jets.  This allows the individual discriminants to
separate differing contributions of signal and background processes
more effectively.  However, this introduces the need to evaluate the
systematic uncertainties carefully in each jet category, as discussed
in Section~\ref{sec:theory}.
The $e^+ \nu e^-\nu$ and $\mu^+ \nu \mu^-\nu$ channels use decision trees discriminants to
reduce the dominant Drell-Yan background, while the $e^\pm \nu \mu^\mp \nu$
channel uses \met-related variables to remove backgrounds.  Decision
trees are used as the final discriminants, including input kinematic as well
as topological variables (e.g. $b$-tagging information in the
case of sub-channels with jets). 
The events in the $e^+ \nu e^-\nu$ and $\mu^+ \nu \mu^-\nu$ channels in the
zero and one jet categories are further subdivided in
two samples each using dedicated decision trees that
enhance or reduce the contribution from the non-resonant
$WW$ background.  All sub-samples are
used in the limit setting, with the additional channels significantly
constraining the uncertainty on the $WW$ cross-section.  
In addition, the integrated luminosity of the sample
used in the $e^+ \nu e^-\nu$ channel 
has been increased from 8.6 to 9.7 fb$^{-1}$, including
data which had not been considered in the previous analysis.
These changes led to an improvement in sensitivity of approximately
5--10\% relative to the previous result~\cite{M12dzHWWlnulnu}.  

Decays involving tau leptons are included in two ways. A dedicated
analysis~\cite{dzHWWmnutnu} ($\mu\tau_{\rm{had}}$) using 7.3~\ifb\ of
data studying the final state involving a muon and a hadronic tau
decay plus up to one jet.  Final states involving other tau decays and
mis-identified hadronic tau decays are included in the \hwwlvlv\ analyses.  The
$\mu\tau_{\rm{had}}$ channel uses neural networks as the final
discriminant. In addition, the trilepton search for $\tau\tau\mu$,
discussed below, is primarily sensitive to $H \rightarrow \tau^+\tau^-$.

For \vhvww~production, we consider final states with three charged
leptons~\cite{dzVVVeemu,dzVVVttmu} ($ee\mu$, $\mu\mu e$, and
$\tau\tau\mu$), as well as the dilepton final state containing an
electron and muon with the same charge~\cite{dzSSem}
($e^\pm\mu^\pm+X$), which benefits greatly from the suppression of
Drell-Yan background.  The $ee\mu$ and $\mu \mu e$ analyses use
decision trees as final discriminants, while the $\tau\tau\mu$
analysis uses the scalar sum of the $p_T$ from all objects.  The
$\mu\mu e$ search has improved the signal-to-background discrimination
by splitting the sample in three orthogonal sub-channels (without $Z$ boson
candidate,  or with $Z$ boson candidate and high/low missing transverse
energy significance) and including jet-related information in the multivariate
analysis. This leads to an improvement in sensitivity of approximately 10--20\%
relative to the previous result~\cite{M12dzVVVeemu}.
The $e^\pm\mu^\pm+X$ analysis uses a two step multivariate approach.
First, a decision tree is used to remove most of the dominant
backgrounds from multijet and $W$+jets/$\gamma$ events.  Then a final
decision tree is used to discriminate signal from the remaining
background.

We also include analyses that search for \hww\ with one or both $W$
bosons decaying hadronically.  These are the
\hwwlnuqq~\cite{dzHWWlnuqq} and \lnuqqqq~\cite{dzWHl} analyses, both
of which are much like the \whl\ search except that the jets are not
$b$-tagged and the \lnuqqqq\ analysis requires at least four jets.
The \lnuqqqq\ analysis represents the first search for the SM Higgs boson
in this final state signature.

Finally, we include an analysis that searches for Higgs bosons
decaying to two photons~\cite{dzHgg}. All three dominant production mechanisms, gluon-gluon
fusion, vector boson fusion, and associated production, are considered in this search.
The contribution of jets misidentified as photons is reduced by
combining information sensitive to differences in the energy
deposition from these particles in the tracker, calorimeter, and
preshower in a neural network.  The output of boosted
decision trees, rather than the diphoton invariant mass, is used as
the final discriminating variable.
Relative to the previous result~\cite{M12dzHgg}, improved vertexing and
energy calibrations have been incorporated.  Additionally, the impact
of systematic uncertainties is now reduced by inclusion of
photon-dominated and jet-dominated sub-channels in the limit setting
procedure. The overall sensitivity improvement is
approximately 30\%.

For all analyses, the backgrounds from multijet production are
measured in data. The other backgrounds were generated by {\sc
  pythia}~\cite{pythia}, {\sc alpgen}~\cite{alpgen}, {\sc
  sherpa}~\cite{sherpa}, or {\sc singletop}~\cite{singletop}, with {\sc
  pythia} providing parton-showering and hadronization.  Drell-Yan,
$W$, and diboson background cross sections are normalized either to
next-to-leading order (NLO) calculations from {\sc mcfm}~\cite{mcfm}
or, when possible, to data control samples.  Top pair and single top
production are normalized to approximate
next-to-next-to-NLO~\cite{xsecsTT} and
next-to-next-to-NLO~\cite{xsecsT} calculations, respectively.

\section{Signal Predictions and Uncertainties}
\label{sec:theory}

A common approach to the signal predictions and associated
uncertainties is followed by both the CDF and \DZ Collaborations. An
outline of the procedures followed is given here; a more complete
discussion can be found in Ref.~\cite{Tevcomb}.  

The Monte Carlo signal simulation is provided by the leading-order
(LO) generator {\sc pythia} (with {\sc CTEQ6L1}~\cite{cteq}
LO parton distribution functions), which includes a parton
shower and fragmentation and hadronization models.  We reweight the
Higgs boson $p_T$ spectra in the {\sc pythia} Monte Carlo samples to
that predicted by {\sc hqt}~\cite{hqt} when making predictions of
differential distributions of gluon-gluon fusion signal events.  To evaluate the
impact of the scale uncertainty on the differential spectra, we use
the {\sc resbos}~\cite{resbos} generator, and apply the
scale-dependent differences in the Higgs boson $p_T$ spectrum to the
{\sc hqt} prediction, and propagate these to our final discriminants
as a systematic uncertainty on the shape of the final variable
distribution, which is included in the calculation of the limits.

We normalize the Higgs boson signal predictions to the most recent
higher-order calculations available.  The $gg\rightarrow H$ production
cross section is calculated at next-to-next-to leading order
(NNLO) in QCD with a next-to-next-to leading log (NNLL) resummation of
soft gluons; the calculation also includes two-loop electroweak
effects and handling of the running $b$ quark
mass~\cite{anastasiou,grazzinideflorian}. The numerical values in
Table~\ref{tab:higgsxsec} are updates~\cite{grazziniprivate} of these
predictions with $m_t$ set to 173.1\gev~\cite{tevtop10}, and an
exact treatment of the massive top and bottom loop corrections up to
NLO + next-to-leading-log accuracy.  The factorization and
renormalization scale choice for this calculation is
$\mu_F=\mu_R=m_H$. These calculations are refinements of the earlier
NNLO calculations of the $gg\rightarrow H$ production cross
section~\cite{harlanderkilgore2002,anastasioumelnikov2002,ravindran2003}.
Electroweak corrections were computed in
Refs.~\cite{actis2008,aglietti}. Soft gluon resummation was introduced
in the prediction of the $gg\rightarrow H$ production cross
section~\cite{catani2003}.  The $gg\rightarrow H$ production cross
section depends strongly on the gluon parton density function, and the
accompanying value of $\alpha_s(q^2)$.  The cross sections used here
are calculated with the MSTW~2008 NNLO PDF set~\cite{mstw2008}, as
recommended by the PDF4LHC working group~\cite{pdf4lhc}.  The
inclusive (over jet multiplicity) Higgs boson production cross sections are listed in
Table~\ref{tab:higgsxsec}.

\begin{table}
\begin{center}
\caption{
The production cross sections (in fb) and decay branching fractions
(in \%) for each SM Higgs boson mass (in\gev) assumed for the combination.}
\vspace{0.2cm}
\label{tab:higgsxsec}
\begin{ruledtabular}
{\footnotesize
\begin{tabular}{ccccccccccc}
$m_H$ & $\sigma_{gg\rightarrow H}$  & $\sigma_{WH}$  & $\sigma_{ZH}$  & $\sigma_{VBF}$  & $B(H\rightarrow b{\bar{b}})$ & $B(H\rightarrow c{\bar{c}})$ & $B(H\rightarrow \tau^+{\tau^-})$ & $B(H\rightarrow W^+W^-)$ & $B(H\rightarrow ZZ)$ & $B(H\rightarrow \gamma \gamma)$\\ 
\hline
   100 &   1821.8   &   281.10    &   162.7      &    100.1 &  79.1   & 3.68     & 8.36    & 1.11   & 0.113  & 0.159    \\ 
   105 &   1584.7   &   238.70    &   139.5      &     92.3 &  77.3   & 3.59     & 8.25    & 2.43   & 0.215  & 0.178    \\
   110 &   1385.0   &   203.70    &   120.2      &     85.1 &  74.5   & 3.46     & 8.03    & 4.82   & 0.439  & 0.197    \\
   115 &   1215.9   &   174.50    &   103.9      &     78.6 &  70.5   & 3.27     & 7.65    & 8.67   & 0.873  & 0.213    \\
   120 &   1072.3   &   150.10    &    90.2      &     72.7 &  64.9   & 3.01     & 7.11    & 14.3   & 1.60   & 0.225    \\
   125 &    949.3   &   129.50    &    78.5      &     67.1 &  57.8   & 2.68     & 6.37    & 21.6   & 2.67   & 0.230    \\
   130 &    842.9   &   112.00    &    68.5      &     62.1 &  49.4   & 2.29     & 5.49    & 30.5   & 4.02   & 0.226    \\
   135 &    750.8   &    97.20    &    60.0      &     57.5 &  40.4   & 1.87     & 4.52    & 40.3   & 5.51   & 0.214    \\
   140 &    670.6   &    84.60    &    52.7      &     53.2 &  31.4   & 1.46     & 3.54    & 50.4   & 6.92   & 0.194    \\
   145 &    600.6   &    73.70    &    46.3      &     49.4 &  23.1   & 1.07     & 2.62    & 60.3   & 7.96   & 0.168    \\
   150 &    539.1   &    64.40    &    40.8      &     45.8 &  15.7   & 0.725    & 1.79    & 69.9   & 8.28   & 0.137    \\
   155 &    484.0   &    56.20    &    35.9      &     42.4 &  9.18   & 0.425    & 1.06    & 79.6   & 7.36   & 0.100    \\
   160 &    432.3   &    48.50    &    31.4      &     39.4 &  3.44   & 0.159    & 0.397   & 90.9   & 4.16   & 0.0533   \\
   165 &    383.7   &    43.60    &    28.4      &     36.6 &  1.19   & 0.0549   & 0.138   & 96.0   & 2.22   & 0.0230   \\
   170 &    344.0   &    38.50    &    25.3      &     34.0 &  0.787  & 0.0364   & 0.0920  & 96.5   & 2.36   & 0.0158   \\
   175 &    309.7   &    34.00    &    22.5      &     31.6 &  0.612  & 0.0283   & 0.0719  & 95.8   & 3.23   & 0.0123   \\
   180 &    279.2   &    30.10    &    20.0      &     29.4 &  0.497  & 0.0230   & 0.0587  & 93.2   & 6.02   & 0.0102   \\
   185 &    252.1   &    26.90    &    17.9      &     27.3 &  0.385  & 0.0178   & 0.0457  & 84.4   & 15.0   & 0.00809  \\
   190 &    228.0   &    24.00    &    16.1      &     25.4 &  0.315  & 0.0146   & 0.0376  & 78.6   & 20.9   & 0.00674  \\
   195 &    207.2   &    21.40    &    14.4      &     23.7 &  0.270  & 0.0125   & 0.0324  & 75.7   & 23.9   & 0.00589  \\
   200 &    189.1   &    19.10    &    13.0      &     22.0 &  0.238  & 0.0110   & 0.0287  & 74.1   & 25.6   & 0.00526  \\ 
\end{tabular}	
}
\end{ruledtabular}	
\end{center}	
\end{table}

For analyses that consider inclusive $gg\rightarrow H$ production, but
do not split the signal prediction into separate channels based on the
number of reconstructed jets, we use the inclusive uncertainties from
the simultaneous variation of the factorization and renormalization
scales up and down by a factor of two.  We use the prescription of the
PDF4LHC working group for evaluating PDF uncertainties on the
inclusive production cross section.  QCD scale uncertainties that
affect the cross section via their impact on the PDFs are included as
a correlated part of the total scale uncertainty.  The remainder of
the PDF uncertainty is treated as uncorrelated with the QCD scale
uncertainty.

For analyses seeking $gg\rightarrow H$ production that divide events
into categories based on the number of reconstructed jets, we employ
an approach for evaluating the impact of the scale uncertainties
following Ref.~\cite{bnlaccord}.  We treat the QCD scale uncertainties
obtained from the NNLL inclusive~\cite{grazzinideflorian,anastasiou},
NLO with one or more jets~\cite{anastasiouwebber}, and NLO with two or
more jets~\cite{campbellh2j} cross section calculations as
uncorrelated with one another.  We then obtain QCD scale uncertainties
for the exclusive $gg\rightarrow H+0$~jet, 1~jet, and 2~or more jet
categories by propagating the uncertainties on the inclusive cross
section predictions through the subtractions needed to predict the
exclusive rates.  For example, the $H$+0~jet cross section is obtained
by subtracting the NLO $H+1$~or more jet cross section from the
inclusive NNLL+NNLO cross section. Therefore, we assign three
separate, uncorrelated scale uncertainties which lead to correlated
and anticorrelated uncertainty contributions between exclusive jet
categories. The procedure in Ref.~\cite{anastasiouwebber} is used to
determine PDF model uncertainties.  These are obtained separately for
each jet bin and treated as 100\% correlated between jet bins.

Another source of uncertainty in the prediction of
$\sigma(gg\rightarrow H)$ is the extrapolation of the QCD corrections
computed for the heavy top quark loops to the light-quark loops
included as part of the electroweak corrections.  It has been
argued~\cite{anastasiou} that the factorization of QCD corrections is
known to work well for Higgs boson masses much larger than the masses
of the particles contributing to the loop. 
A 4\% change in the predicted cross section is seen when all QCD corrections 
are removed from the diagrams containing light-flavored quark loops, which
would represent an overestimate of the uncertainty.  For
the $b$ quark loop~\cite{anastasiou}, the QCD corrections are much
smaller than for the top loop, further giving confidence that it does
not introduce large uncertainties. Uncertainties at the level of 1-2\% due to
these effects are included in the predictions we use~\cite{grazzinideflorian,anastasiou}.  

We consider all significant Higgs boson production modes in our
searches.  Besides gluon-gluon fusion through virtual quark loops, we consider Higgs
boson production in association with a $W$ or $Z$ vector boson, and
vector boson fusion.  We use the $WH$ and $ZH$ production cross
sections computed at NNLO~\cite{djouadibaglio}.  This calculation
starts with the NLO calculation of {\sc v2hv}~\cite{v2hv} and includes
NNLO QCD contributions~\cite{vhnnloqcd}, as well as one-loop
electroweak corrections~\cite{vhewcorr}.  We use the vector-boson fusion cross section
computed at NNLO in QCD~\cite{vbfnnlo}.  Electroweak corrections to
the vector-boson fusion production cross section are computed with the {\sc hawk}
program~\cite{hawk}, and are very small (0.03 fb and less) for the
Higgs boson mass range considered here.

The Higgs boson decay branching ratio predictions are calculated with
{\sc hdecay}~\cite{hdecay}, and are also listed in
Table~\ref{tab:higgsxsec}.  We use {\sc hdecay} Version 3.53.  While
the $HWW$ coupling is well predicted, $B(H\rightarrow W^+W^-)$ depends
on the partial widths of all other Higgs boson decays. The partial
width $\Gamma(H\rightarrow b{\bar{b}})$ is sensitive to $m_b$ and
$\alpha_s$, $\Gamma(H\rightarrow c{\bar{c}})$ is sensitive to $m_c$
and $\alpha_s$, and $\Gamma(H\rightarrow gg)$ is sensitive to
$\alpha_s$.  The impacts of these uncertainties on $B(H\rightarrow
W^+W^-)$ depend on $m_H$ due to the fact that $B(H\rightarrow
b{\bar{b}})$, $B(H\rightarrow c{\bar{c}})$, $B(H\rightarrow gg)$
become very small for Higgs boson masses above 160\gev, while they
have a larger impact for lower $m_H$.  We use the uncertainties on the
branching fraction $B(H\rightarrow W^+W^-)$ from
Ref.~\cite{bagliodjouadilittlelhc}.  At $m_H=130$\gev, for example,
the $m_b$ variation gives a $^{-4.89}_{+1.70}\%$ relative variation in
$B(H\rightarrow W^+W^-)$, $\alpha_s$ gives a $^{-1.02}_{+1.09}\%$
variation, and $m_c$ gives a $^{-0.45}_{+0.51}\%$ variation.  At
$m_H=165$\gev, all three of these uncertainties drop below 0.1\%.

\section{Limit Calculations}

We combine results using the $CL_s$ method with a negative
log-likelihood ratio (LLR) test statistic~\cite{cls}. The value of
$CL_s$ is defined as $CL_s = CL_{s+b}/CL_b$ where $CL_{s+b}$ and
$CL_b$ are the confidence levels for the signal-plus-background
hypothesis and the background-only hypothesis, respectively.  These
confidence levels are evaluated by integrating corresponding LLR
distributions populated by simulating outcomes via Poisson
statistics. Separate channels and bins are combined by summing LLR
values over all bins and channels. This method provides a robust means
of combining individual channels while maintaining individual channel
sensitivities and incorporating systematic uncertainties. Systematic
uncertainties are treated as Gaussian uncertainties on the expected
number of signal and background events, not the outcomes of the limit
calculations. This approach ensures that the uncertainties and their
correlations are propagated to the outcome with their proper
weights. The $CL_s$ approach used in this combination utilizes binned
final-variable distributions rather than a single-bin (fully
integrated) value for each contributing analysis. The exclusion
criteria are determined by increasing the signal cross section until
$CL_s = 1-\alpha$, which defines a signal cross section excluded at
95\% confidence level for $\alpha=0.95$.

\subsection{Final Variable Distributions}

Searches are performed assuming different values of the Higgs boson mass 
between 100\gev\ and 200\gev, in steps of 5\gev.
For each tested Higgs boson mass, each analysis provides
binned distributions of the final discriminants for each sub-channel.
These input distributions can be found in the
corresponding references (see Table~\ref{tab:chans}).
The limit calculation uses the individual inputs, however, for
visualization purposes, it can be useful to collect all of the inputs
into a single distribution.  To preserve the sensitivity from the bins
with high signal-to-background ($s/b$) ratios, only bins with similar
$s/b$ are combined.  Therefore, the aggregate distribution is made by
reordering all of the bins from the input distributions according to
$s/b$.  The range of $s/b$ is quite large, so $\log_{10}(s/b)$ is
used.  Figure~\ref{fig:sbInputs} shows the aggregate distributions for
test Higgs boson masses of 115\gev, 125\gev, and 165\gev, indicating
good agreement between data and predictions over many orders of
magnitude.  Figure~\ref{fig:subtractInputs} shows the same
distributions after subtracting the expected background from the data.
Integrating the distributions in Fig.~\ref{fig:sbInputs} from {\it
  right to left} ({\it i.e.}, starting with the highest $s/b$ events)
allows one to see how the data compare to the background-only and
signal+background hypotheses as the most significant events are
accumulated.  Figure~\ref{fig:integralInputs} shows these cumulative
distributions for the $\approx 150$ most significant events as a
function of the integrated number of predicted signal events.  For a
Higgs boson mass of 125\gev, the highest $s/b$ bins contain an excess
of signal like candidate events, while for a mass of 165\gev, the data
clearly follow the background-only expectation.

\begin{figure}[htbp]
\begin{centering}
\includegraphics[width=0.45\textwidth]{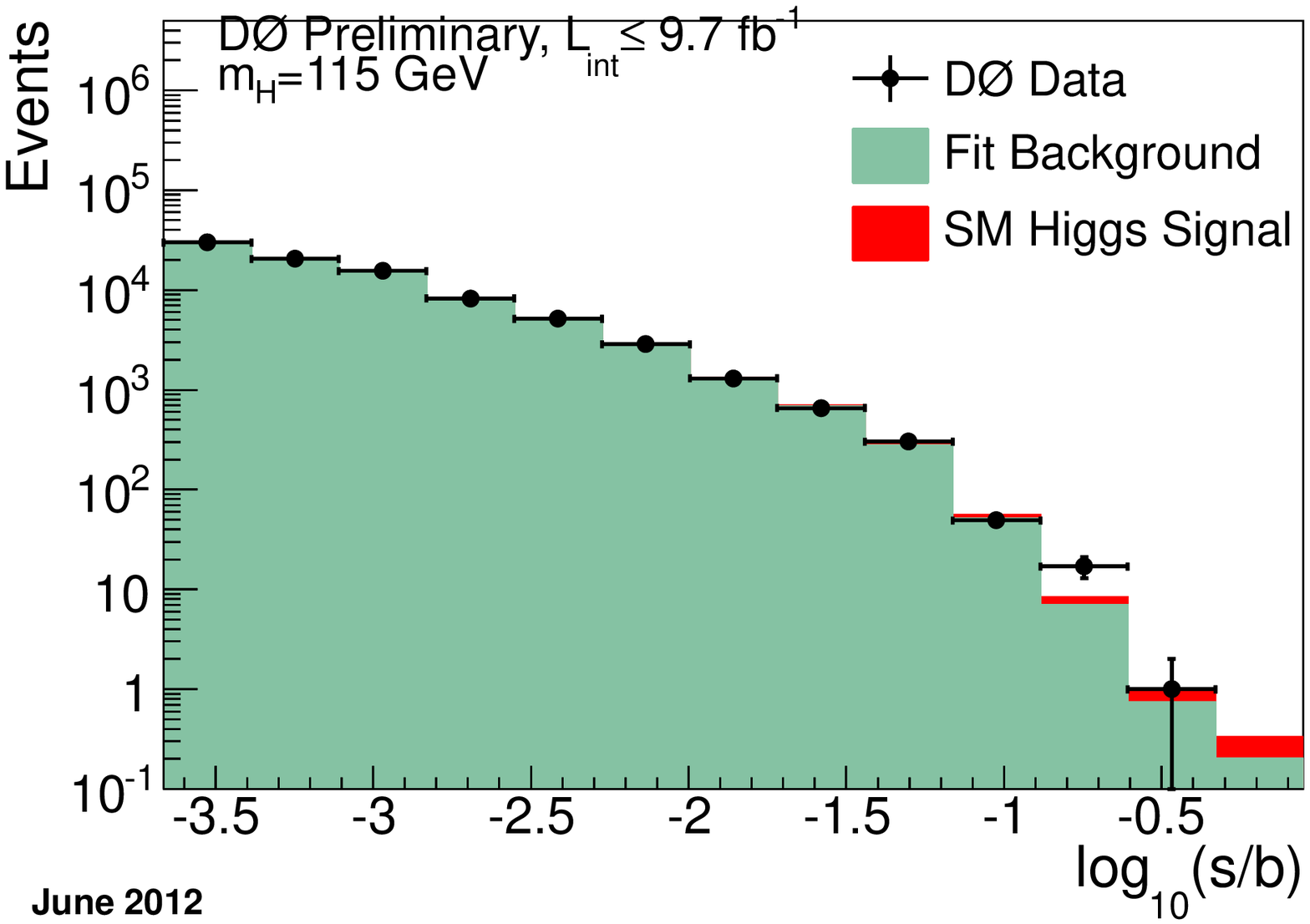}
\includegraphics[width=0.45\textwidth]{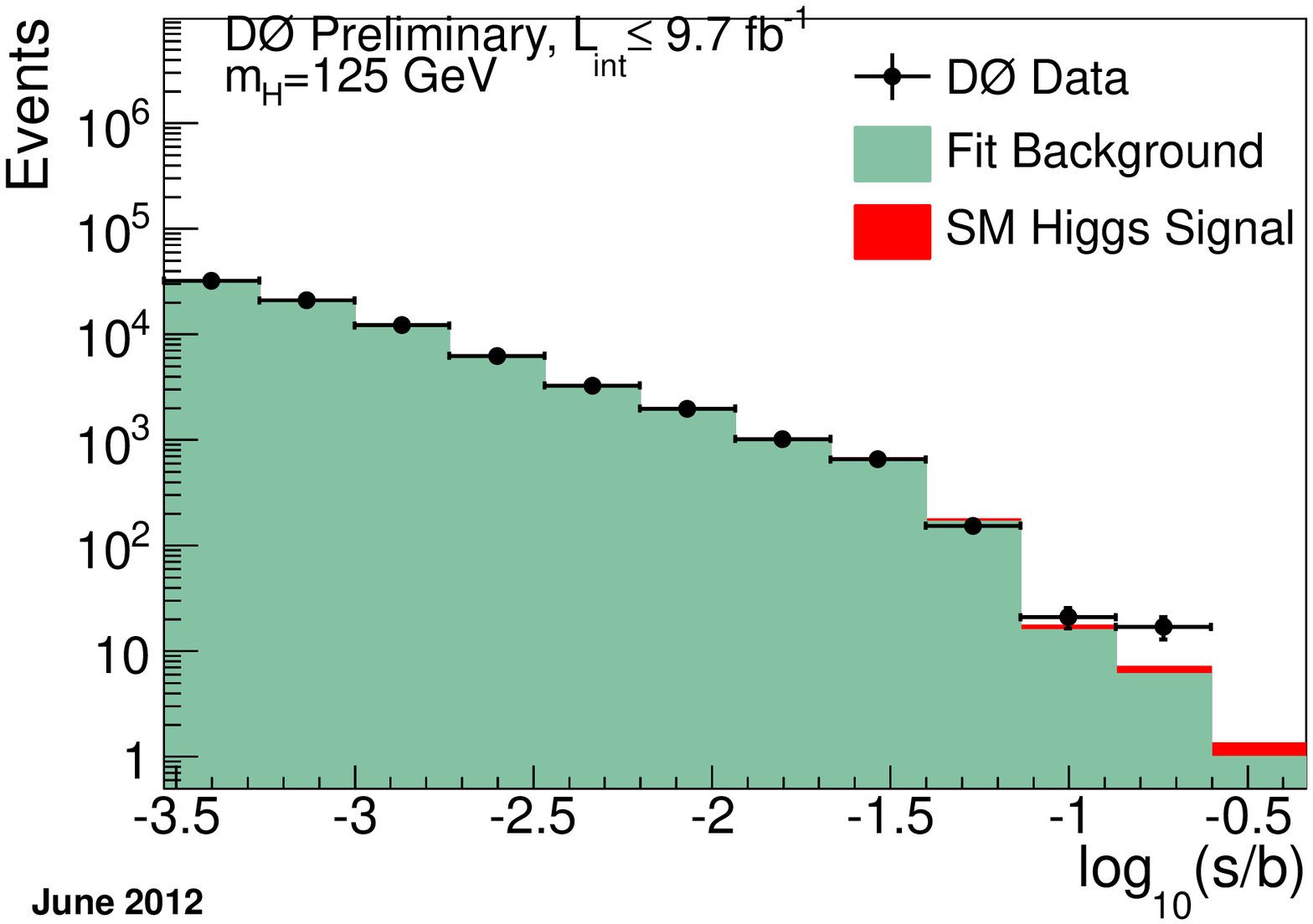}
\includegraphics[width=0.45\textwidth]{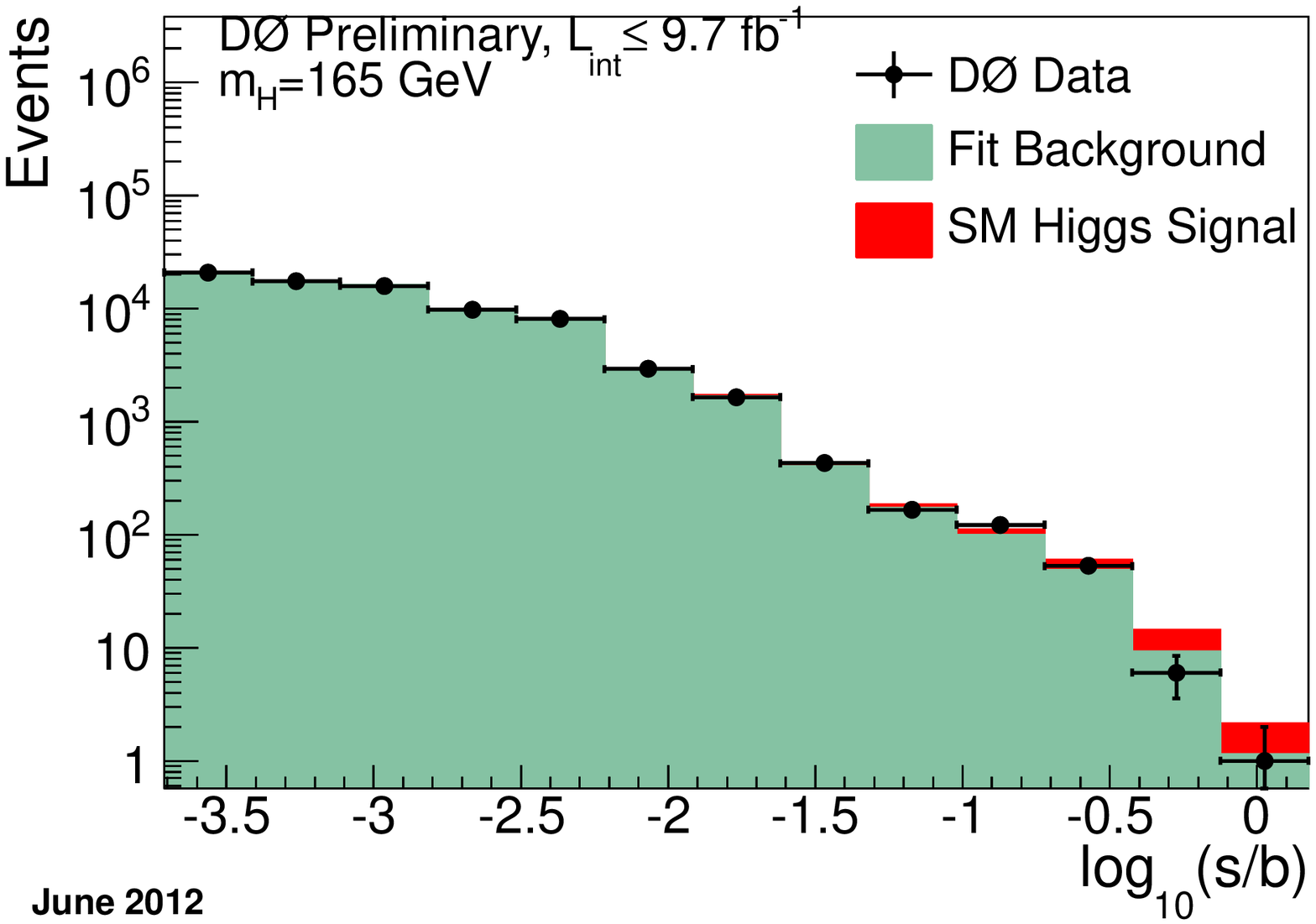}
\caption{
\label{fig:sbInputs} Distributions of $\log_{10}(s/b)$ for the data from all
contributing channels for Higgs boson masses of 115\gev, 125\gev, and
165\gev.  The data are shown with points and the expected signal is
shown stacked on top of the backgrounds. Only statistical uncertainties
on the data points are shown. Systematic uncertainties on the background
prediction are displayed in Fig.~\ref{fig:subtractInputs}.}
\end{centering}
\end{figure}

\begin{figure}[htbp]
\begin{centering}
\includegraphics[width=0.45\textwidth]{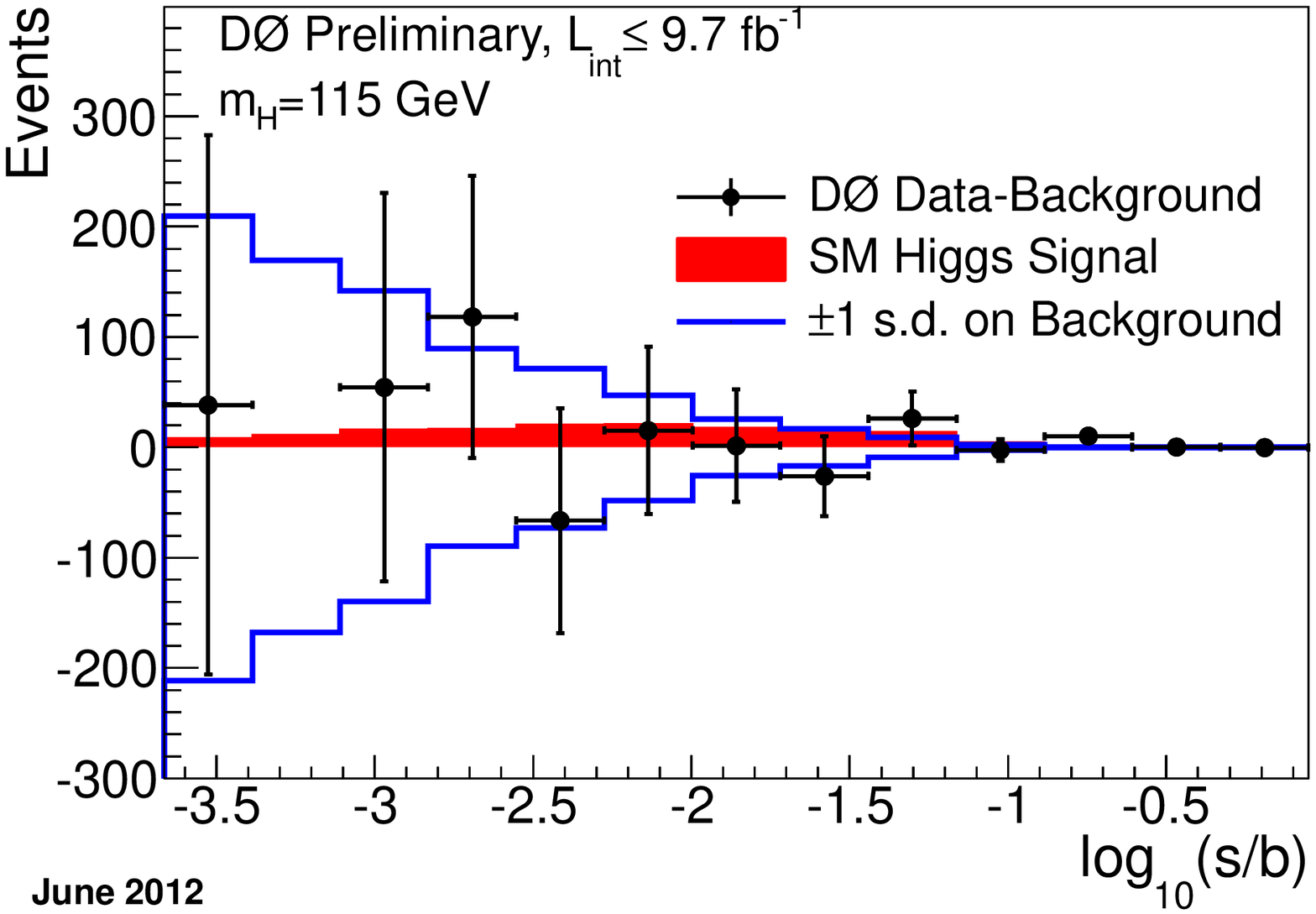}
\includegraphics[width=0.45\textwidth]{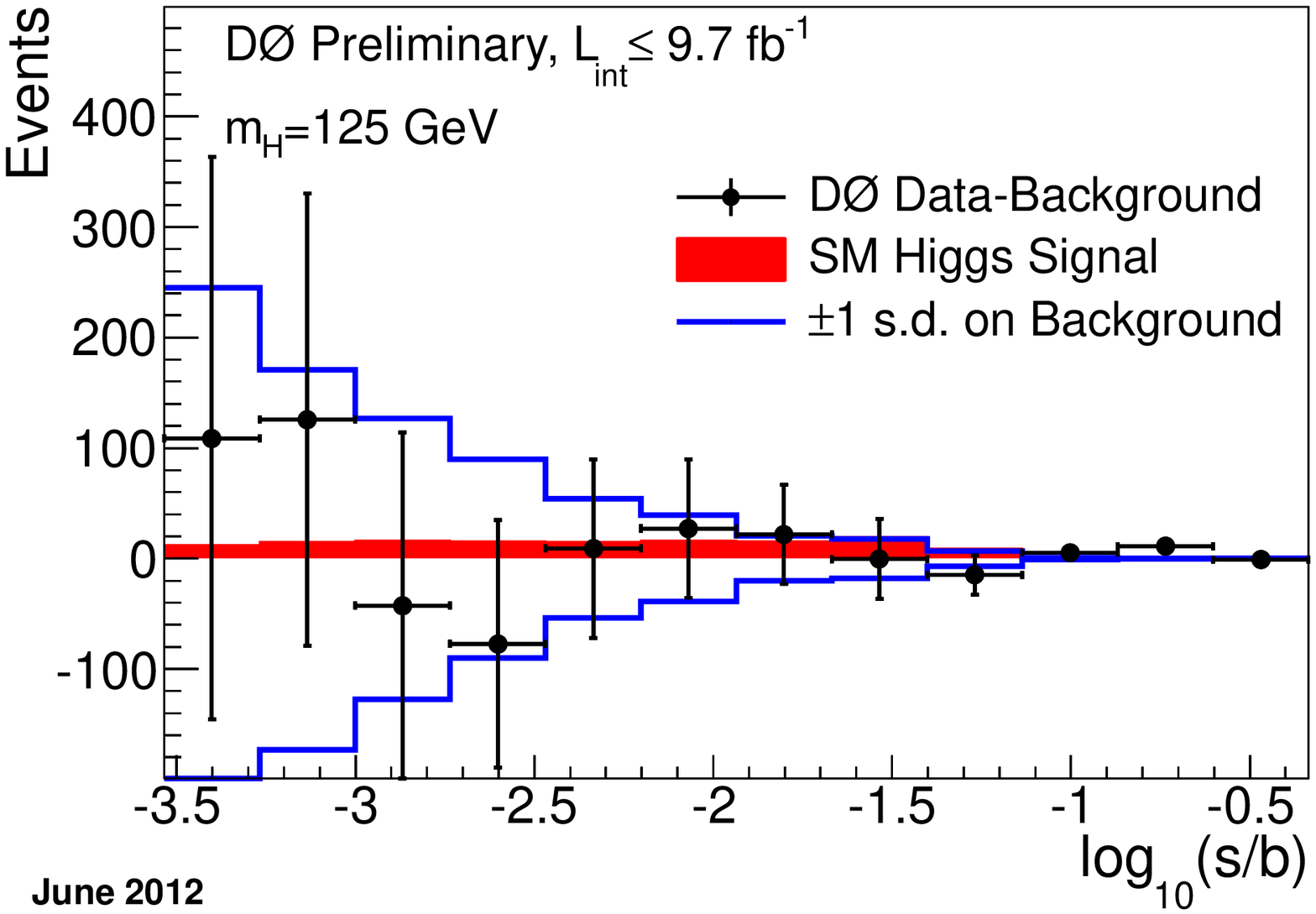}
\includegraphics[width=0.45\textwidth]{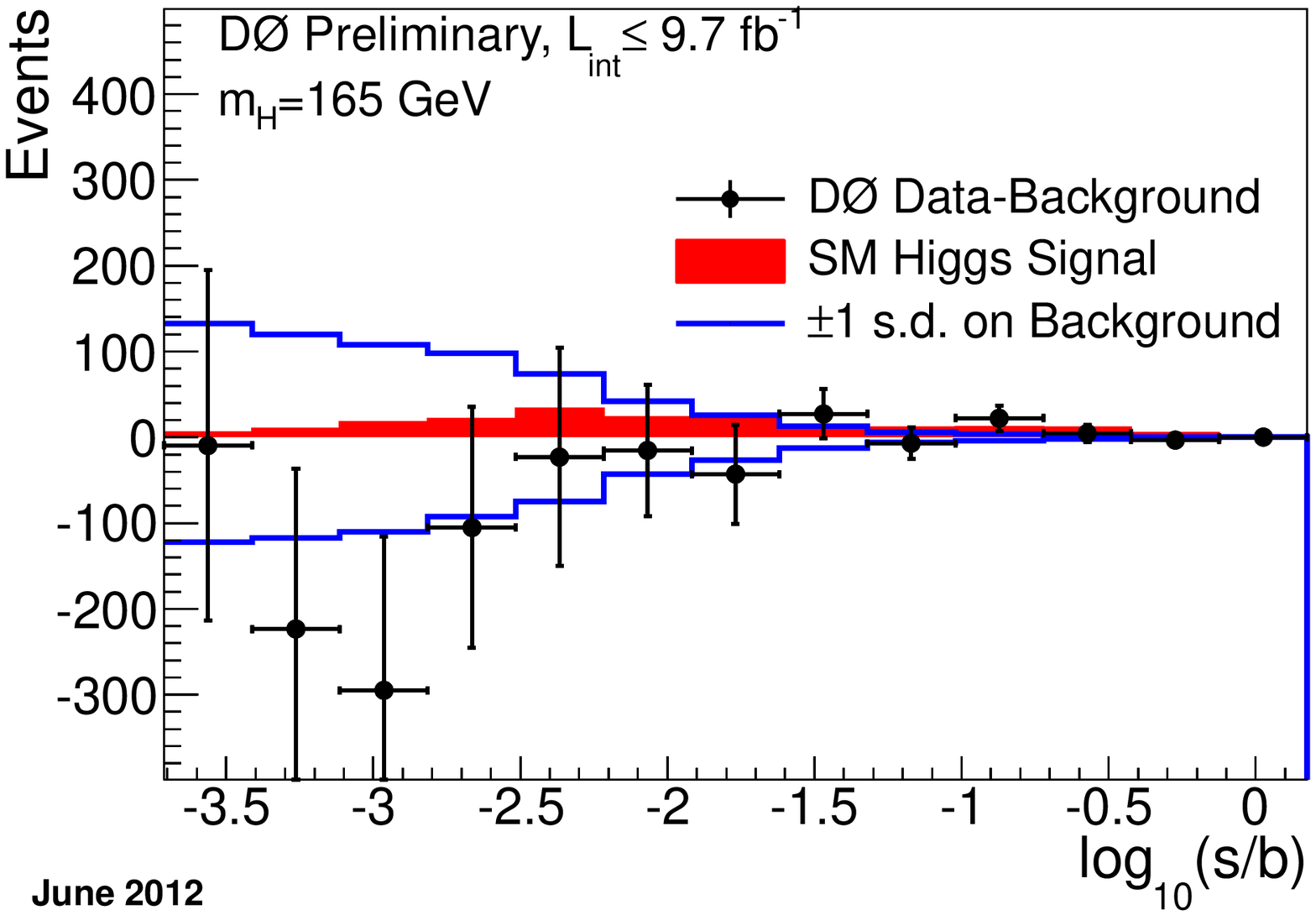}
\caption{
\label{fig:subtractInputs} Background subtracted distributions of $\log_{10}(s/b)$ for the data from all
contributing channels for Higgs boson masses of 115\gev, 125\gev, and
165\gev.  The background subtracted data are shown as points and the
expected signal is the red histogram. The blue lines indicate the uncertainty on the background prediction.}
\end{centering}
\end{figure}

\begin{figure}[htbp]
\begin{centering}
\includegraphics[width=0.45\textwidth]{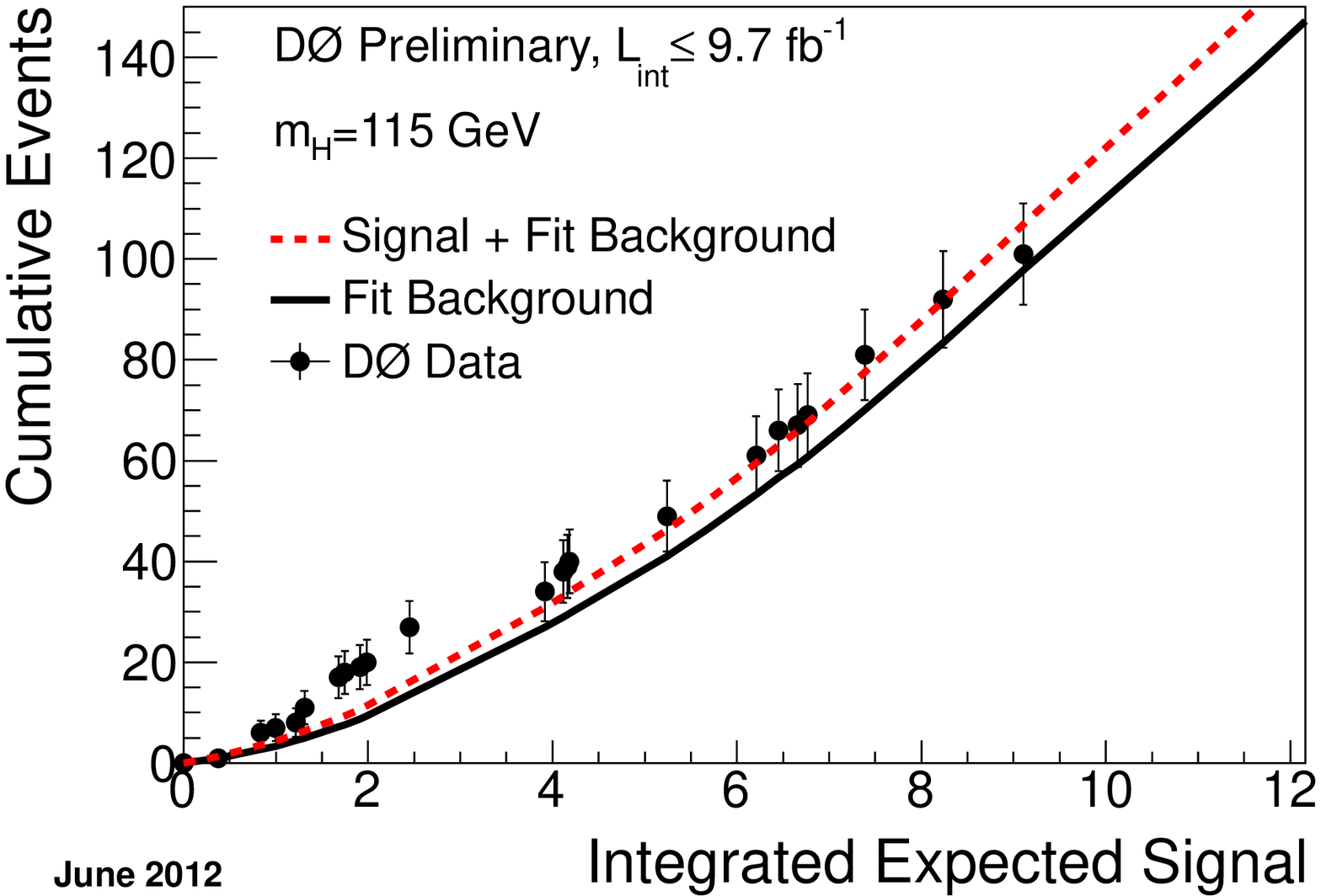}
\includegraphics[width=0.45\textwidth]{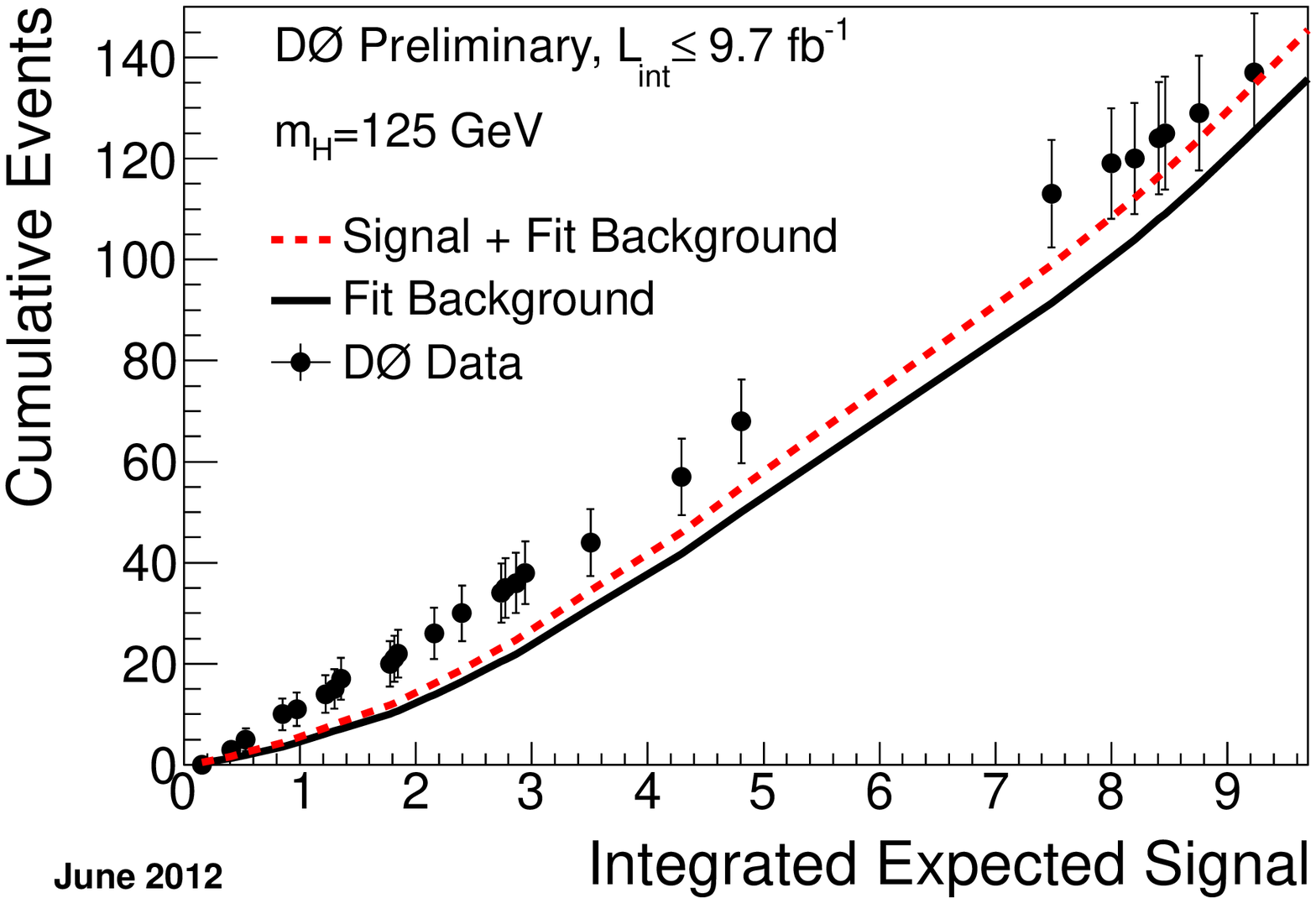}
\includegraphics[width=0.45\textwidth]{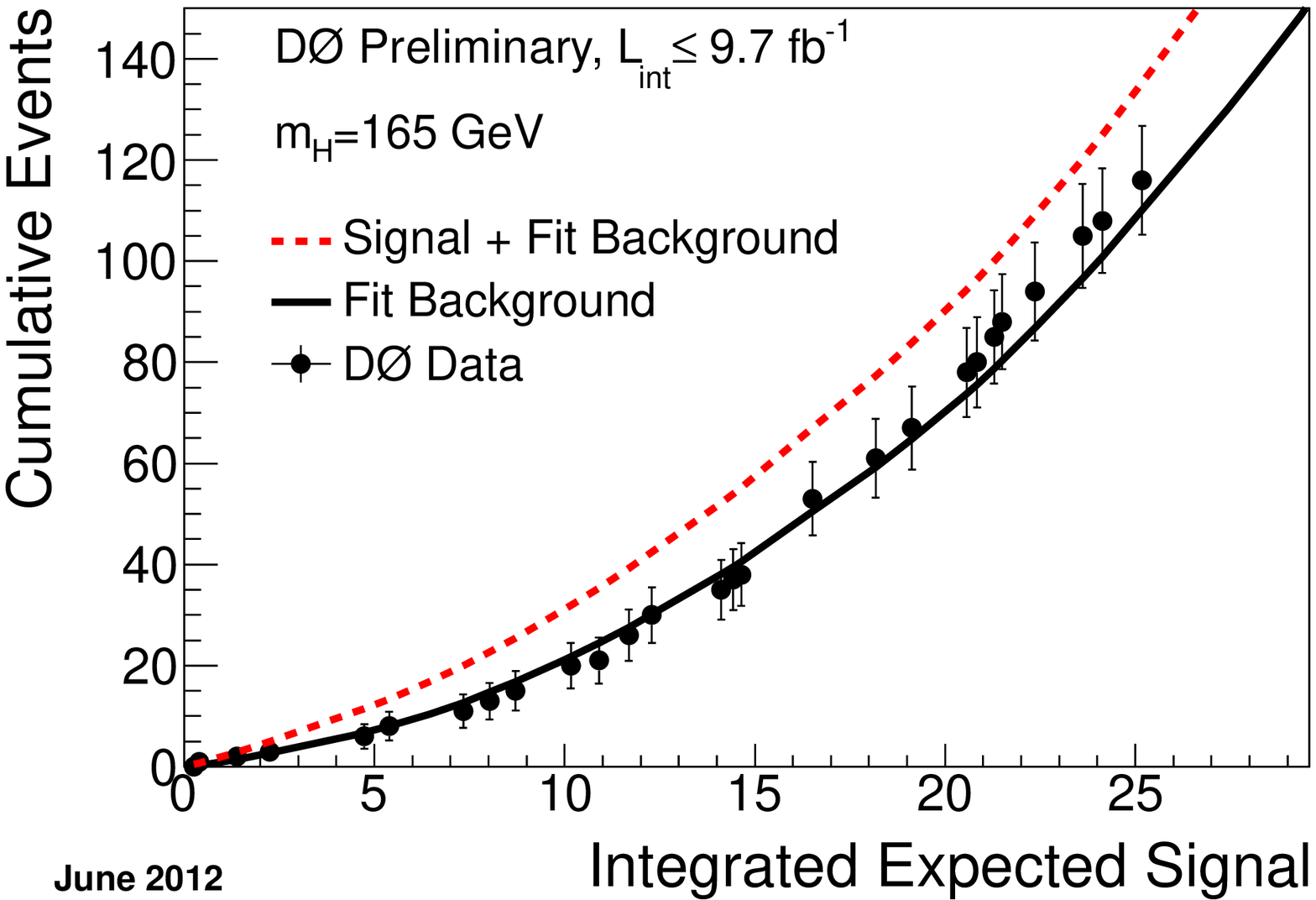}
\caption{
\label{fig:integralInputs} Cumulative number of events for the highest $s/b$ bins from all
contributing channels for Higgs boson masses of 115\gev, 125\gev, and
165\gev.  The integrated background-only and signal+background
predictions are shown as a function of the accumulated number of
signal events.  The points show the integrated number of observed
events, including only the statistical uncertainty, which is correlated point-to-point.
Systematic uncertainties on the integrated background-only and signal+background
predictions are not displayed.}
\end{centering}
\end{figure}

\subsection{Systematic Uncertainties}
\label{sec:systs}

The systematic uncertainties differ between analyses for both the
signals and backgrounds~\citeall.  Here we summarize only the largest
contributions.  Most analyses carry an uncertainty on the integrated
luminosity of 6.1\%~\cite{lumi}, while the overall normalization of
other analyses is determined from the NNLO $Z/\gamma^*$ cross section
in data events near the peak of $Z\rightarrow  \ell\ell$ decays.
The \hbb~analyses have an uncertainty on the $b$-tagging rate of 1-10\%.
These analyses also have an uncertainty on the jet measurement and acceptances of $\sim 7\%$.
All analyses include uncertainties associated with lepton measurement
and acceptances, which range from 1-9\% depending on the final state.
The largest contribution for all analyses is the uncertainty on the
background cross sections at 4-30\% depending on the analysis channel
and specific background. These values include both the uncertainty on
the theoretical cross section calculations and the uncertainties on
the higher-order correction factors. The uncertainty on the expected
multijet background is dominated by the statistics of the data sample
from which it is estimated, and is considered separately from the
other cross section uncertainties.
All analyses take into account the uncertainties on the theoretical production
cross sections for the different signal processes due to PDF 
model and scale choice.  
The $H\rightarrow W^+W^- \rightarrow \ell^{+} \nu \ell^{-} \nu$
($\ell=e, \mu$) analyses divide the data by jet multiplicity and, as
discussed, apply different uncertainties on the gluon-gluon fusion
Higgs boson production cross section for each jet multiplicity final
state. In addition, several analyses incorporate uncertainties that
alter the differential distributions and kinematics of the dominant
backgrounds in the analyses.  These shapes are derived from the
potential variations of the final variables due to generator and
background modeling uncertainties. Further details on the systematic
uncertainties are given in Appendix~\ref{app:syst}.

In much of the phase space, the systematic uncertainties for
background rates are several times larger than the signal expectation
itself and are an important factor in the calculation of limits.  Each
systematic uncertainty is folded into the signal and background
expectations in the limit calculation via Gaussian
distributions. These Gaussian values are sampled for each Monte Carlo
(MC) trial (pseudo-experiment) using Poisson distributions for the
number of signal and background events.  Several of the systematic
uncertainties, for example the jet energy scale uncertainty, typically
impact the shape of the final variable.  These variations in the final
variable distributions were preserved in the description of systematic
fluctuations for each Poisson trial. Correlations between systematic
sources are carried through in the calculation.  For example, the
uncertainty on the integrated luminosity is held to be correlated
between all signals and backgrounds and, thus, the same fluctuation in
the luminosity is common to all channels for a single
pseudo-experiment. All systematic uncertainties originating from a
common source are held to be correlated, as detailed in
Table~\ref{tab:corr}.

To minimize the degrading effects of systematic uncertainties on the
search sensitivity, the individual background contributions are fitted
to the data observation by maximizing a likelihood function for each
hypothesis~\cite{pflh}. The likelihood is a joint Poisson probability
over the number of bins in the calculation and is a function of the
nuisance parameters in the system and their associated uncertainties,
which are given an additional Gaussian constraint associated with
their prior predictions.  The maximization of the likelihood function
is performed over the nuisance parameters. A fit is performed to both
the background-only (b) and signal-plus-background (s+b) hypothesis
separately for each Poisson MC trial.

\begin{table}[htpb]
\caption{\label{tab:corr}The correlation matrix for the analysis
channels. All uncertainties within a group are considered 100\%
correlated across channels.  The correlated systematic uncertainty on
the background cross section ($\sigma$) is itself subdivided according
to the different background processes in each analysis. }
\begin{ruledtabular}
\begin{tabular}{lcccccccc}\\
Source                     & \whl\      & \zhv\        & \zhl\      &  \hwwlvlv\  \\ \hline
Luminosity                 & $\times$   & $\times$   &            &             \\
Jet Energy Scale           & $\times$   & $\times$   & $\times$   & $\times$    \\
Jet ID                     & $\times$   & $\times$   & $\times$   & $\times$    \\
Tau Energy Scale/ID        &            &            &            &             \\
Electron ID/Trigger        & $\times$   & $\times$   & $\times$   & $\times$    \\
Muon ID/Trigger            & $\times$   & $\times$   & $\times$   & $\times$    \\
Photon ID/Trigger          &            &            &            &             \\
$b$ Jet Tagging            & $\times$   & $\times$   & $\times$   &             \\
Background $\sigma$        & $\times$   & $\times$   & $\times$   & $\times$    \\
Background Modeling        &            &            &            &             \\
Multijet                   &            &            &            &             \\
Signal $\sigma$            & $\times$   & $\times$   &  $\times$  & $\times$    \\
Signal modeling            &            &            &            & $\times$    \\
\hline
\\
Source                     & \vhvww\    & \lnuqqqq    &  \hwwlnuqq\ & \hwwmvtv\     & \hgg\   \\ \hline
Luminosity                 &            & $\times$   & $\times$   & $\times$   & $\times$    \\
Jet Energy Scale           & $\times$   & $\times$   & $\times$   & $\times$   &             \\
Jet ID                     & $\times$   & $\times$   & $\times$   & $\times$   &             \\
Tau Energy Scale/ID        &            &            &            & $\times$   &             \\
Electron ID/Trigger        & $\times$   & $\times$   & $\times$   & $\times$   &             \\
Muon ID/Trigger            & $\times$   & $\times$   & $\times$   & $\times$   &             \\
Photon ID/Trigger          &            &            &            &            & $\times$    \\
$b$ Jet Tagging            &            &            &            &            &             \\
Background $\sigma$        & $\times$   & $\times$   & $\times$   & $\times$   &             \\
Background Modeling        &            &            &            &            &             \\
Multijet                   &            &            &            &            &             \\
Signal $\sigma$            & $\times$   & $\times$   & $\times$   & $\times$   & $\times$    \\
Signal modeling            & $\times$   & $\times$   & $\times$   & $\times$   & $\times$    \\
\end{tabular}
\end{ruledtabular}
\end{table}

\section{Results}

We derive limits on SM Higgs boson production $\sigma \times BR(H
\rightarrow$$ b\bar{b}/ W^{+}W^{-}/ \tau^{+}\tau^{-}/\gamma\gamma$)
via individual channels~\citeall. 
The relative contributions of the different production and decay modes
are set to the SM prediction.
To facilitate model
transparency and to accommodate analyses with different degrees of
sensitivity, we present our results in terms of the ratio of 95\% C.L. upper
cross section limits to the SM predicted cross section as a function of
Higgs boson mass. The SM prediction for Higgs boson production would
therefore be considered excluded at 95\% C.L. when this limit ratio falls
below unity.

The individual analyses described in Table~\ref{tab:chans} are grouped to
evaluate combined limits over the range $100 \leq m_{H} \leq 200$\gev.  
The \zhl, \zhv, \whl~and \hgg~ analyses contribute for $m_{H} \leq
150$\gev, the \lll,~\ttm\ and \lnuqqqq\ analyses contribute for $m_{H} \geq
100$\gev, the \ssem~and \hwwlnulnu~analyses contribute for $m_{H} \geq 115$\gev,
and the \hwwlnuqq~analysis contributes for $m_{H} \geq 155$\gev.

Figure~\ref{fig:allHI} shows the expected and observed 95\% C.L. cross
section limits as a ratio to the SM cross section and for the probed
mass region ($100 \leq m_H \leq 200$\gev), with all analyses
combined. These results are also summarized in
Table~\ref{tab:limits}. The LLR distributions for the full combination
are shown in Fig.~\ref{fig:allLLR}. Included in these figures are the
median LLR values for the signal-plus-background hypothesis
(LLR$_{s+b}$), background-only hypothesis (LLR$_{b}$), and the
observed data (LLR$_{obs}$). The shaded bands represent the 1 and 2
standard deviation ({\rm $\sigma$}) departures for LLR$_{b}$. These
distributions can be interpreted as follows:

\begin{itemize}
\item The separation between LLR$_{b}$ and LLR$_{s+b}$ provides a
measure of the discriminating power of the search.  This is the ability of
the analysis to separate the $s+b$ and $b-$only hypotheses.

\item The width of the LLR$_{b}$ distribution (shown here as one and
two standard deviation ({\rm $\sigma$}) bands) provides an estimate of
how sensitive the analysis is to a signal-like background fluctuation
in the data, taking account of the presence of systematic
uncertainties.  For example, the analysis sensitivity is limited when
a 1{\rm $\sigma$} background fluctuation is large compared to the
signal expectation.

\item The value of LLR$_{obs}$ relative to LLR$_{s+b}$ and LLR$_{b}$
indicates whether the data distribution appears to be more like
signal-plus-background or background-only.  As noted above, the
significance of any departures of LLR$_{obs}$ from LLR$_{b}$ can be
evaluated by the width of the LLR$_{b}$ distribution.

\end{itemize}

\noindent Figure~\ref{fig:allCLS} illustrates the exclusion criterion
$1-CL_{s}$ for the region $100\leq m_{H} \leq 200$\gev.  We provide in
Fig.~\ref{fig:allCLSB} the values for the observed $CL_{s+b}$ and its
expected distribution as a function of $m_H$. The quantity $CL_{s+b}$
is the $p$-value for the signal-plus-background
hypothesis. Figure~\ref{fig:allCLB} contains the values for the
observed 1-$CL_{b}$, which is the $p$-value for the background-only
hypothesis.  These probabilities are local $p$-values, corresponding
to searches for each value of $m_H$ separately, thus they do not
include the look-elsewhere-effect (LEE).  These two $p$-values
($CL_{s+b}$ and 1-$CL_{b}$) each provide information on the
compatibility of their respective hypothesis with the observed data.
Small values indicate rejection of the hypothesis and values near
unity indicate general agreement between the hypothesis in question
and the data.  As can be seen in Figure~\ref{fig:allCLSB}, the
observed value of $CL_{s+b}$ drops to $\approx$1\% for Higgs boson
masses near 160\gev, indicating very small compatibility with the
signal-plus-background hypothesis in this mass range. In contrast, the
observed value of $CL_{s+b}$ is close to 0.5 for 120\gev$ \le m_{H}
\le 140$\gev, favoring the hypothesis of a signal in that mass range.
At $m_{H}=$ 135~(120)\gev, the local $p$-value of 1-$CL_{b}$ is
\getvar{135localpval}\ (\getvar{120localpval}), corresponding to
\getvar{135localzval}\ (\getvar{120localzval}) Gaussian standard
deviations above the background-only prediction.

We estimate the LEE effect as discussed below (see Ref.~\cite{Tevcomb} for
more details).  In the mass range
100--135\gev, where the low-mass $H\rightarrow b{\bar{b}}$ searches
dominate, the reconstructed mass resolution is 10--15\%.  We therefore
estimate a trials factor of approximately two for the low-mass region.
For the high-mass searches, the $H\rightarrow W^+W^-$ searches
dominate the sensitivity.  There is little-to-no resolution in
reconstructing $m_H$ in these channels due to the presence of two
neutrinos in the final state of the most sensitive analyses.  We
expect a trials factor of approximately two for the high-mass
searches.  In total, we expect that there are roughly four possible
independent locations for uncorrelated excesses to appear in our
analysis.  The global $p$-value is therefore $1-(1-p_{\rm{min}})^4$,
using the Dunn-\^Sid\'ak correction~\cite{dunn}, where $p_{\rm{min}}$ is
the smallest local $p$-value found as a function of $m_H$.  The global
significance for such an excress anywhere in the full mass range is
estimated to be approximately $\getvar{135LEEzval}$\ standard deviations.

As a further investigation of this deviation from the background-only
hypothesis, we present in Figure~\ref{fig:comboXsec} the distribution
of the best-fit Higgs boson signal cross section ratio to the SM
prediction ($\sigma^{\rm Fit}/\sigma^{\rm SM}$).  This value is
obtained by performing a maximum likelihood fit over all search
channels simultaneously, in which the fit is allowed to vary all
nuisance parameters within their priors and with the Higgs boson
signal rate as a free parameter.  The result of this fit, shown along
with the $\pm1$ standard deviation distribution from the fit, yields a
best-fit signal rate of roughly 1.5 times the SM Higgs boson predicted
cross section for masses between 120\gev\ and 140\gev.  And as
expected from Figs~\ref{fig:allHI}-\ref{fig:allCLB}, there is also an
excursion from zero cross section near $m_{H}=200$\gev.  
However, the excursion from the background hypothesis at $m_{H}=200$\gev\ 
has a shape incompatible with that expected from a SM Higgs signal given the 
mass resolution, and is less significant than the excess between 120\gev\ and 140\gev,
so it will not be discussed further in the following.
We also explore the compatibility of the excess with the presence
of a signal.  Fig. \ref{fig:SignalInjection} compares the LLR obtained from 
the data to the expectation from the signal+background hypothesis for $m_H=125$\gev,
at the rate predicted by the SM, and at the best fit signal rate.  This test produces
a broad negative excursion in the LLR that is similar to the observation in the data.

The low mass excesses can be studied by separating the contributing
sources by Higgs boson decay: $H\rightarrow b\bar{b}$ and
$H\rightarrow W^+W^-$.  Figures~\ref{fig:hbbLLR} and~\ref{fig:hwwLLR}
show the LLR value for $H\rightarrow b\bar{b}$ and $H\rightarrow
W^+W^-$ final states, respectively.  Figure~\ref{fig:hbbLLR} includes
contributions from $ZH\rightarrow \ell\ell b\bar{b}$, $ZH\rightarrow
\nu\nu b\bar{b}$ and $WH\rightarrow \ell\nu b\bar{b}$ searches, and
illustrates a small data excess ($\sim1$ standard deviation above
expected background) that is nonetheless compatible with the SM Higgs
boson rate for $120 \leq m_{H} \leq 135$\gev.  Figure~\ref{fig:hwwLLR}
includes contributions from $H\rightarrow W^+W^- \rightarrow
\ell\nu\ell\nu$, $H\rightarrow W^+W^- \rightarrow \ell\nu jj$ and $VH
\rightarrow W^+W^-/ZZ$ searches, and shows a general excess of data
somewhat larger than the background prediction for $m_{H} \leq
140$\gev.  Figures~\ref{fig:hbbLimit} and~\ref{fig:hwwLimit} show the
expected and observed 95\% C.L. cross section limits as a ratio to the
SM cross section for the probed mass region
%($100 \leq m_H \leq 200$\gev), 
for $H\rightarrow b\bar{b}$ and $H\rightarrow W^+W^-$ final
states, respectively.

\begin{table}[tp]
\caption{Combined 95\% C.L. expected (median) and observed limits on
  $\sigma \times BR(H$$\rightarrow$$ X)$ for SM Higgs boson
  production. The limits are reported in units of the SM production
  cross section times branching fraction.
\label{tab:limits}}
\begin{ruledtabular}
\begin{tabular}{lccccccccccccccccccccc}
$m_{H}$ &100 &105 &110 &115 &120 &125 &130 &135 &140 &145 &150 &155 &160 &165 &170 &175 &180 &185 &190 &195 &200 \\
Expected: &1.14 &1.24 &1.37 &1.46 &1.61 &1.70 &1.74 &1.67 &1.56 &1.39 &1.23 &1.07 &0.78 &0.72 &0.90 &1.06 &1.28 &1.62 &1.98 &2.31 &2.62 \\
Observed: &1.04 &1.52 &1.39 &2.11 &2.84 &2.94 &3.10 &3.14 &2.83 &2.05 &1.57 &1.39 &0.84 &0.73 &0.98 &1.57 &1.59 &1.95 &2.59 &3.22 &4.38 \\
\end{tabular}
\end{ruledtabular}
\end{table}

Appendix~\ref{app:moriond_compare} presents a comparison between the results documented in this note
and the previous results~\cite{M12dzcombo}. In general both sets of results are found to be consistent
given the improvements to sensitivity in the various contributing analyses.

\needspace{8\baselineskip}
\section{Conclusions}
We have presented a combination of searches for the standard model
Higgs boson at the D\O\ experiment using the full Run II data set of
\getvar{lumimax}~\ifb of \pp\ collisions at \tevE. These searches are carried 
out for Higgs boson masses ($m_H$) in the range $100\leq m_H \leq 200$\gev.
In most of the searched region, no significant departure of the data from the 
background estimation is found, and upper limits on the standard model
Higgs boson production cross section are derived as a function of $m_H$.
The observed 95\% C.L. upper limits are found to be a factor of
\getvar{obs115}\ (\getvar{obs165}) times the predicted standard model
cross section at $m_H=115~(165)$\gev, while the expected limit is
found to be a factor of \getvar{exp115}\ (\getvar{exp165}) times the
standard model prediction for the same mass(es). We exclude at the
95\% C.L. the region $\getvar{exclmin} <m_H<
\getvar{exclmax}$\gev~with an {\it a priori} expected exclusion of
$\getvar{exclminexp} <m_H<\getvar{exclmaxexp}$\gev.
In the mass range 120--140\gev, the data exhibit an excess above
the background prediction with a local significance of approximately 
two Gaussian standard deviations.  
The results presented here supersede the previous \DZ combination
results~\cite{M12dzcombo}.

\begin{acknowledgments}
% acknowledgement.tex                            9 February 2012 
%
We thank the staffs at Fermilab and collaborating institutions,
and acknowledge support from the
DOE and NSF (USA);
CEA and CNRS/IN2P3 (France);
MON, Rosatom and RFBR (Russia);
CNPq, FAPERJ, FAPESP and FUNDUNESP (Brazil);
DAE and DST (India);
Colciencias (Colombia);
CONACyT (Mexico);
NRF (Korea);
FOM (The Netherlands);
STFC and the Royal Society (United Kingdom);
MSMT and GACR (Czech Republic);
BMBF and DFG (Germany);
SFI (Ireland);
The Swedish Research Council (Sweden);
and
CAS and CNSF (China).

\end{acknowledgments}

\newpage

\begin{figure}[p]
\psfrag{m}{{\boldmath $M$}}
\begin{centering}
\includegraphics[width=11.5cm]{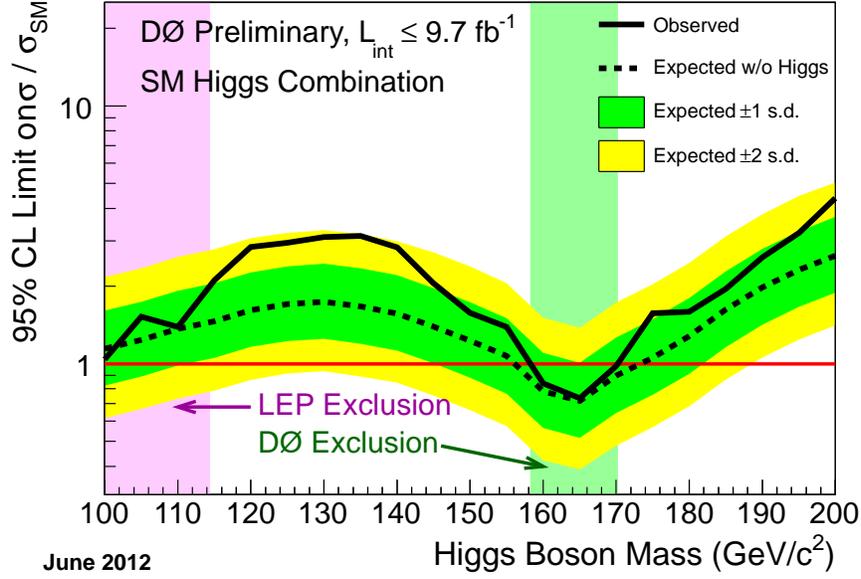}
\caption{
\label{fig:allHI}
Expected (median) and observed 95\% C.L. cross section upper limit
ratios for the combined $WH/ZH/H,
H$$\rightarrow$$b\bar{b}/W^+W^-/\gamma\gamma/\tau^+\tau^-$ analyses
over the $100 \leq m_H \leq 200$\gev~mass range.  The green and yellow
bands correspond to the regions enclosing 1 and 2 standard deviation
fluctuations of the background, respectively.}
\end{centering}
\end{figure}

\begin{figure}[p]
\psfrag{m}{{\boldmath $M$}}
\begin{centering}
\includegraphics[width=11.5cm]{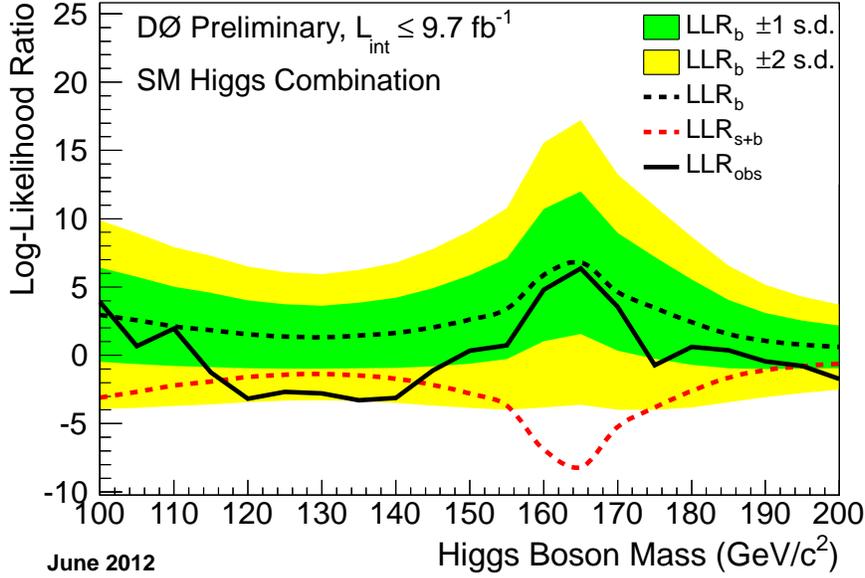}
\caption{
\label{fig:allLLR}
Log-likelihood ratio distribution for the combined $WH/ZH/H, H$$\rightarrow$$ b\bar{b}/W^+W^-/\gamma\gamma/\tau^+\tau^-$ analyses over the $100 \leq m_H \leq 200$\gev~mass range.  The green and yellow bands correspond to the
regions enclosing 1 and 2 standard deviation
fluctuations of the background, respectively.}
\end{centering}
\end{figure}

\begin{figure}[p]
\psfrag{m}{{\boldmath $M$}}
\begin{centering}
\includegraphics[width=11.5cm]{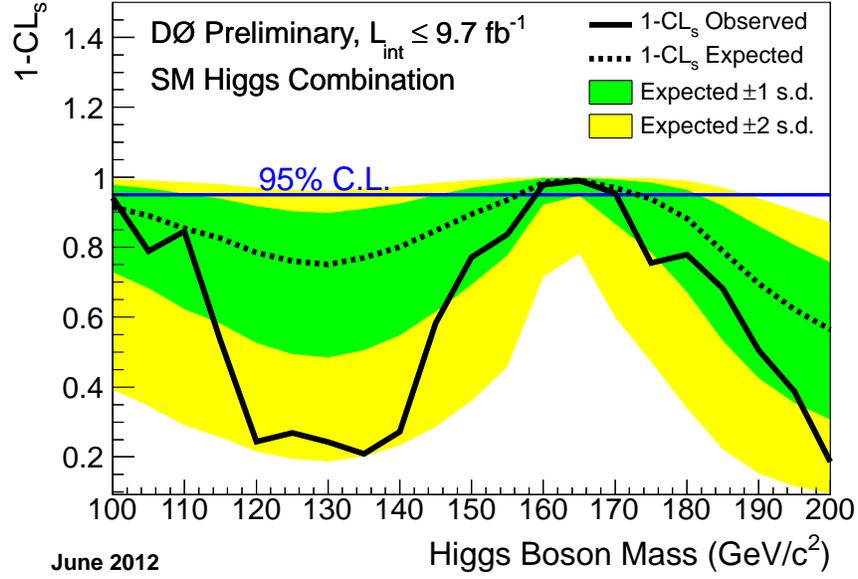}
\caption{
\label{fig:allCLS}
The $1-CL_{S}$ (exclusion probability) distribution for the combined
$WH/ZH/H, H$$\rightarrow$$ b\bar{b}/W^+W^-/\gamma\gamma/\tau^+\tau^-$
analyses over the $100 \leq m_H \leq 200$\gev~mass range.  The green and yellow bands correspond to the
regions enclosing 1 and 2 standard deviation
fluctuations of the background, respectively.}
\end{centering}
\end{figure}

\begin{figure}[p]
\psfrag{m}{{\boldmath $M$}}
\begin{centering}
\includegraphics[width=11.5cm]{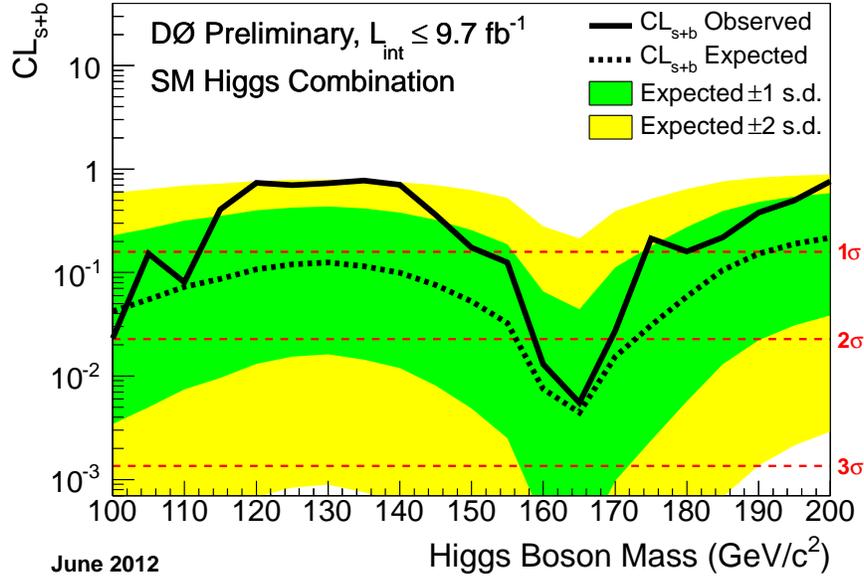}
\caption{
\label{fig:allCLSB}
The $CL_{s+b}$ (signal-plus-background $p$-value) distribution for the combined
$WH/ZH/H, H$$\rightarrow$$ b\bar{b}/W^+W^-/\gamma\gamma/\tau^+\tau^-$
analyses over the $100 \leq m_H \leq 200$\gev~mass range.  The green and yellow bands correspond to the
regions enclosing 1 and 2 standard deviation
fluctuations of the background, respectively. The three
horizontal dashed lines indicate the p-values corresponding to
significances of 1, 2 and 3 standard deviations.}
\end{centering}
\end{figure}

\begin{figure}[p]
\psfrag{m}{{\boldmath $M$}}
\begin{centering}
\includegraphics[width=11.5cm]{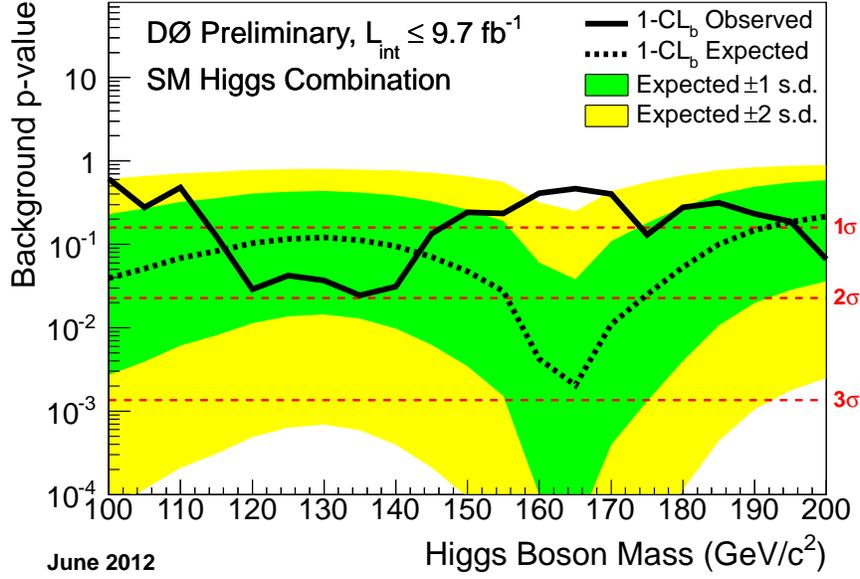}
\caption{
\label{fig:allCLB}
The $1-CL_{b}$ (background $p$-value) distribution for the combined
$WH/ZH/H, H$$\rightarrow$$ b\bar{b}/W^+W^-/\gamma\gamma/\tau^+\tau^-$
analyses over the $100 \leq m_H \leq 200$\gev~mass range. Also shown
is the expected background $p$-value for the SM Higgs boson signal
(dotted line). The three horizontal dashed lines indicate the p-values
corresponding to significances of 1, 2 and 3 standard deviations.}
\end{centering}
\end{figure}

\begin{figure}[p]
\psfrag{m}{{\boldmath $M$}}
\begin{centering}
\includegraphics[width=11.5cm]{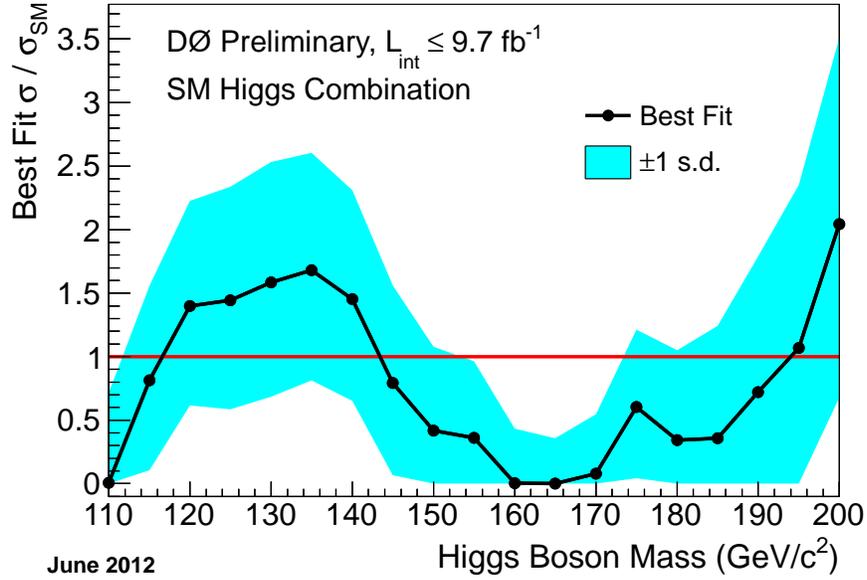}
\caption{
\label{fig:comboXsec} 
The best-fit signal cross section ratio to the standard model Higgs
boson prediction ($\sigma^{Fit}/\sigma^{SM}$) for the combined
$WH/ZH/H, H$$\rightarrow$$ b\bar{b}/W^+W^-/\gamma\gamma/\tau^+\tau^-$
analyses over the $100 \leq m_H \leq 200$\gev~mass range. This value
indicates the value of the Higgs boson cross section that would best
match the observed data in a global fit over all nuisance parameters.
The Higgs boson cross section is treated as a free parameter, bounded
at zero. The light-blue band indicates the $\pm1$ standard deviation
region from the fit.}
\end{centering}
\end{figure}

\begin{figure}[p]
\psfrag{m}{{\boldmath $M$}}
\begin{centering}
\includegraphics[width=11cm]{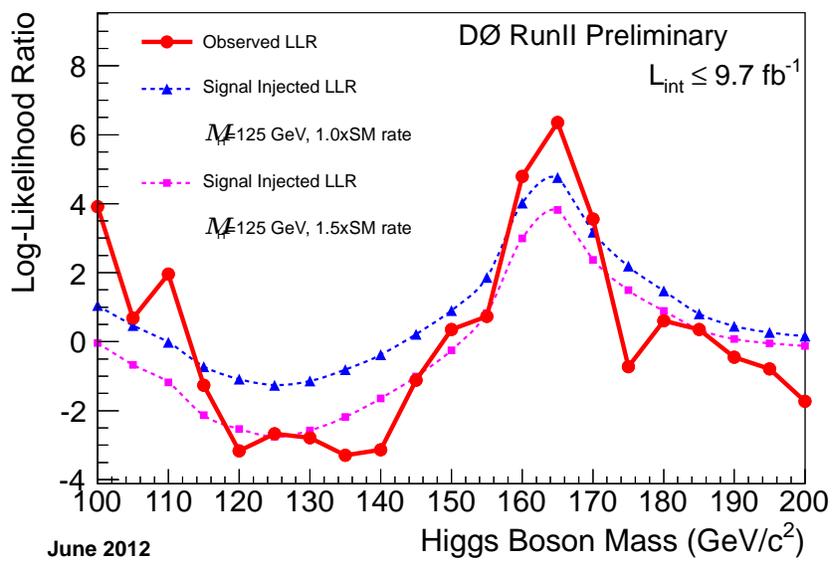}
\caption{Comparison of the LLR observed in the data with the LLR expected
in the presence of a Higgs boson with a mass of 125\gev~at the rate
predicted by the SM, and at a rate equal to 1.5 times the SM
prediction.}
\label{fig:SignalInjection}
\end{centering}
\end{figure}

\clearpage

\begin{figure}[p]
\psfrag{m}{{\boldmath $M$}}
\begin{centering}
\includegraphics[width=11.5cm]{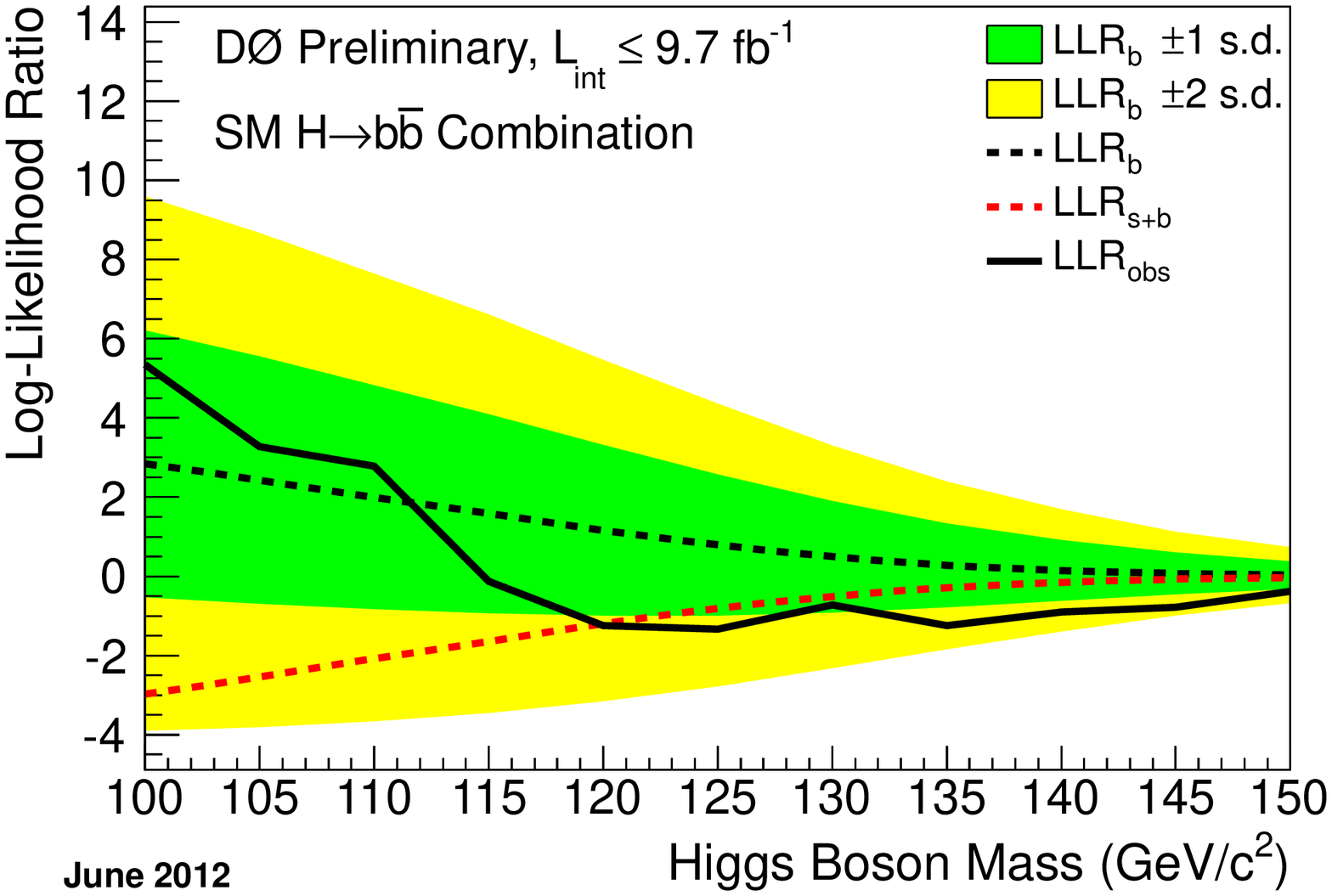}
\caption{Log-likelihood ratio distribution for the combined $WH/ZH,
  H$$\rightarrow$$ b\bar{b}$ analyses over the $100 \leq m_H \leq
  150$\gev~mass range.  The green and yellow bands correspond to the
  regions enclosing 1 and 2 standard deviation fluctuations of the
  background, respectively.
\label{fig:hbbLLR}}
\end{centering}
\end{figure}

\begin{figure}[p]
\psfrag{m}{{\boldmath $M$}}
\begin{centering}
\includegraphics[width=11.5cm]{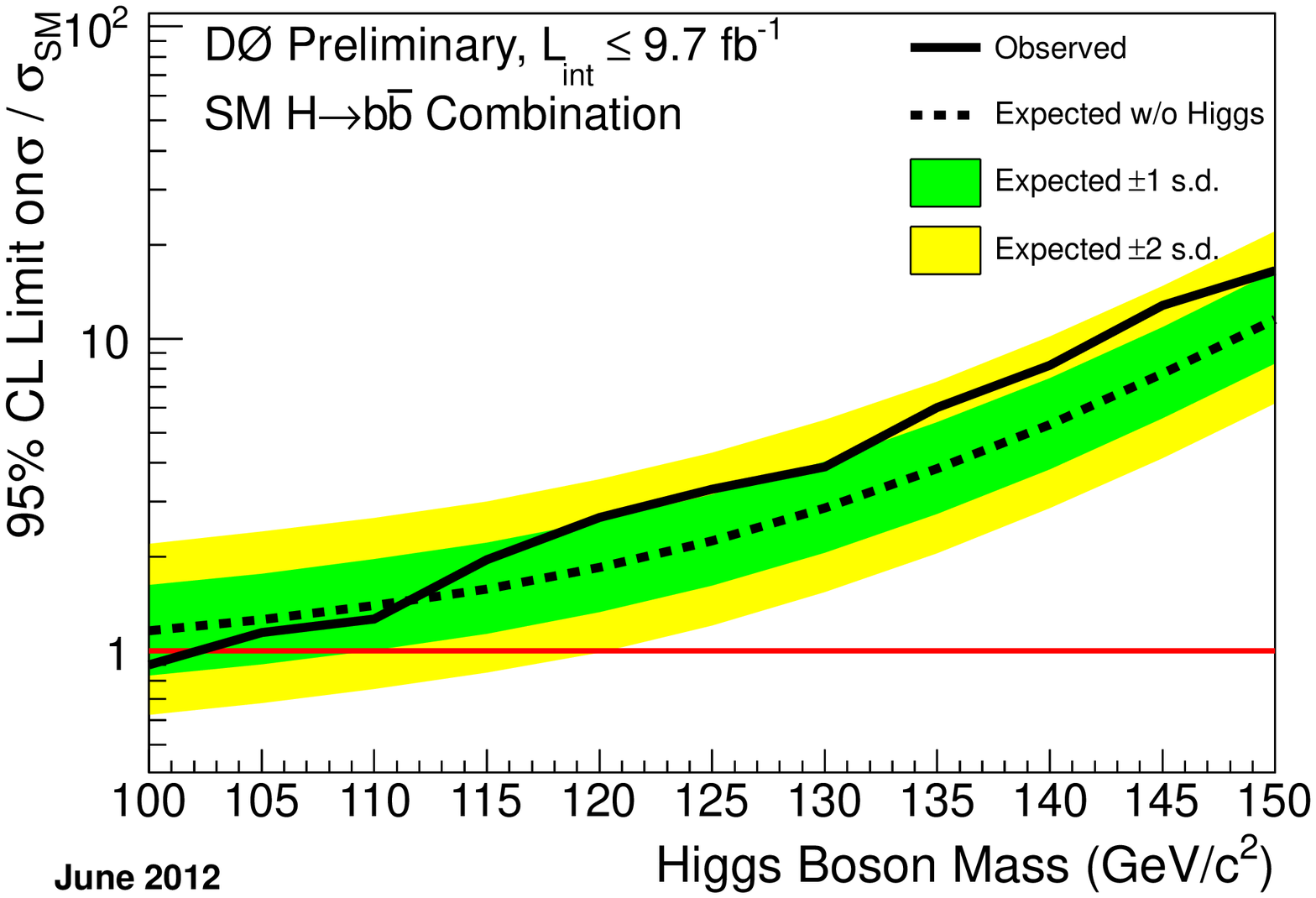}
\caption{
\label{fig:hbbLimit}
Expected (median) and observed 95\% C.L. cross section upper limit
ratios for the combined $WH/ZH, H$$\rightarrow$$ b\bar{b}$ analyses
over the $100 \leq m_H \leq 150$\gev~mass range.  The green and yellow
bands correspond to the regions enclosing 1 and 2 standard deviation
fluctuations of the background, respectively.}
\end{centering}
\end{figure}

\begin{table}[p]
\caption{Expected (median) and observed 95\% C.L. cross section upper
  limit ratios for the combined $WH/ZH, H$$\rightarrow$$ b\bar{b}$
  analyses over the $100 \leq m_H \leq 150$\gev~mass range.
\label{tab:HBBlimits}}
\begin{ruledtabular}
\begin{tabular}{lccccccccccc}
$m_{H}$ &100 &105 &110 &115 &120 &125 &130 &135 &140 &145 &150 \\
Expected: &1.16 &1.26 &1.40 &1.58 &1.85 &2.25 &2.87 &3.82 &5.31 &7.72 &11.53 \\
Observed: &0.90 &1.14 &1.26 &1.96 &2.67 &3.30 &3.89 &6.01 &8.23 &12.81 &16.52 \\
\end{tabular}
\end{ruledtabular}
\end{table}

\clearpage

\begin{figure}[p]
\psfrag{m}{{\boldmath $M$}}
\begin{centering}
\includegraphics[width=11cm]{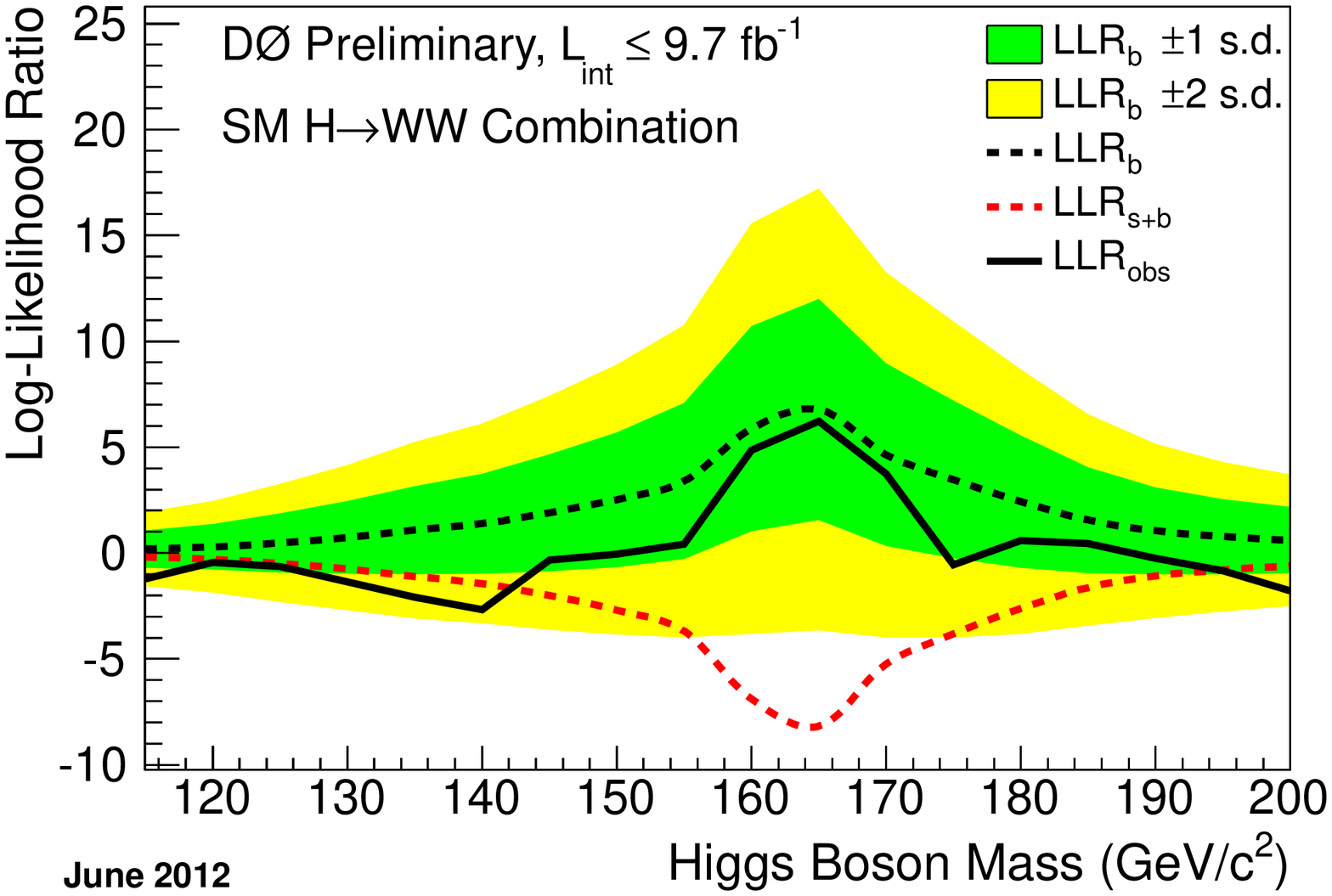}
\caption{Log-likelihood ratio distribution for the combined $WH/ZH/H,
  H$$\rightarrow$$ W^+W^-$ analyses over the $115 \leq m_H \leq
  200$\gev~mass range.  The green and yellow bands correspond to the
  regions enclosing 1 and 2 standard deviation fluctuations of the
  background, respectively.
\label{fig:hwwLLR}}
\end{centering}
\end{figure}

\begin{figure}[p]
\psfrag{m}{{\boldmath $M$}}
\begin{centering}
\includegraphics[width=11cm]{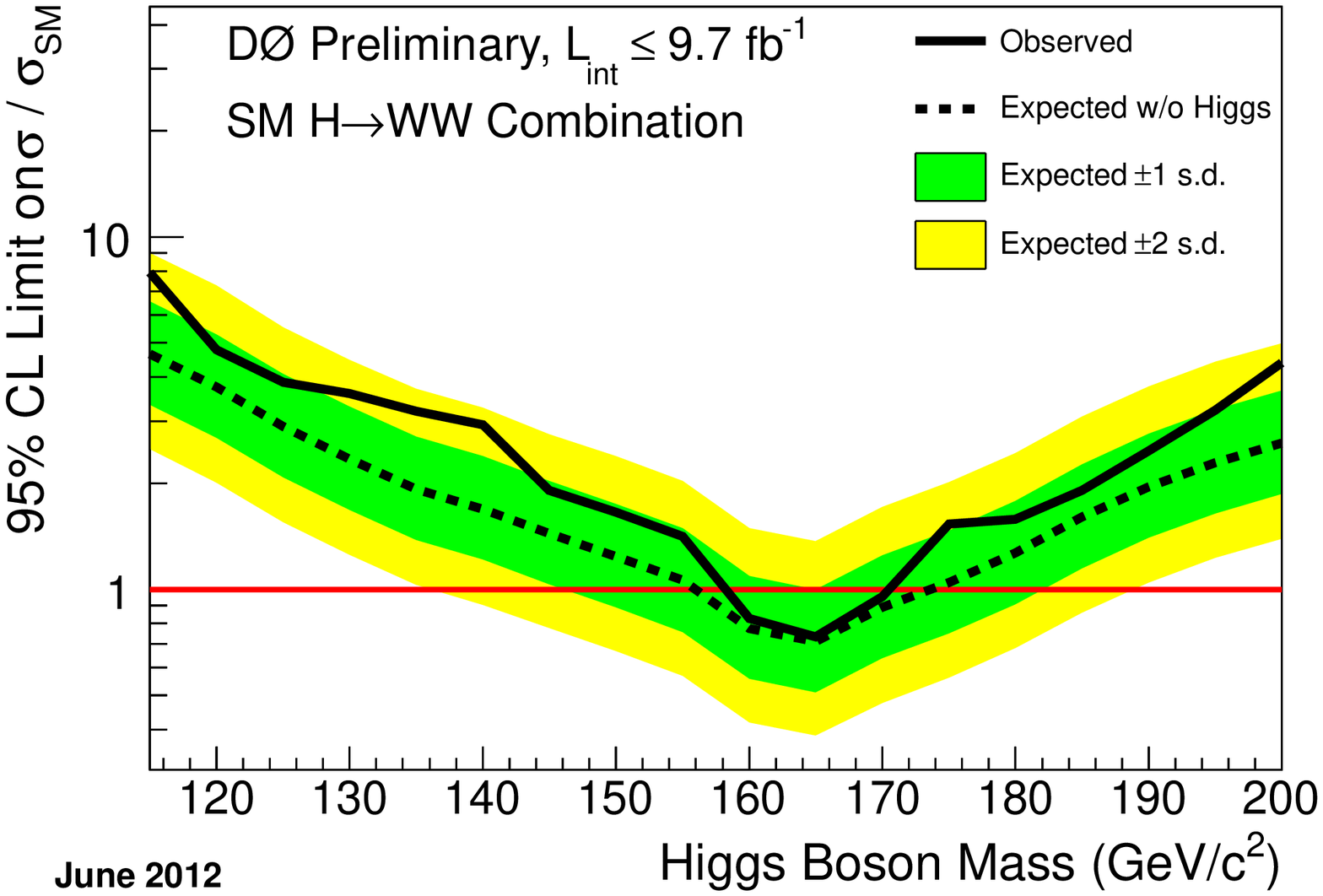}
\caption{
\label{fig:hwwLimit}
Expected (median) and observed 95\% C.L. cross section upper limit
ratios for the combined $WH/ZH/H, H$$\rightarrow$$ W^+W^-$ analyses
over the $115 \leq m_H \leq 200$\gev~mass range.  The green and yellow
bands correspond to the regions enclosing 1 and 2 standard deviation
fluctuations of the background, respectively.}
\end{centering}
\end{figure}

%\vspace{-1in}
\begin{table}[p]
\caption{Expected (median) and observed 95\% C.L. cross section upper
  limit ratios for the combined $WH/ZH/H, H$$\rightarrow$$ W^+W^-$
  analyses over the $115 \leq m_H \leq 200$\gev~mass range.
\label{tab:HWWlimits}}
\begin{ruledtabular}
\begin{tabular}{lccccccccccccccccccccc}
$m_{H}$  &115 &120 &125 &130 &135 &140 &145 &150 &155 &160 &165 &170 &175 &180 &185 &190 &195 &200 \\
Expected:  &5.81 &4.37 &3.20 &2.57 &2.09 &1.81 &1.54 &1.31 &1.10 &0.79 &0.72 &0.91 &1.07 &1.32 &1.68 &2.05 &2.43 &2.80 \\
Observed:  &10.59 &5.87 &4.59 &3.18 &3.42 &2.76 &1.89 &1.63 &1.41 &0.80 &0.74 &0.99 &1.60 &1.35 &1.87 &2.37 &3.02 &3.98 \\
\end{tabular}
\end{ruledtabular}
\end{table}

\clearpage
\appendix
\appendixpage
\addappheadtotoc

\section{Systematic Uncertainties}
\label{app:syst}

%%%%%%%%%%%% ZH LLBB   D0
\begin{table}[hp]
\begin{center}
\caption{\label{tab:d0llbb1}Systematic uncertainties on the contributions for 
D0's $ZH\rightarrow \ell^+\ell^-b{\bar{b}}$ channels.
Systematic uncertainties are listed by name; see the original references for a 
detailed explanation of their meaning and on how they
are derived.
Systematic uncertainties for $ZH$  shown in this table are obtained for $m_H=125$ GeV/$c^2$.
Uncertainties are relative, in percent, and are symmetric unless otherwise indicated. Shape uncertainties are 
labeled with an ``(S)''. }
%\label{tab:d0llbb1}
%\begin{footnotesize}

\vskip 0.5cm
{\centerline{$ZH \rightarrow \ell\ell b \bar{b}$ Single Tag (ST) channel relative uncertainties (\%)
in the $t\bar t$ depleted region}}
%{\centerline{D0: ~ $ZH \rightarrow \ell\ell b \bar{b}$ analysis relative uncertainties (\%)}}
\vskip 0.099cm
\begin{ruledtabular}
\begin{tabular}{  l  c  c  c  c  c  c  c  c }   
Contribution              & $ZH$  & Multijet& $Z$+l.f.  &  $Z+$\bb & $Z+$\cc & Dibosons & Top\\ \hline
Jet Energy Scale (S)      &  0.6  &   --    &  3.1      &  2.3   &  2.3  &  4.8   &  0.3  \\ %
Jet Energy Resolution (S) &  0.7  &   --    &  2.7      &  1.3   &  1.6  &  1.0   &  1.1  \\ %
Jet ID (S)                &  0.6  &   --    &  1.5      &  0.0   &  0.5  &  0.7   &  0.7  \\ %
Taggability (S)           &  2.0  &   --    &  1.9      &  1.7   &  1.7  &  1.8   &  2.2  \\ %
$Z p_T$ Model (S)         &   --  &   --    &  1.6      &  1.7   &  1.5  &   --   &   --  \\ %
HF Tagging Efficiency (S) &  0.5  &   --    &   --      &  1.6   &  3.9  &   --   &  0.7  \\ %
LF Tagging Efficiency (S) &   --  &   --    &   68      &   --   &   --  &  2.9   &   --  \\ %
$ee$ Multijet Shape (S)   &   --  &   45    &   --      &   --   &   --  &   --   &   --  \\ %
Multijet Normalization    &   --  &   10    &   --      &   --   &   --  &   --   &   --  \\ %
$Z$+jets Jet Angles (S)   &   --  &   --    &  1.7      &  1.7   &  1.7  &   --   &   --  \\ %
Alpgen MLM (S)            &   --  &   --    &  0.2      &   --   &   --  &   --   &   --  \\ %
Alpgen Scale (S)          &   --  &   --    &  0.3      &  0.5   &  0.5  &   --   &   --  \\ %
Underlying Event (S)      &   --  &   --    &  0.4      &  0.4   &  0.4  &   --   &   --  \\ %
Trigger (S)               & 0.4-2 &   --    &  0.03-2   &  0.2-2 &  0.2-2&  0.2-2&  0.5-2 \\ %
Cross Sections            & 6     &   --    &   --      &  20    &  20   &  7     &  10   \\ %
Signal Branching Fraction & 1-9   &   --    &   --      &  --    &  --   &  --    &  --   \\ %
Normalization             & 5     &   --    &  4        &  4     &  4    &  6     &  5    \\ %
PDFs                      & 0.6   &   --    &  1.0      &  2.4   &  1.1  &  0.7   &  5.9  \\ %

\end{tabular}
\end{ruledtabular}

\vskip 0.5cm
{\centerline{$ZH \rightarrow \ell\ell b \bar{b}$ Double Tag (DT) channel relative uncertainties (\%)
in the $t\bar t$ depleted region}}
\vskip 0.099cm
\begin{ruledtabular}
\begin{tabular}{  l  c c  c  c  c  c  c  c }  \\
Contribution             & $ZH$ & Multijet& $Z$+l.f.  &  $Z+$\bb & $Z+$\cc & Dibosons & Top\\  \hline
Jet Energy Scale (S)     &  0.5  &   --     &  4.6   &  3.0   &  1.3   &  4.5   &  1.4  \\ %
Jet Energy Resolution(S) &  0.4  &   --     &  7.0   &  1.8   &  2.9   &  0.9   &  0.9  \\ %
JET ID (S)               &  0.6  &   --     &  7.9   &  0.3   &  0.5   &  0.5   &  0.5  \\ %
Taggability (S)          &  1.7  &   --     &  7.0   &  1.5   &  1.5   &  3.0   &  1.7  \\ %
$Z_{p_T}$ Model (S)      &   --  &   --     &  2.9   &  1.4   &  1.9   &   --   &   --  \\ %
HF Tagging Efficiency (S)&  4.4  &   --     &   --   &  5.0   &  5.6   &   --   &  3.8  \\ %
LF Tagging Efficiency (S)&   --  &   --     &  75    &   --   &   --   &  4.7   &   --  \\ %
$ee$ Multijet Shape (S)  &   --  &    66    &   --   &   --   &   --   &   --   &   --  \\ %
Multijet Normalization   &   --  &   10     &   --   &   --   &   --   &   --   &   --  \\ %
$Z$+jets Jet Angles (S)  &   --  &   --     &  1.9   &  3.5   &  3.8   &   --   &   --  \\ %
Alpgen MLM (S)           &   --  &   --     &  0.2   &   --   &   --   &   --   &   --  \\ %
Alpgen Scale (S)         &   --  &   --     &  0.4   &  0.5   &  0.5   &   --   &   --  \\ %
Underlying Event(S)      &   --  &   --     &  0.5   &  0.4   &  0.4   &   --   &   --  \\ %
Trigger (S)              &  0.4-2&   --     &  0.6-6 &  0.3-2 &  0.3-3 &  0.4-2 &  0.6-5\\ %
Cross Sections           &  6    &   --     &   --   &  20    &  20    &  7     &  10   \\ %
Signal Branching Fraction& 1-9   &   --     &   --   &  --    &  --    &  --    &  --   \\ %
Normalization             & 5     &   --    &  4        &  4     &  4    &  6     &  5    \\ %
PDFs                     & 0.6   &   --     &  1.0   &  2.4   &  1.1   &  0.7   &  5.9  \\ %
\end{tabular}
\end{ruledtabular}

\end{center}
\end{table}

\begin{table}
\begin{center}
\vskip 0.5cm
{\centerline{$ZH \rightarrow \ell\ell b \bar{b}$ Single Tag (ST) channel relative uncertainties (\%)
in the $t\bar t$ enriched region}}
%{\centerline{D0: ~ $ZH \rightarrow \ell\ell b \bar{b}$ analysis relative uncertainties (\%)}}
\vskip 0.099cm
\begin{ruledtabular}
\begin{tabular}{  l  c  c  c  c  c  c  c  c }   
Contribution              & $ZH$  & Multijet& $Z$+l.f.  &  $Z+$\bb & $Z+$\cc & Dibosons & Top\\ \hline
Jet Energy Scale (S)      &  7.5  &   --    &  4.6      &  1.7   &  3.9   &  11   &  2.5  \\ %
Jet Energy Resolution (S) &  0.2  &   --    &  4.5      &  0.7   &  3.1  &  3.9   &  0.7  \\ %
Jet ID (S)                &  1.2  &   --    &  2.1      &  1.0   &  1.2  &  0.9   &  0.7  \\ %
Taggability (S)           &  2.1  &   --    &  7.3      &  2.7   &  3.0  &  2.0   &  3.2  \\ %
$Z p_T$ Model (S)         &   --  &   --    &  3.3      &  1.5   &  1.4  &   --   &   --  \\ %
HF Tagging Efficiency (S) &  0.5  &   --    &   --      &  1.3   &  4.8  &   --   &  0.8  \\ %
LF Tagging Efficiency (S) &   --  &   --    &   73      &   --   &   --  &  4.1   &   --  \\ %
$ee$ Multijet Shape (S)   &   --  &   59    &   --      &   --   &   --  &   --   &   --  \\ %
Multijet Normalization    &   --  &   10    &   --      &   --   &   --  &   --   &   --  \\ %
$Z$+jets Jet Angles (S)   &   --  &   --    &  1.7      &  2.3   &  2.7  &   --   &   --  \\ %
Alpgen MLM (S)            &   --  &   --    &  0.4      &   --   &   --  &   --   &   --  \\ %
Alpgen Scale (S)          &   --  &   --    &  0.7      &  0.7   &  0.7  &   --   &   --  \\ %
Underlying Event (S)      &   --  &   --    &  0.9      &  1.1   &  1.1  &   --   &   --  \\
Trigger (S)               & 1-4   &   --    &  1-4      &  0.7-4 &  0.7-4&  1-8   &  1-8  \\ %
Cross Sections            & 6     &   --    &   --      &  20    &  20   &  7     &  10   \\ %
Signal Branching Fraction & 1-9   &   --    &   --      &  --    &  --   &  --    &  --   \\ %
Normalization             & 5     &   --    &  4        &  4     &  4    &  6     &  5    \\ %
PDFs                      & 0.6   &   --    &  1.0      &  2.4   &  1.1  &  0.7   &  5.9  \\ %

\end{tabular}
\end{ruledtabular}

\vskip 0.5cm
{\centerline{$ZH \rightarrow \ell\ell b \bar{b}$ Double Tag (DT) channel relative uncertainties (\%)
in the $t\bar t$ enriched region}}
\vskip 0.099cm
\begin{ruledtabular}
\begin{tabular}{  l  c c  c  c  c  c  c  c }  \\
Contribution             & $ZH$ & Multijet& $Z$+l.f.  &  $Z+$\bb & $Z+$\cc & Dibosons & Top\\  \hline
Jet Energy Scale (S)     &  6.6  &   --    &  0.8     &  1.6   &  2.2   &  5.9   &  1.5  \\ %
Jet Energy Resolution(S) &  1.4  &   --    &  267     &  1.4   &  2.1   &  4.0   &  0.4  \\ %
JET ID (S)               &  0.9  &   --    &  0.6     &  0.5   &  3.6   &  2.8   &  0.6  \\ %
Taggability (S)          &  2.0  &   --    &  0.9     &  1.6   &  1.9   &  3.1   &  2.1  \\ %
$Z_{p_T}$ Model (S)      &   --  &   --    &  1.8     &  1.4   &  1.5   &   --   &   --  \\ %
HF Tagging Efficiency (S)&  4.0  &   --    &   --     &  5.1   &  6.6   &   --   &  4.2  \\ %
LF Tagging Efficiency (S)&   --  &   --    &  72      &   --   &   --   &   --   &   --  \\ %
$ee$ Multijet Shape (S)  &   --  &    91   &   --     &   --   &   --   &   --   &   --  \\ %
Multijet Normalization   &   --  &   10    &   --     &   --   &   --   &   --   &   --  \\ %
$Z$+jets Jet Angles (S)  &   --  &   --    &  1.4     &  3.7   &  2.3   &   --   &   --  \\ %
Alpgen MLM (S)           &   --  &   --    &  0.5     &   --   &   --   &   --   &   --  \\ %
Alpgen Scale (S)         &   --  &   --    &  0.8     &  0.5   &  0.4   &   --   &   --  \\ %
Underlying Event(S)      &   --  &   --    &  0.9     &  0.7   &  0.5   &   --   &   --  \\
Trigger (S)              &  1-3  &   --    &  1-3     &  0.6-3 &  0.7-4 &  0.7-4 &  1-3  \\ %
Cross Sections           &  6    &   --    &   --     &  20    &  20    &  7     &  10   \\ %
Signal Branching Fraction& 1-9   &   --    &   --     &  --    &  --    &  --    &  --   \\ %
Normalization             & 5     &   --    &  4        &  4     &  4    &  6     &  5    \\ %
PDFs                     & 0.6   &   --    &  1.0   &  2.4   &  1.1   &  0.7   &  5.9  \\ %
\end{tabular}
\end{ruledtabular}

\end{center}
%\end{footnotesize}
%\end{center}
\end{table}

\begin{table}[hp]
\caption{\label{tab:d0vvbb} Systematic uncertainty ranges on the signal and background 
contributions and the error on the total background
for D0's $ZH\rightarrow\nu\nu b{\bar{b}}$ medium-tag and tight-tag channels.
Systematic uncertainties are listed by name, see the original
references for a detailed explanation of their meaning and on how they
are derived.  Systematic uncertainties for $VH$ ($WH$+$ZH$) shown in
this table are obtained for $m_H=115$ GeV/$c^2$. Uncertainties are
relative, in percent, and are symmetric unless otherwise indicated.
Shape uncertainties are labeled with an ``(S)'', and ``SH'' represents
shape only uncertainty.}
%\begin{footnotesize}
\vskip 0.2cm
%%%%%%%
{\centerline{$ZH \rightarrow\nu\nu b\bar{b}$ medium-tag channel relative uncertainties (\%)}}
\vskip 0.099cm
\begin{ruledtabular}
\begin{tabular}{l c c c c c c }\\
Contribution            & Top  & $V+b\bar{b}/c\bar{c}$ & $V$+l.f. & Dibosons  & Total Bkgd & $VH$  \\
\hline
Jet ID/Reco Eff (S)             &   2.0 &   2.0  & 2.0    &  2.0     &  1.9 & 2.0  \\
Jet Energy Scale (S)            &   1.3 &   1.5  & 2.8    &  1.5    &1.9  & 0.3  \\
Jet Resolution (S)              &   0.5 &   0.4  & 0.5    &  0.8     & 0.5  & 0.9  \\
Vertex Conf. / Taggability (S)  &   3.4 &   2.2  & 2.0    &  2.3     & 2.2  & 2.1  \\
b Tagging (S)                   &   1.5 &   2.6  & 8.0    &  3.6     & 3.7  & 0.6  \\
Lepton Identification           &   1.5 &   0.9  & 0.8    &  0.9     & 0.9  & 0.9  \\
Trigger                         &   2.0 &   2.0  & 2.0    &  2.0     & 1.9  & 2.0  \\
Heavy Flavor Fractions          & --   &  20.0  &  --    &  --      & 8.4  & --   \\
Cross Sections                  &  10.0 & 10.2   & 10.2   &  7.0     &  9.8 & 7.0  \\
Signal Branching Fraction       & --   &  --    &  --    &  --      &  --  & 1-9  \\
Luminosity                      &   6.1 &  6.1   & 6.1    &  6.1     & 5.8  & 6.1  \\
Multijet Normalilzation         & --   &  --    &  --    &  --      & 1.1  & --   \\
ALPGEN MLM (S)                  & --   &  --    &  SH    &  --      & --   & --   \\
ALPGEN Scale (S)                & --   &  SH    &  SH    &  --      & --   & --   \\
Underlying Event (S)            & --   &  SH    &  SH    &  --      & --   & --   \\
PDF, reweighting (S)            & SH   &  SH    &  SH    &  SH      & SH   & SH   \\
Total uncertainty     &  12.8 & 23.8   &  15.1  &  10.8    & 14.2 & 10.0  \\
%%%%%%%%%%%%%%%%%%%%%%%%%%%%%%%%%%%%%%%%%%%%%%%%%%%%%%%%%%%%%%%%%%%%%%%%%%%%%%%%%%%%%%%%%
\end{tabular}
%\end{ruledtabular}
\vskip 0.5cm
{\centerline{$ZH \rightarrow\nu\nu b\bar{b}$ tight-tag channel relative uncertainties (\%)}}
\vskip 0.099cm
%%\begin{ruledtabular}
\begin{tabular}{ l c c c c c c  }   \\
Contribution            & Top  & $V+b\bar{b}/c\bar{c}$ & $V$+l.f. & Dibosons  & Total Bkgd & $VH$  \\
\hline
Jet ID/Reco Eff (S)             & 2.0 & 2.0    & 2.0    & 2.0      &  2.0 & 2.0  \\
Jet Energy Scale (S)            & 1.0 & 1.6    & 3.9    & 1.6      &  1.6 & 0.5  \\
Jet Resolution (S)              & 0.7 & 0.6    & 2.6    & 1.4      &  0.8 & 1.3  \\
Vertex Conf. / Taggability (S)  & 3.0  & 1.9    & 2.4    & 2.0      & 2.3  & 1.9  \\
b Tagging (S)                   & 8.9  & 7.3    & 12.5    & 6.4      & 7.4  & 7.8  \\
Lepton Identification           & 1.9  & 0.8    & 0.3    &  0.7     & 1.1  & 0.8  \\
Trigger                         & 2.0  & 2.0    & 2.0    &  2.0     & 2.0  & 2.0  \\
Heavy Flavor Fractions          & --   &  20.0  &  --    &  --      & 11.0  & --   \\
Cross Sections                  &  10.0 & 10.2   & 10.2   & 7.0      & 10.0  & 7.0  \\
Signal Branching Fraction       & --   &  --    &  --    &  --      &  --  & 1-9  \\
Luminosity                      & 6.1  &  6.1   &  6.1   &  6.1     & 6.1  & 6.1  \\
Multijet Normalilzation         & --   &  --    &  --    &  --      & 0.2  & --   \\
ALPGEN MLM (S)                  & --   &  --    &  SH    &  --      & --   & --   \\
ALPGEN Scale (S)                & --   &  SH    &  SH    &  --      & --   & --   \\
Underlying Event (S)            & --   &  SH    &  SH    &  --      & --   & --   \\
PDF, reweighting (S)            & SH   &  SH    &  SH    &  SH      & SH   & SH   \\
Total uncertainty     &  15.5 & 24.7   & 18.3   &  12.0    & 16.8 & 12.7  \\
\end{tabular}
\end{ruledtabular}

\end{table}

%%%%%%%%       D0 WH and tau channels

\begin{table}[p]
\begin{center}
\caption{\label{tab:d0systwh1} Systematic uncertainties on the signal and background
contributions for D0's $WH\rightarrow\ell\nu b{\bar{b}}$ single and double tag channels.
Systematic uncertainties are listed by name, see the original
references for a detailed explanation of their meaning and on how they are derived.
Systematic uncertainties for $WH$ shown in this table are obtained for $m_H=115$ GeV/$c^2$. Uncertainties are
relative, in percent, and are symmetric unless otherwise indicated.   Shape uncertainties are labeled with an ``(S)'',
and ``SH'' represents shape only uncertainty.
}
%\begin{footnotesize}
\vskip 0.2cm
%%%%%%%
{\centerline{$WH \rightarrow\ell\nu b\bar{b}$ Single Tag (TST) channels relative uncertainties (\%)}}
\vskip 0.099cm
\begin{ruledtabular}
\begin{tabular}{l c c c c c c c }\\
Contribution             &~Dibosons~ & $W+b\bar{b}/c\bar{c}$& $W$+l.f. & $~~~t\bar{t}~~~$ &single top&Multijet& ~~~$WH$~~~\\
\hline
Luminosity                &  6.1  &  6.1  &  6.1  &  6.1  &  6.1  &   --    &  6.1  \\ 
Electron ID/Trigger eff. (S)& 1--5  & 2--4  &  2--4 & 1--2  & 1--2  &   --    &  2--3 \\       
Muon Trigger eff. (S)     & 1    &  1    &  1    &  1    &  1    &   --    &  1 \\       
Muon ID/Reco eff./resol.  & 4.1  &   4.1 &   4.1 &   4.1 &   4.1 &   --    &   4.1 \\        
Jet ID/Reco eff.          & 2    &  2    &  2    &  2    &  2    &   --    &  2    \\ 
Jet Resolution    (S)     & 1--2 &  2--4 &  2--3 &  2--5 &  1--2 &   --    &  2 \\       
Jet Energy Scale  (S)     & 4--7 &  1--5 &  2--5 &  2--7 &  1--2 &   --    &  2--6 \\       
Vertex Conf. Jet  (S)     & 4--6 & 3--4 & 2--3 & 6--10 & 2--4 &   --    &  3--7 \\       
$b$-tag/taggability (S)   & 1--3 &  1--4 & 7--10  &  1--6 &  1--2 &   --    &  2--9 \\ 
Heavy-Flavor K-factor     & --   &    20 &      -- &   --    &   --    &   --    &   --    \\       
Inst.-WH $e\nu b\bar{b}$ (S) & 1--2 & 2--4 & 1--3 & 1--2  &  1--3 &  15  &  1--2 \\ 
Inst.-WH $\mu\nu b\bar{b} $  &  --  &   2.4 &   2.4 &   --    &   --    &  20  &   --    \\ 
Cross Section             & 6    &     9 &     6 &    7 &    7 &   --    &     6.1 \\ 
Signal Branching Fraction & --   &   --  &  --  &  --  &  --  &  --  & 1-9 \\      
ALPGEN MLM pos/neg(S)     & --   &   --  &    SH  &     --    &   --    &   --    &   --    \\       
ALPGEN Scale (S)          & --   &  SH   &    SH  &     --    &   --    &   --    &   --    \\       
Underlying Event (S)      & --   &  SH   &    SH  &     --    &   --    &   --    &   --    \\       
PDF, reweighting          &  2   &  2    & 2     & 2     &  2    &   --    &  2    \\
%Total uncertainty        &  12.1 &  24.3 &  17.2 &  14.5 &  14.5 &  26.0 &  12.1 \\ \hlin  
%%%%%%%%%%%%%%%%%%%%%%%%%%%%%%%%%%%%%%%%%%%%%%%%%%%%%%%%%%%%%%%%%%%%%%%%%%%%%%%%%%%%%%%%%
\end{tabular}
\end{ruledtabular}

\end{center}
\end{table}

\begin{table}[p]
\begin{center}
\vskip 0.5cm
%\vskip 0.2cm
{\centerline{$WH \rightarrow\ell\nu b\bar{b}$ Loose Double Tag (LDT) channels relative uncertainties (\%)}}
\vskip 0.099cm
\begin{ruledtabular}
\begin{tabular}{ l c c c c c c c }   \\
Contribution  &~Dibosons~&$W+b\bar{b}/c\bar{c}$&$W$+l.f.&$~~~t\bar{t}~~~$&single top&Multijet& ~~~$WH$~~~\\
\hline
Luminosity                &  6.1  &  6.1  &  6.1  &  6.1  &  6.1  &   --    &  6.1  \\ 
Electron ID/Trigger eff. (S)  & 2--5  & 2--3  &  2--3 & 1--2  & 1--2  &   --    &  1--2 \\       
Muon Trigger eff. (S)     &  1    &  1    &  1    &  1    &  1    &   --    &  1 \\       
Muon ID/Reco eff./resol.  &   4.1 &   4.1 &   4.1 &   4.1 &   4.1 &   --    &   4.1 \\        
Jet ID/Reco eff.          &  2    &  2    &  2    &  2    &  2    &   --    &  2    \\ 
Jet Resolution    (S)     &  1--7 &  2--7 &  2--3 &  2--7 &  2--4 &   --    &  1--5 \\       
Jet Energy Scale  (S)     &  2--11 & 2--5 &  2--7 &  2--7 &  2--5 &   --    &  2--8 \\       
Vertex Conf. Jet  (S)     &  2--11 & 2--12 & 2--3 & 4--15 & 2--3 &   --    &  3--7 \\       
$b$-tag/taggability (S)   & 2--15  &  2--6 & 6--10 & 2--5 & 2--3 &   --    &  1--5 \\ 
Heavy-Flavor K-factor     &   --    &    20 &      -- &   --    &   --    &   --    &   --    \\       
Inst.-WH $e\nu b\bar{b}$ (S) & 1--2 & 2--4 & 1--3 & 1--2  &  1--3 &  15  &  1--2 \\ 
Inst.-WH $\mu\nu b\bar{b} $  &  --  &   2.4 &   2.4 &   --    &   --    &  20  &   --    \\ 
Cross Section             &     6 &     9 &     6 &    7 &    7 &   --    &     6.1 \\
Signal Branching Fraction &   --  &   --  &  --  &  --  &  --  &  --  & 1-9 \\      
ALPGEN MLM pos/neg(S)     &   --    &   --  &     SH &   --    &   --    &   --    &   --    \\       
ALPGEN Scale (S)          &   --    &   SH  &    SH &   --    &   --    &   --    &   --    \\       
Underlying Event (S)      &   --    &   SH  &    SH &   --    &   --    &   --    &   --    \\       
PDF, reweighting          &  2    &  2    & 2     & 2     &  2    &   --    &  2    \\
%Total uncertainty        &  12.9 &  24.7 &  24.0 &  15.2 &  15.2 &  26.0 &  12.9 \\ \hline
\end{tabular}
\end{ruledtabular}

\end{center}
\end{table}

\begin{table}[p]
\begin{center}
\vskip 0.5cm
%\vskip 0.2cm
{\centerline{$WH \rightarrow\ell\nu b\bar{b}$ Medium Double Tag (MDT) channels relative uncertainties (\%)}}
\vskip 0.099cm
\begin{ruledtabular}
\begin{tabular}{ l c c c c c c c }   \\
Contribution  &~Dibosons~&$W+b\bar{b}/c\bar{c}$&$W$+l.f.&$~~~t\bar{t}~~~$&single top&Multijet& ~~~$WH$~~~\\
\hline
Luminosity                &  6.1  &  6.1  &  6.1  &  6.1  &  6.1  &   --    &  6.1  \\ 
Electron ID/Trigger eff. (S)  & 2--5  & 2--3  &  2--3 & 1--2  & 1--2  &   --    &  1--2 \\       
Muon Trigger eff. (S)     &  2--5    &  1--3    &  1--3    &  1--5    &  2--3    &   --    &  1--3 \\            
Muon ID/Reco eff./resol.  &   4.1 &   4.1 &   4.1 &   4.1 &   4.1 &   --    &   4.1 \\        
Jet ID/Reco eff.          &  2    &  2    &  2    &  2    &  2    &   --    &  2    \\    
Jet Resolution    (S)     &  2--15 &  2--10 &  5--20 &  1--3 &  1--3 &   --    &  1--10 \\       
Jet Energy Scale  (S)     &  2--10 &  2--20 &  1--8 &  1--5 &  1--5 &   --    &  2--10 \\       
Vertex Conf. Jet  (S)     & 1--5  &  2--3 & 2--7  & 5--7  & 2--3  &   --    &  2--4 \\       
$b$-tag/taggability (S)   & 3--15 &  4--15& 10--15 & 4--10 & 3--9 &   --    &  2--5 \\ 
Heavy-Flavor K-factor     &   --    &    20 &      -- &   --    &   --    &   --    &   --    \\       
Inst.-WH $e\nu b\bar{b}$ (S) & 1--2 & 2--4 & 1--3 & 1--2  &  1--3 &  15  &  1--2 \\ 
Inst.-WH $\mu\nu b\bar{b} $  &  --  &   2.4 &   2.4 &   --    &   --    &  20  &   --    \\ 
Cross Section             &     6 &     9 &     6 &    7 &    7 &   --    &     6.1 \\
Signal Branching Fraction &   --  &   --  &  --  &  --  &  --  &  --  & 1-9 \\      
ALPGEN MLM pos/neg(S)     &   --    &   --  &    SH &   --    &   --    &   --    &   --    \\       
ALPGEN Scale (S)          &   --    &   SH  &    SH &   --    &   --    &   --    &   --    \\       
Underlying Event (S)      &   --    &   SH  &    SH &   --    &   --    &   --    &   --    \\       
PDF, reweighting          &  2    &  2    & 2     & 2     &  2    &   --    &  2    \\
%Total uncertainty        &  12.9 &  24.7 &  24.0 &  15.2 &  15.2 &  26.0 &  12.9 \\ \hline
\end{tabular}
\end{ruledtabular}
\end{center}
\end{table}

\begin{table}[p]
\begin{center}
\vskip 0.5cm
%\vskip 0.2cm
{\centerline{$WH \rightarrow\ell\nu b\bar{b}$ Tight Double Tag (TDT) channels relative uncertainties (\%)}}
\vskip 0.099cm
\begin{ruledtabular}
\begin{tabular}{ l c c c c c c c }   \\
Contribution  &~Dibosons~&$W+b\bar{b}/c\bar{c}$&$W$+l.f.&$~~~t\bar{t}~~~$&single top&Multijet& ~~~$WH$~~~\\
\hline
Luminosity                &  6.1  &  6.1  &  6.1  &  6.1  &  6.1  &   --    &  6.1  \\ 
Electron ID/Trigger eff. (S)  & 2--5  & 2--3  &  2--3 & 1--2  & 1--2  &   --    &  1--2 \\       
Muon Trigger eff. (S)     &  1    &  1    &  1    &  1    &  1    &   --    &  1 \\            
Muon ID/Reco eff./resol.  &   4.1 &   4.1 &   4.1 &   4.1 &   4.1 &   --    &   4.1 \\        
Jet ID/Reco eff.          &  2    &  2    &  2    &  2    &  2    &   --    &  2    \\    
Jet Resolution    (S)     &  2--5 &  4--7 &  2--6 &  1--4 &  2--6 &   --    &  2--9 \\       
Jet Energy Scale  (S)     &  2--15 &  2--8 &  1--8 &  2--7 &  1--4 &   --    &  1--9 \\       
Vertex Conf. Jet  (S)     & 2--3  &  2--4 & 2--5  & 5--6  & 2--3  &   --    &  2--4 \\       
$b$-tag/taggability (S)   & 3--15 &  5--10& 5--15 & 6--10 & 5--10 &   --    &  5--12 \\ 
Heavy-Flavor K-factor     &   --    &    20 &      -- &   --    &   --    &   --    &   --    \\       
Inst.-WH $e\nu b\bar{b}$ (S) & 1--2 & 2--4 & 1--3 & 1--2  &  1--3 &  15  &  1--2 \\ 
Inst.-WH $\mu\nu b\bar{b} $  &  --  &   2.4 &   2.4 &   --    &   --    &  20  &   --    \\ 
Cross Section             &     6 &     9 &     6 &    7 &    7 &   --    &     6.1 \\
Signal Branching Fraction &   --  &   --  &  --  &  --  &  --  &  --  & 1-9 \\      
ALPGEN MLM pos/neg(S)     &   --    &   --  &    SH &   --    &   --    &   --    &   --    \\       
ALPGEN Scale (S)          &   --    &   SH  &    SH &   --    &   --    &   --    &   --    \\       
Underlying Event (S)      &   --    &   SH  &    SH &   --    &   --    &   --    &   --    \\       
PDF, reweighting          &  2    &  2    & 2     & 2     &  2    &   --    &  2    \\
%Total uncertainty        &  12.9 &  24.7 &  24.0 &  15.2 &  15.2 &  26.0 &  12.9 \\ \hline
\end{tabular}
\end{ruledtabular}
\end{center}
\end{table}

\begin{table}[hp]
\begin{center}
\caption{\label{tab:d0systww}
Systematic uncertainties on the signal and background contributions
for D0's $H\rightarrow W^+W^- \rightarrow\ell^{\pm}\ell^{\mp}$
channels.  Systematic uncertainties are listed by name; see the
original references for a detailed explanation of their meaning and on
how they are derived.  Shape uncertainties are labeled with the ``s''
designation. Systematic uncertainties given in this table are obtained
%for the $m_H=165$\,GeV~Higgs selection.
for the $m_H=165$ GeV/c$^2$ Higgs selection.
  Cross section uncertainties on
the $gg\to H$ signal depend on the jet multiplicity, as described in
the main text. Uncertainties are relative, in percent, and are
symmetric unless otherwise indicated.
}\vskip 0.1cm
{\centerline{$H\rightarrow W^+W^- \rightarrow\ell^{\pm}\ell^{\mp}$ channels relative uncertainties (\%)}}
\vskip 0.099cm
\begin{ruledtabular}
\begin{tabular}{ l  c  c  c  c  c  c  c  c  }  \\
Contribution & Dibosons & ~~$Z/\gamma^* \rightarrow \ell\ell$~~&$~~W+$jet$/\gamma$~~ &~~~~$t\bar{t}~~~~$ & ~~Multijet~~ & $gg\to H$ & $qq\to qqH$ & $VH$ \\
\hline
Luminosity/Normalization                   & 4   & --  & 4   & 4   & 4   & 4          & 4     & 4      \\
Cross Section (Scale/PDF)                  & 5-7 & --  & --  & 7   & --  & 13-33/8-30 & 5     & 6      \\
$Z/\gamma^*\rightarrow\ell\ell$ n-jet norm & --  & 2-15& --  & --  & --  & --         & --    & --     \\
$Z/\gamma^*\rightarrow\ell\ell$ MET model  & --  & 5-19& --  & --  & --  & --         & --    & --     \\
$W+$jet$/\gamma$ norm                      & --  & --  & 6-30& --  & --  & --         & --    & --     \\
$W+$jet$/\gamma$ ISR/FSR model (s)         & --  & --  & 2-20& --  & --  & --         & --    & --     \\
Vertex Confirmation (s)                    & 1-5 & 1-5 & 1-5 & 5-6 & --  & 1-5        & 1-5   & 1-5    \\
Jet identification (s)                     & 1   & 1  & 1    & 1   & --  & 1          & 1     & 1      \\
Jet Energy Scale (s)                       & 1-5 & 1-5& 1-5  & 1-4 & --  & 1-5        & 1-5   & 1-4    \\
Jet Energy Resolution(s)                   & 1-4 & 1-4& 1-4  & 1-4 & --  & 1-3        & 1-4   & 1-3    \\
B-tagging (s)                              & --  & -- & --   & 1-5 & --  & --         & --    & --     \\
\end{tabular}
\end{ruledtabular}
\end{center}
\end{table}

\begin{table}[hp]
\begin{center}
\caption{\label{tab:d0systwwtau} Systematic uncertainties on the signal and background contributions for D0's
$H\rightarrow W^+ W^- \rightarrow \mu\nu \tau_{\rm{had}}\nu $ channel.  Systematic uncertainties are listed by
name; see the original references for a detailed explanation of their meaning and on how they are derived.
Shape uncertainties are labeled with the shape designation (S). Systematic uncertainties shown in this table are obtained for the $m_H=165$ GeV/c$^2$ Higgs selection.
Uncertainties are relative, in percent, and are symmetric unless otherwise indicated.}
\vskip 0.1cm
{\centerline{$H\rightarrow W^+ W^- \rightarrow \mu\nu \tau_{\rm{had}}\nu $ channel relative uncertainties (\%)}}
\vskip 0.099cm
\begin{ruledtabular}
%\begin{tabular}{ l  c  c  c  c  c  c  c}  \\
%Contribution       & Diboson    & ~~$Z/\gamma^* \rightarrow
%\ell\ell$~~ &$~~W+\tm{jets}$~~ &~~~~$t\bar{t}~~~~$    & ~~Multijet~~
%& ~~~~$H$~~~~ \\
%\hline
%Luminosity ($\sigma_{\tm{inel}}(p\bar{p})$) &4.6   & 4.6   & -  & 4.6    &    -   &   4.6    \\
%Luminosity Monitor    &  4.1  & 4.1           & -          & 4.1         &-        &   4.1    \\
%Trigger     &  5.0            &5.0             & -          & 5.0        &-        &   5.0    \\
%Lepton ID    &  3.7   &3.7           & -           & 3.7              &-        &   3.7    \\
%EM veto        &  5.0         &-         & -         & 5.0        &-        &   5.0    \\
%Tau Energy Scale (s)    &  1.0        &1.1          & -        & $<$1     &-        &   $<$1   \\
%Jet Energy Scale (s)    &  8.0     &  $<$1    & -       & 1.8     &-        &   2.5    \\
%Jet identification (s)  &  $<$1       & $<$1       & -         & 7.5        &-        &   5.0    \\
%Multijet  (s)  ~~~~~    &  --      & -    & -     & -       &20-50    &   -      \\
%Cross Section     &  7.0       & 4.0      & -      & 10       & -        &   10     \\
%Modeling    ~~~~~       &  1.0     &-      & 10    & -      &-        &   3.0    \\
%
\begin{tabular}{ l  c  c  c  c  c  c  c  c  c}  \\
Contribution       & Diboson    & ~~$Z/\gamma^* \rightarrow \ell\ell$~~ & ~~$W$+$\rm{jets}$~~ &
~~~~$t\bar{t}$~~~~ & ~~Multijet~~ & ~~~~$gg \rightarrow H$~~~~ &
~~~~$qq \rightarrow qqH$~~~~ & ~~~~$VH$~~~~\\
\hline
Luminosity ($\sigma_{\rm{inel}}(p\bar{p})$) &4.6   & 4.6   & -  & 4.6    &    -   &   4.6 &   4.6 &   4.6    \\
Luminosity Monitor    &  4.1  & 4.1           & -          & 4.1         &-        &   4.1 &   4.1 &   4.1    \\
Trigger     &  5.0            &5.0             & -          & 5.0        &-        &   5.0  &   5.0 &   5.0  \\
Lepton ID    &  3.7   &3.7           & -           & 3.7              &-        &   3.7 &   3.7 &   3.7    \\
EM veto        &  5.0         &-         & -         & 5.0        &-        &   5.0  &   5.0 &   5.0   \\
Tau Energy Scale (S)    &  1.0        &1.1          & -        & $<$1     &-        &   $<$1 &   $<$1 &   $<$1   \\
Jet Energy Scale (S)    &  8.0     &  $<$1    & -       & 1.8     &-        &   2.5 &   2.5 &   2.5     \\
Jet identification (S)  &  $<$1       & $<$1       & -         & 7.5        &-         &   5.0 &   5.0 &   5.0    \\
Multijet  (S)  ~~~~~    &  -      & -    & -     & -       &20-50    &   -    &   - &   -  \\
Cross Section (scale/PDF)     &  7.0       & 4.0      & -      & 10       & -        &   7/8 & 4.9 & 6.1    \\
Signal Branching Fraction & -  & -        & -      &-         &-         &0-7.3  &0-7.3 &0-7.3 \\
Modeling    ~~~~~       &  1.0     &-      & 10    & -      &-        &   3.0  &   3.0 &   3.0   \\

\end{tabular}
\end{ruledtabular}
\end{center}
\end{table}

\begin{table}[hp]
\begin{center}
\caption{\label{tab:d0systwww-em}
Systematic uncertainties on the signal and background contributions for D0's
$VH\rightarrow e^\pm \nu_e \mu^\pm \nu_\mu$($V=W,Z$) channels. Systematic uncertainties are
listed by name; see the original references for a detailed explanation of their meaning and on how
they are derived. Shape uncertainties are labeled with the ``shape'' designation. Systematic uncertainties
shown in this table are obtained 
%for the $m_H=165$\gev~Higgs selection. 
for the $m_H=165$ GeV/c$^2$ Higgs selection. Uncertainties are relative,
in percent, and are symmetric unless otherwise indicated. }
\vskip 0.2cm
%%%%%%%%%%
{\centerline{$VH \rightarrow e^\pm \nu_e \mu^\pm \nu_\mu$ like charge electron muon pair channel relative uncertainties (\%)}}
\begin{ruledtabular}
\begin{tabular}{l c c c c c c}  \\
\hline
Contribution 				& VH		& $Z+jet/\gamma$ 	& $W+jet/\gamma$ 	& $t\bar{t}$	& Diboson 	& Multijet 	 \\
\hline
Cross section				& 6.2		& -- 				& -- 				& 6 			& 7 			& --		 \\
Luminosity/Normalization		& 4   		& -- 				& 4  				& 4  			& 4  			& --		 \\
Multijet              				& --  		& -- 				& -- 				& -- 			& -- 			& 30		\\
Trigger            				& 2 		& 2 				& 2 				& 2 			& 2 			& 2		 \\
Charge flip           			& --  		& 50 				& -- 				& 50  		& 50  		& --		 \\
W+jets/$\gamma$ 			& --  		& --  				& 10  			& --  			& --  			& --		 \\
$W-p_T$ model 			& --  		& --  				& shape			& --  			& --  			& --		\\
$Z-p_T$ model  			& --  		& shape 			& --   				& --  			& --  			& --		 \\
W+jets/$\gamma$ ISR/FSR model	& --  	& --  				& shape			& --  			& --  			& --		 \\
\hline
\end{tabular}
\end{ruledtabular}
\end{center}
\end{table}

\begin{table}
\begin{center}
\caption{\label{tab:d0systlll} 
Systematic uncertainties on the signal and background contributions
for D0's $VH\rightarrow VWW \rightarrow ee\mu, \mu\mu e$ channels.  
Systematic uncertainties are listed by name; see the
original references for a detailed explanation of their meaning and on
how they are derived.  Shape uncertainties are labeled with the ``s''
designation. Systematic uncertainties given in this table are obtained
for the $m_H=145$~GeV~Higgs selection. Uncertainties are relative, 
in percent, and are symmetric unless otherwise indicated.  Jet shape 
uncertainties are applied to the $\mu \mu e$ channel only.
}\vskip 0.1cm
{\centerline{$VH\rightarrow VWW \rightarrow$ Trilepton channels relative uncertainties (\%)}}
\vskip 0.099cm
\begin{ruledtabular}
\begin{tabular}{ l  c  c  c  c  c  c  c  c  }  \\
Contribution                                 & Dibosons         & ~~$Z/\gamma^* \rightarrow \ell\ell$~~&$~~W+$jet$/\gamma$~~ &~~~~$t\bar{t}~~~~$    & ~~$Z\gamma$~~  & $VH$ 		 & $gg\to H$       & $qq\to qqH$  \\ 
\hline
Luminosity    				& 6.1                  & 6.1  	                                                    & 6.1  	                          	  & 6.1 	                         & --  	                 & 6.1 		 & 6.1  	      & 6.1  \\
Cross Section (Scale/PDF)   	& 6     	                & 6     	                                                     & 6     	                          	  & 7    	                         & --  	                 & 6.2    		 & 7	 	       & 4.9  \\
PDF              				& 2.5                  & 2.5  	                                                     & 2.5  	                          	  & 2.5 	                         & --  	                 & 2.5 		 & 2.5   	       & 2.5  \\
Electron Identification       	          & 2.5  	                & 2.5  	                                                     & 2.5  	                          	  & 2.5 	                         & --   	                 & 2.5 		 & 2.5   	       & 2.5  \\
Muon Identification     		& 4     	                & 4     	                                                     & 4     	                          	  & 4    	                         & --    	                 & 4    		 & 4      	       & 4     \\
Trigger				& 3.5	                & 3.5	                                                     & 3.5	                                  & 3.5	                         & -- 	                 & 3.5		 & 3.5	                  & 3.5  \\
$Z\gamma$ 				& --                     & --		                                          & --		                        & --		                 	   & 9.5		      &	--		 &	--	       &	--  \\
$V+jets$ lepton fake rate	           & --	                 & 30		                                          & 30		                        & -- 	                                    & --		        	      & --		 & --		       &	--  \\
Z-$p_{T}$ reweighting (s)            & --                     & $\pm 1\sigma$                                         & --                                        & --                                      & --                            & --       		 & --                     & --  \\
Electron smearing (s)  		& $\pm 1\sigma$ & $\pm 1\sigma$                                         & $\pm 1\sigma$                   & $\pm 1\sigma$                 & -- 			     & $\pm 1\sigma$ & $\pm 1\sigma$ & $\pm 1\sigma$ \\
Muon smearing (s) 			& $\pm 1\sigma$ & $\pm 1\sigma$ 			          & $\pm 1\sigma$ 		  & $\pm 1\sigma$ 		   & -- 			     & $\pm 1\sigma$ & $\pm 1\sigma$ & $\pm 1\sigma$ \\
\hline
\multicolumn{8}{c}{Jet Shape systematics below applied to $\mu \mu e$ channel only} \\
\hline
Jet Energy Scale (s)                     & $\pm 1\sigma$ & $\pm 1\sigma$ 			          & $\pm 1\sigma$ 		  & $\pm 1\sigma$ 		   & -- 			     & $\pm 1\sigma$ & $\pm 1\sigma$ & $\pm 1\sigma$ \\
Jet Energy Resolution (s)             & $\pm 1\sigma$ & $\pm 1\sigma$ 			          & $\pm 1\sigma$ 		  & $\pm 1\sigma$ 		   & -- 			     & $\pm 1\sigma$ & $\pm 1\sigma$ & $\pm 1\sigma$ \\
Jet Indentification (s)                   & $-1\sigma$      & $-      1\sigma$ 			          & $ -1\sigma$ 		  & $-1\sigma$ 		   & -- 			     & $-1\sigma$       & $-1\sigma$      & $- 1\sigma$ \\
Vertex Confirmation (s)                & $-1\sigma$      & $-      1\sigma$ 			          & $ -1\sigma$ 		  & $-1\sigma$ 		   & -- 			     & $-1\sigma$       & $-1\sigma$      & $- 1\sigma$ \\
\end{tabular}
\end{ruledtabular}
\end{center}
\end{table}

\begin{table}[hp]
\begin{center}
\caption{\label{tab:d0systttm} 
Systematic uncertainties on the signal and background contributions
for D0's $\tau\tau\mu$ +X 
channel.  Systematic uncertainties are listed by name; see the
original references for a detailed explanation of their meaning and on
how they are derived.  Shape uncertainties are labeled with the ``s''
designation. Cross section uncertainties on
the $gg\to H$ signal depend on the jet multiplicity, as described in
the main text. Uncertainties are relative, in percent, and are
symmetric unless otherwise indicated.
}\vskip 0.1cm
%{\centerline{{$\tau\tau\mu$ +X  channel relative uncertainties (\%)}}
{\centerline{$\tau\tau\mu$ +X  channels relative uncertainties (\%)}}
\vskip 0.099cm
\begin{ruledtabular}
\begin{tabular}{ l  c  c  c  c  c  c  c  c  }  \\
Contribution & Dibosons & ~~$Z/\gamma^*$~~&~~~~$t\bar{t}~~~~$    & ~~Instrumental~~  & $gg\to H$ & $qq\to qqH$ & $VH$ \\ 
\hline
Luminosity/Normalization    &  6      &   6                     & 6            &  24   &   6  &   6  &   6  \\
Trigger &  3      &   3                     & 3           & --   &   3  &   3  &   3  \\
Cross Section (Scale/PDF)      &  7        &   6           & 10                       &  --  &   13-33/7.6-30 & 4.9 & 6.2  \\
PDF              &      2.5     &   2.5          & 2.5                 &  --   &   2.5   &   2.5     &   2.5      \\
Tau Id per $\tau$  (Type 1/2/3)    &  7/3.5/5  &   7/3.5/5                   & 7/3.5/5      &  --   &   7/3.5/5       &   7/3.5/5        &   7/3.5/5         \\
Tau Energy Scale &  1  &   1                  & 1    &  --   &   1     &   1        &  1      \\
Tau Track Match per $\tau$&  1.4  &   1.4                  & 1.4    &  --   &   1.4     &   1.4        &  1.4      \\
Muon Identification     &  2.9           &  2.9                       & 2.9            &  --   &   2.9      &   2.9     &   2.9       \\
\end{tabular}
\end{ruledtabular}
\end{center}
\end{table}

\begin{table}[hp]
\begin{center}
\caption{\label{tab:d0lvjjjj}
%Systematic uncertainties for the electron and muon channels.
Systematic uncertainties on the signal and background contributions for the
 $VH\rightarrow V W W^{*} \rightarrow \ell\nu jjjj$ analysis.  Systematic uncertainties are listed
 by name; see the original references for a detailed explanation of their meaning and on how they are
 derived.
Signal uncertainties are shown for the total signal contribution at $m_H=125$ GeV/$c^2$ for all channels.  Those affecting the shape of
the RF discriminant are indicated with ``Y.''
Uncertainties are listed as relative changes in normalization,
in percent, except for those also marked by ``S,'' where
the overall normalization is constant, and the value given
denotes the maximum percentage change from nominal in any region of the
distribution.}

\vskip 0.1cm
{\centerline{D0: $VH\rightarrow V W W^{*} \rightarrow \ell\nu jjjj$ Run~II Zero Tag channel relative uncertainties (\%)}}
\vskip 0.099cm
\begin{ruledtabular}
\begin{tabular}{l c c c c c c }\\
Contribution             &~Dibosons~ & $W+b\bar{b}/c\bar{c}$& $W$+l.f. & Top quark &Multijet& Signal \\
\hline
Luminosity                    &  6.1  &  6.1  &  6.1  &  6.1  &   --  &  6.1  \\
Electron ID/Trigger eff. (S)  &   3   &   3   &    3  &   3   &   --  &  3    \\
Muon Trigger eff. (S)         & 1     &  1    &  1    &  1    &   --  &  1    \\
Muon ID/Reco eff./resol.      &   3   &     3 &     3 &     3 &   --  &     3 \\
Jet ID/Reco eff.              & 2     &  2    &  2    &  2    &   --  &  2    \\
Jet Resolution    (S)         & 1--2  &  2--4 &  2--3 &  2--5 &   --  &  2    \\
Jet Energy Scale  (S)         & 5--10 &  1--5 &  2--7 &  2--7 &   --  &  2--6 \\
Vertex Conf. Jet  (S)         & 3--4  & 1--2  & 1--2  & 3--4  &   --  &  3--7 \\
$b$-tag/taggability (S)       & 4--5  &  1--3 & 1--3  &  5--10&   --  &  4--10\\
Heavy-Flavor K-factor         & --    &    20 &    -- &   --  &   --  &   --  \\
Cross Section                 & 6     &     9 &     6 &    7  &   --  &   6.1 \\
Signal Branching Fraction     & --    &   --  &   --  &  --   &   --  &  1--9 \\
ALPGEN MLM pos/neg(S)         & --    &   SH  &   --  &   --  &   --  &   --  \\
ALPGEN Scale (S)              & --    &   SH  &   SH  &   --  &   --  &   --  \\
Underlying Event (S)          & --    &   SH  &   --  &   --  &   --  &   --  \\
PDF, reweighting              &  2    &  2    & 2     & 2     &   --  &  2    \\
\end{tabular}
\end{ruledtabular}
%\end{center}
%\end{table}

\vskip 0.5cm
{\centerline{D0: $VH\rightarrow V W W^{*} \rightarrow \ell\nu jjjj$ Run~II Loose Single Tag channel relative uncertainties (\%)}}
\vskip 0.099cm
\begin{ruledtabular}
\begin{tabular}{l c c c c c c }\\
Contribution             &~Dibosons~ & $W+b\bar{b}/c\bar{c}$& $W$+l.f. & Top quark &Multijet& Signal \\
\hline
Luminosity                    &  6.1  &  6.1  &  6.1  &  6.1  &   --  &  6.1  \\
Electron ID/Trigger eff. (S)  &  3    &  3    &  3    &  3    &   --  &  3    \\
Muon Trigger eff. (S)         & 1     &  1    &  1    &  1    &   --  &  1    \\
Muon ID/Reco eff./resol.      &   3   &     3 &     3 &     3 &   --  &     3 \\
Jet ID/Reco eff.              & 2     &  2    &  2    &  2    &   --  &  2    \\
Jet Resolution    (S)         & 1--2  &  2--4 &  2--3 &  2--5 &   --  &  2    \\
Jet Energy Scale  (S)         & 5--10 &  1--5 &  2--7 &  2--7 &   --  &  2--6 \\
Vertex Conf. Jet  (S)         & 3--4  & 1--2  & 1--2  & 3--4  &   --  &  3--5 \\
$b$-tag/taggability (S)       & 2--8  &  1--3 & 1--2  &  5--10&   --  &  4--10\\
Heavy-Flavor K-factor         & --    &    20 &    -- &   --  &   --  &   --  \\
Cross Section                 & 6     &     9 &     6 &    7  &   --  &   6.1 \\
Signal Branching Fraction     & --    &   --  &   --  &  --   &   --  &  1--9 \\
ALPGEN MLM pos/neg(S)         & --    &   SH  &   --  &   --  &   --  &   --  \\
ALPGEN Scale (S)              & --    &   SH  &   SH  &   --  &   --  &   --  \\
Underlying Event (S)          & --    &   SH  &   --  &   --  &   --  &   --  \\
PDF, reweighting              &  2    &  2    & 2     & 2     &   --  &  2    \\
\end{tabular}
\end{ruledtabular}
\end{center}
\end{table}

\begin{table}[hp]
\begin{center}
\caption{\label{tab:d0lvjj}
%Systematic uncertainties for the electron and muon channels.
Systematic uncertainties on the signal and background contributions for D0's
 $H\rightarrow W W^{*} \rightarrow \ell\nu jj$ electron and muon channels.  Systematic uncertainties are listed
 by name; see the original references for a detailed explanation of their meaning and on how they are
 derived.
Signal uncertainties are shown for $m_H=160$ GeV/$c^2$ for all channels except for $WH$,
shown for $m_H=115$ GeV/$c^2$.  Those affecting the shape of
the RF discriminant are indicated with ``Y.''
Uncertainties are listed as relative changes in normalization,
in percent, except for those also marked by ``S,'' where
the overall normalization is constant, and the value given
denotes the maximum percentage change from nominal in any region of the
distribution.}

\vskip 0.1cm
{\centerline{$H\rightarrow W W^{*} \rightarrow \ell\nu jj$ Run~II channel relative uncertainties (\%)}}
\vskip 0.099cm
\begin{ruledtabular}
\begin{tabular}{llccccccl}

Contribution & Shape & $W$+jets & $Z$+jets & Top & Diboson & $gg\to H$ & $qq\to qqH$ & $WH$ \\ \hline
Jet energy scale & Y & $\binom{+6.7}{-5.4}^S$ & $<0.1$ & $\pm$0.7 & $\pm$3.3 & $\binom{+5.7}{-4.0}$ & $\pm$1.5 &$\binom{+2.7}{-2.3}$  \\
Jet identification & Y & $\pm 6.6^S$ & $<0.1$ & $\pm$0.5 & $\pm$3.8  & $\pm$1.0 & $\pm$1.1 & $\pm$1.0 \\
Jet resolution & Y & $\binom{+6.6}{-4.1}^S$ & $<0.1$ & $\pm$0.5 & $\binom{+1.0}{-0.5}$ & $\binom{+3.0}{-0.5}$ & $\pm 0.8$ & $\pm 1.0$ \\
Association of jets with PV & Y & $\pm 3.2^S$ & $\pm 1.3^S$ & $\pm$1.2 & $\pm$3.2 & $\pm$2.9 & $\pm$2.4 & $\binom{+0.9}{-0.2}$ \\
Luminosity & N & n/a & n/a & $\pm$6.1 & $\pm$6.1 & $\pm$6.1 & $\pm$6.1 &  $\pm$6.1 \\
Muon trigger  & Y & $\pm 0.4^S$ & $<0.1$ & $<0.1$ & $<0.1$ & $<0.1$ & $<0.1$ &  $<0.1$ \\
Electron identification & N & $\pm$4.0  & $\pm$4.0  & $\pm$4.0  & $\pm$4.0  & $\pm$4.0  & $\pm$4.0  & $\pm$4.0 \\
Muon identification  & N & $\pm$4.0  & $\pm$4.0  & $\pm$4.0  & $\pm$4.0  & $\pm$4.0  & $\pm$4.0  & $\pm$4.0  \\
ALPGEN tuning & Y & $\pm 1.1^S$ & $\pm 0.3^S$ & n/a & n/a & n/a & n/a & n/a \\
Cross Section & N & $\pm$6 & $\pm$6 &  $\pm$10 & $\pm$7 & $\pm$10 & $\pm$10 & $\pm$6 \\
Heavy-flavor fraction  & Y & $\pm$20 & $\pm$20 & n/a & n/a & n/a & n/a & n/a  \\
Signal Branching Fraction & N & n/a &n/a &n/a& n/a & 0-7.3 & 0-7.3 &  0-7.3 \\
PDF & Y & $\pm 2.0^S$ & $\pm 0.7^S$ & $<0.1^S$ & $<0.1^S$ & $<0.1^S$ & $<0.1^S$ & $<0.1^S$ \\
 &  &  &  &  &  &  &  &  \\
 &  & \multicolumn{ 3}{c}{Electron channel} & \multicolumn{ 3}{c}{Muon channel} &  \\
Multijet Background & Y  & \multicolumn{ 3}{c}{$\pm$6.5} & \multicolumn{ 3}{c}{$\pm$26} &  \\
\end{tabular}
\end{ruledtabular}
\end{center}
\end{table}

\begin{table}[hp]
\begin{center}
\caption{\label{tab:d0systgg} Systematic uncertainties on the signal and background contributions for D0's
$H\rightarrow \gamma \gamma$ channel. Systematic uncertainties for the Higgs signal shown in this table are
obtained for $m_H=125$ GeV/$c^2$.  Systematic uncertainties are listed by name; see the original references
for a detailed explanation of their meaning and on how they are derived.  Uncertainties are relative, in
percent, and are symmetric unless otherwise indicated.}
\vskip 0.1cm
{\centerline{$H \rightarrow \gamma \gamma$ channel relative uncertainties (\%)}}
\vskip 0.099cm
\begin{ruledtabular}
\begin{tabular}{lcc}\\
Contribution &  ~~~Background~~~  & ~~~Signal~~~    \\
\hline
Luminosity~~~~                            &  6     &  6    \\
Acceptance                                &  --    &  2    \\
electron ID efficiency                    &  2     &  --   \\
electron track-match inefficiency         & 10     &  --   \\
Photon ID efficiency                      &  3     &   3   \\
%Photon energy scale                       &  2     &   1   \\
%Cross Section ($Z$)                       &  4     &  10   \\
Cross Section                             &  4     &  10   \\
Background subtraction                    &  15 &  -       \\
%ONN Shape                              & 1-5  & -       \\
\end{tabular}
\end{ruledtabular}
\end{center}
\end{table}

\clearpage
\section{Comparison With Previous Results}
\label{app:moriond_compare}

Here we document a comparison of these results with those presented in 
Ref.~\cite{M12dzcombo}.  In Fig. \ref{fig:allHI_compare}, we show
comparisons of the expected and observed limits.  The expected limits
have improved by approximately 10\%.  There observed limits have not
changed significantly, and exhibit fewer fluctuations between adjacent
assumed values of $m_H$.  Figure  \ref{fig:sb_compare} compares the
$\log_{10}(s/b)$ distributions of the two results for Higgs boson
masses of 115\gev~and 125\gev.  In this result, the distribution ranges
to larger values of $\log_{10}(s/b)$ than in the previous combination,
indicating an increase in sensitivity.  Figure ~\ref{fig:sb_compare_subtract}
shows the corresponding background subtracted distributions.  In Fig.
\ref{fig:compare_integral}, we compare the cumulative signal
distributions. For $m_H=125$\gev, the highest $s/b$ bins contain an excess
of signal like events in the current result that is more significant
than the excess exhibited in Ref. \cite{M12dzcombo}.  Figures~\ref{fig:HBBsb_compare}-\ref{fig:HBBcompare_integral} and 
\ref{fig:HWWsb_compare}-\ref{fig:HWWcompare_integral} show the same
distributions for the combined 
$WH/ZH, H$$\rightarrow$$ b\bar{b}$
and the combined
$WH/ZH/H, H$$\rightarrow$$ W^+W^-$ 
analyses respectively.

\begin{figure}[bp]
\centering
\includegraphics[height=0.2\textheight]{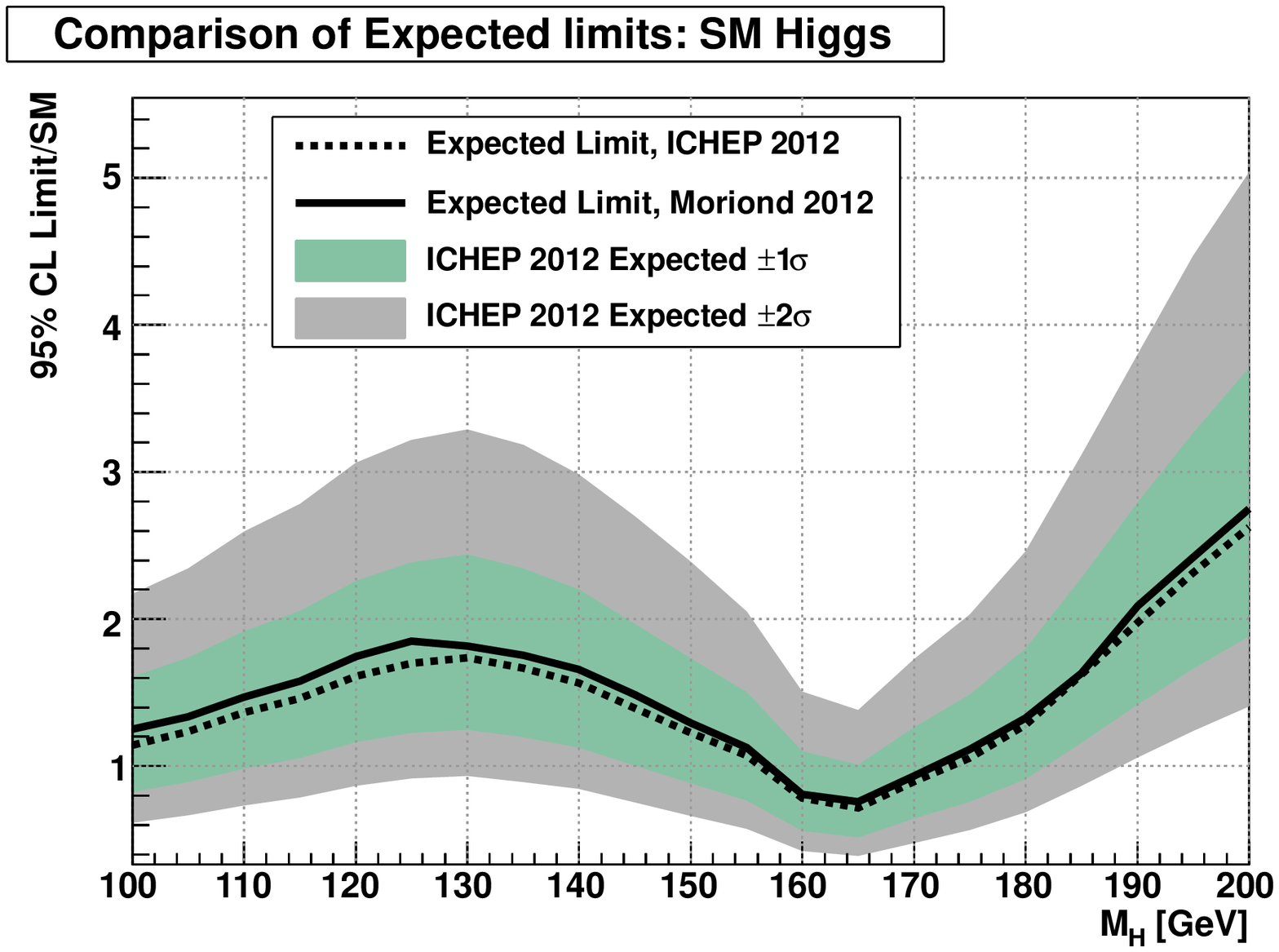}
\includegraphics[height=0.2\textheight]{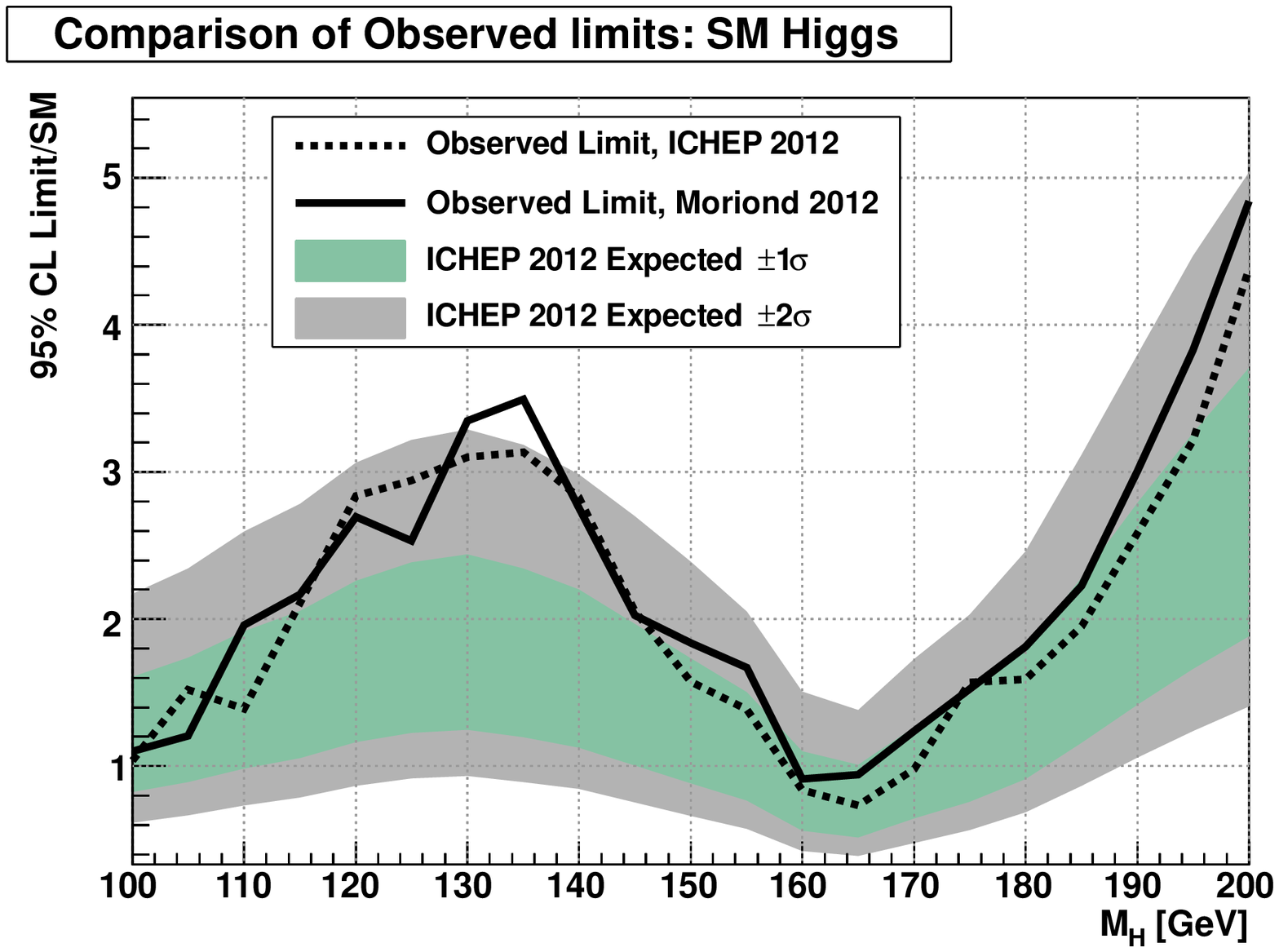}
\caption{\label{fig:allHI_compare} 
Comparison of expected (median), and observed 95\% C.L. cross
section upper limit ratios for this result (right) and the result from 
Ref.~\cite{M12dzcombo} (left) over the $100 \leq m_H \leq 200$\gev~mass range.
The green and grey
bands correspond to the regions enclosing 1 and 2 standard deviation
fluctuations of the background, respectively.}
\end{figure}

\begin{figure}[bp]
\centering
\includegraphics[height=0.2\textheight]{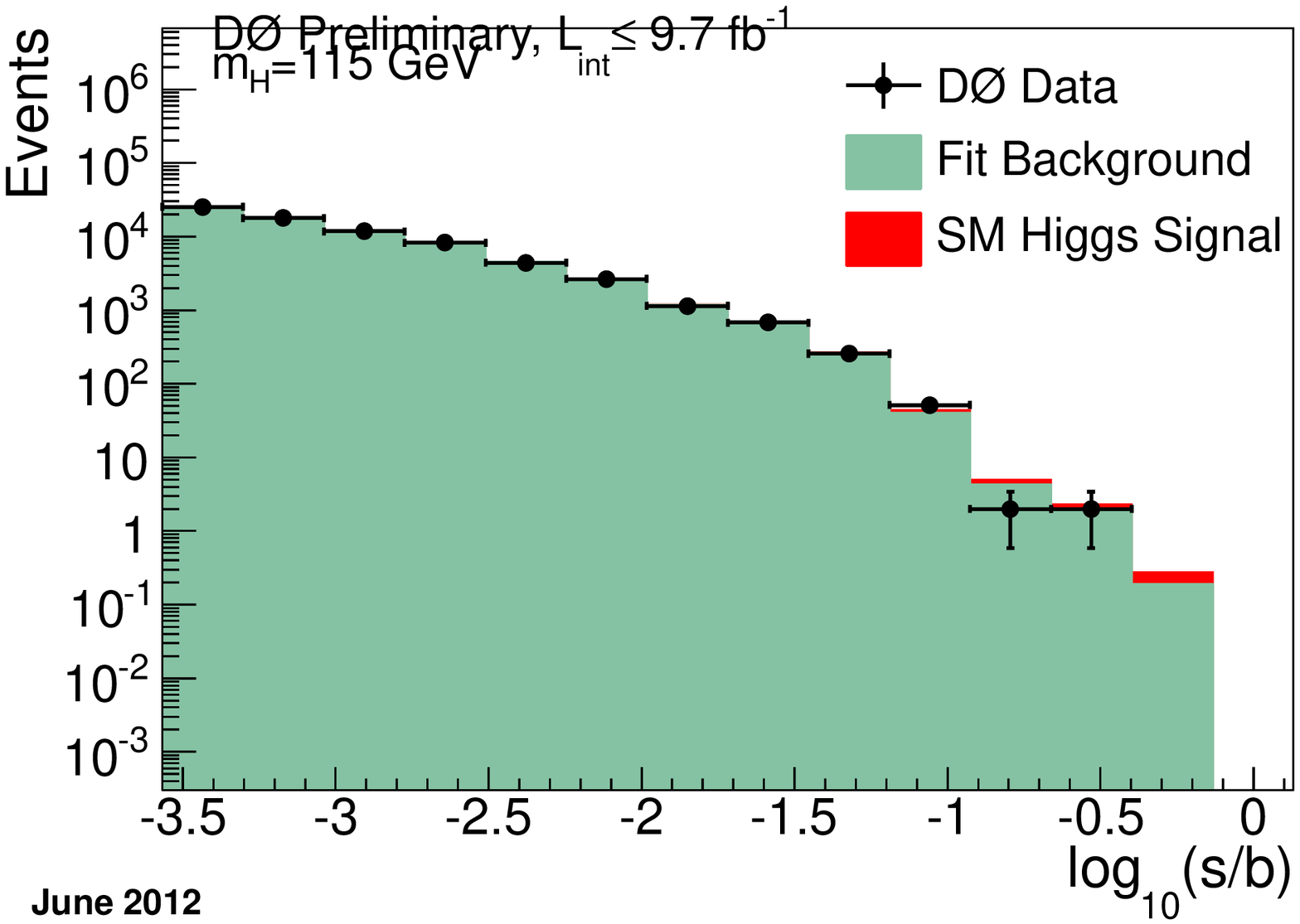} 
\includegraphics[height=0.2\textheight]{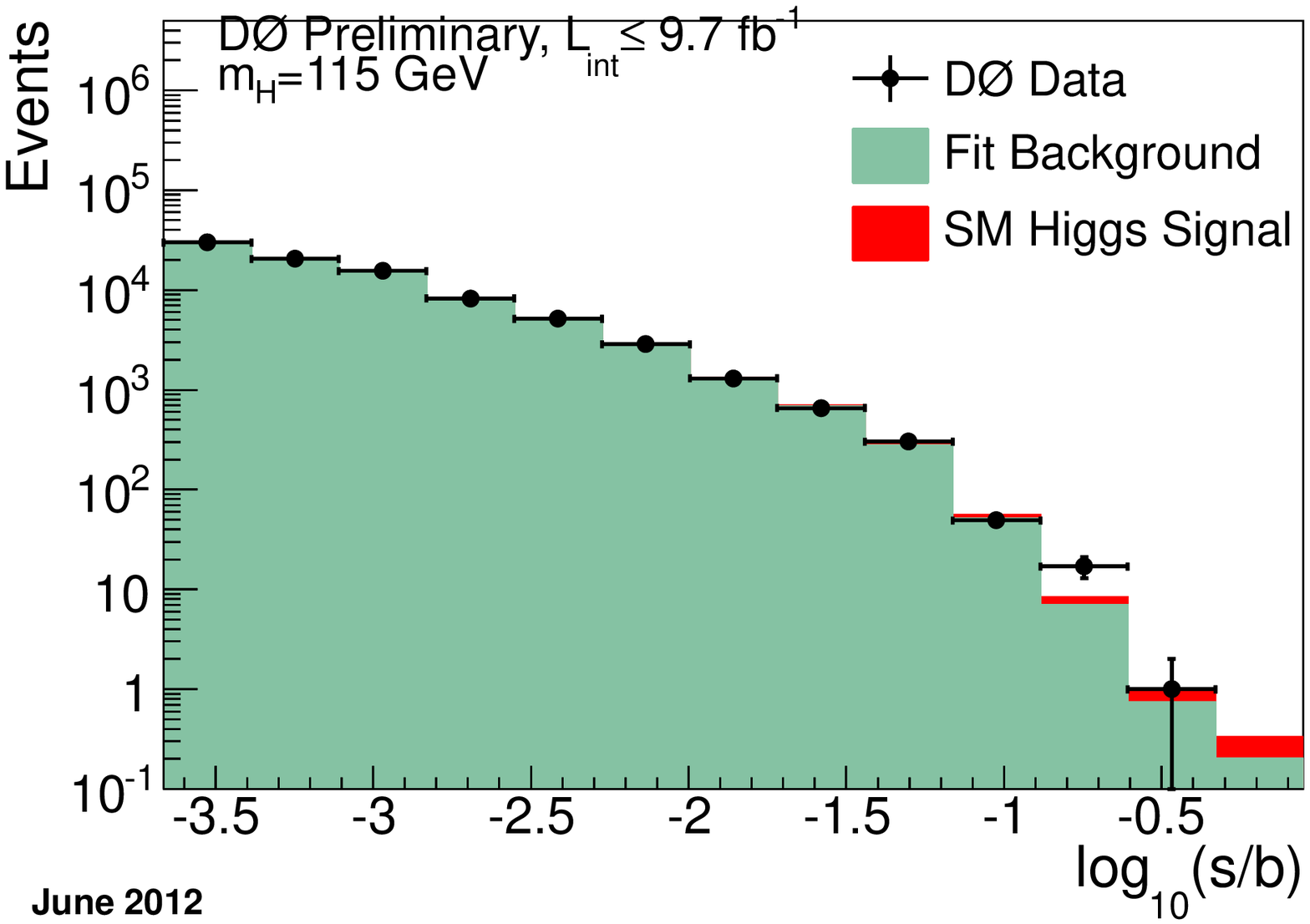}
\includegraphics[height=0.2\textheight]{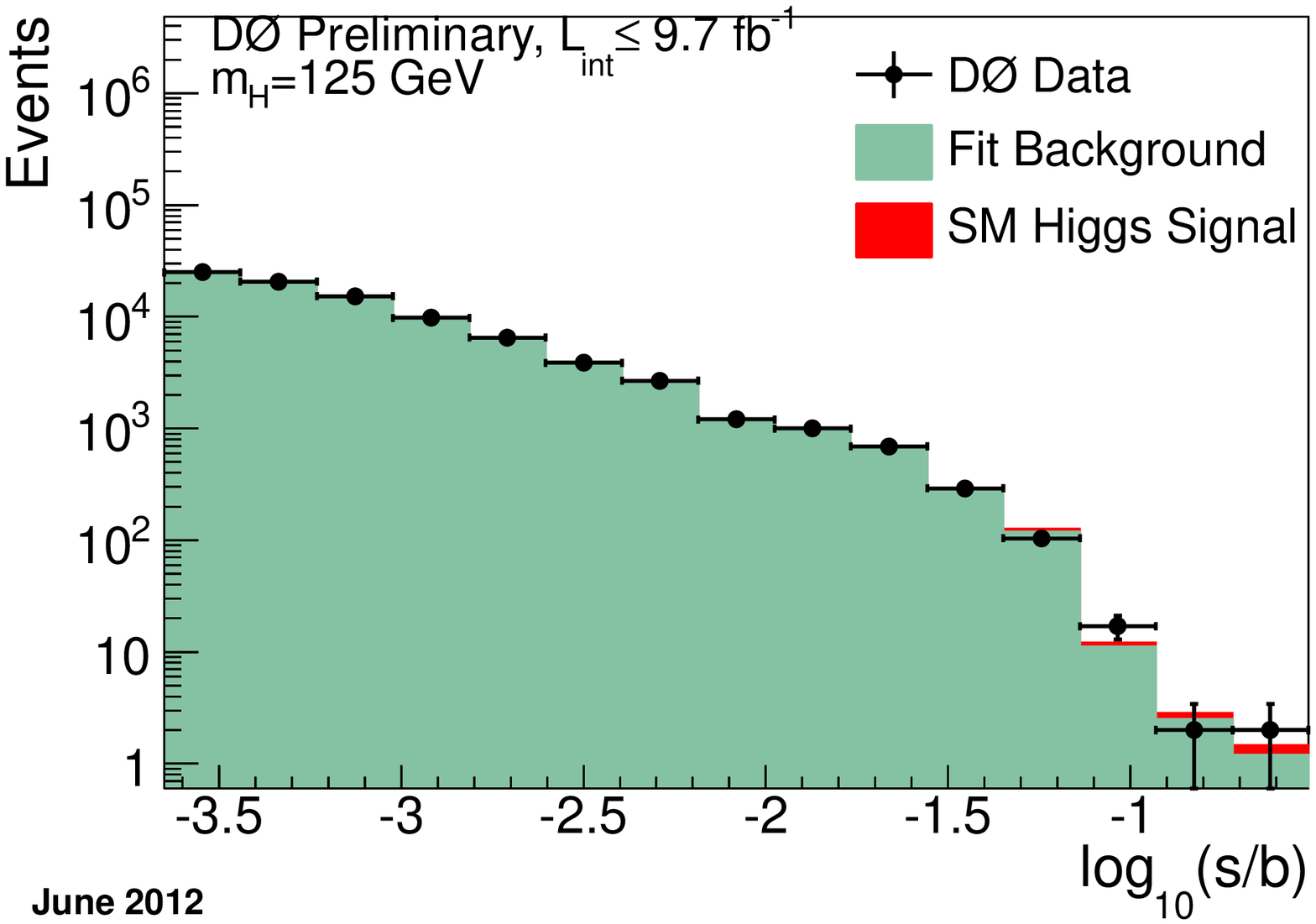} 
\includegraphics[height=0.2\textheight]{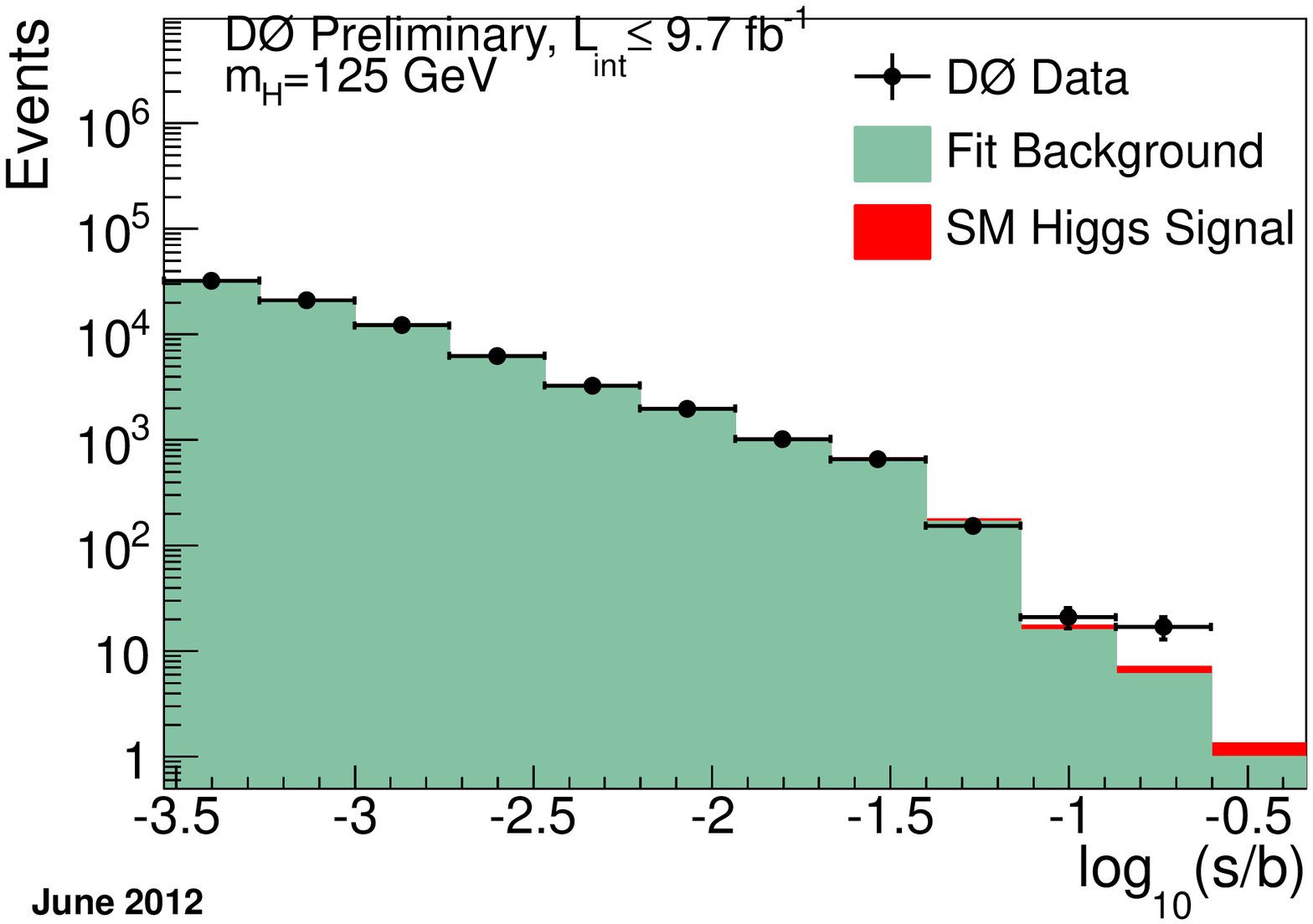}
\caption{\label{fig:sb_compare} Distributions of $\log_{10}(s/b)$ from
  this result (right) and the result in Ref. \cite{M12dzcombo} (left) for
  assumed Higgs boson masses of 115\gev\ and 125\gev.  The data are
  shown with points and the expected signal is stacked on top of the
  sum of backgrounds. Only statistical uncertainties
  on the data points are shown. }
\end{figure}

\begin{figure}[tbp]
\centering
\includegraphics[height=0.2\textheight]{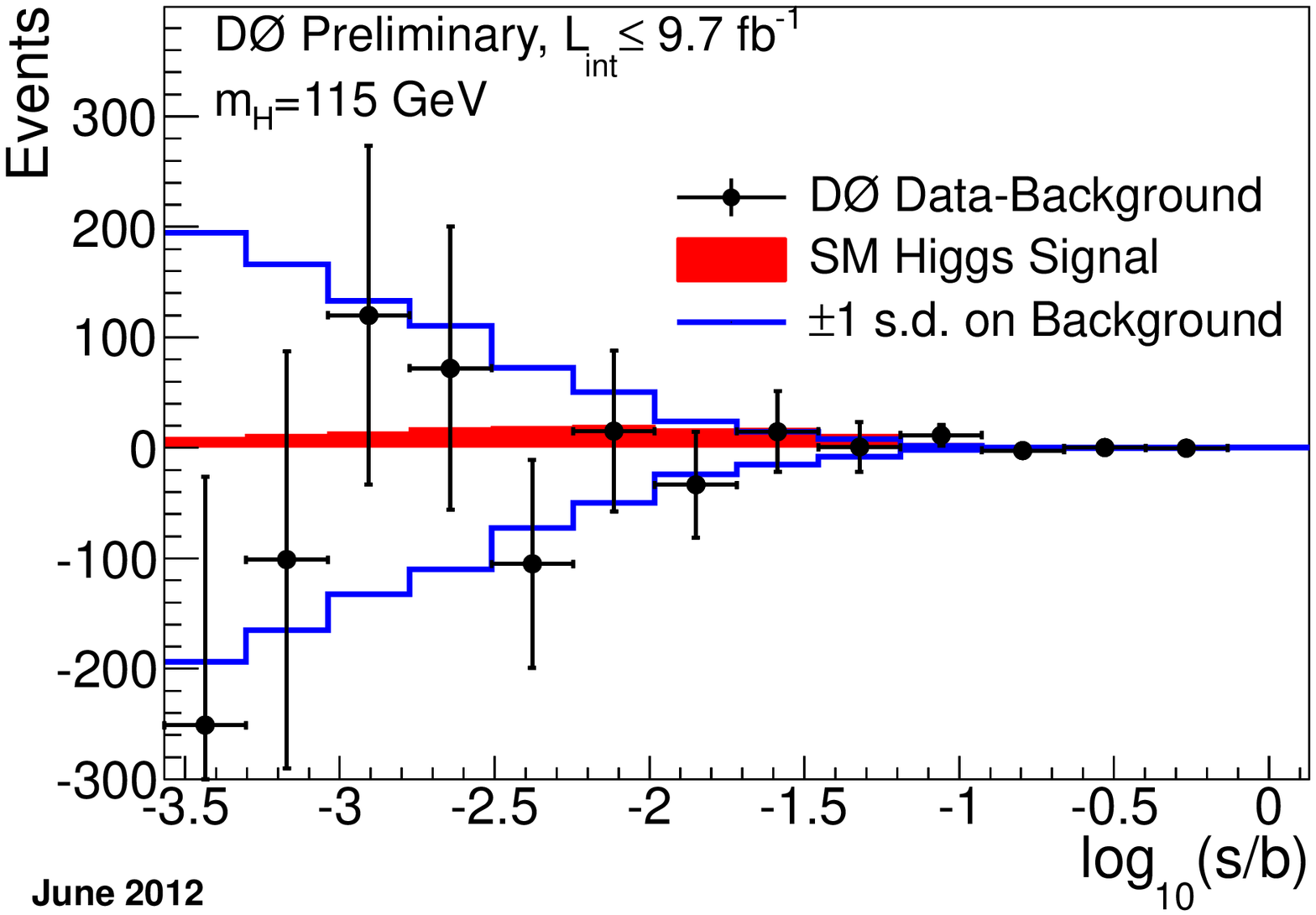} 
\includegraphics[height=0.2\textheight]{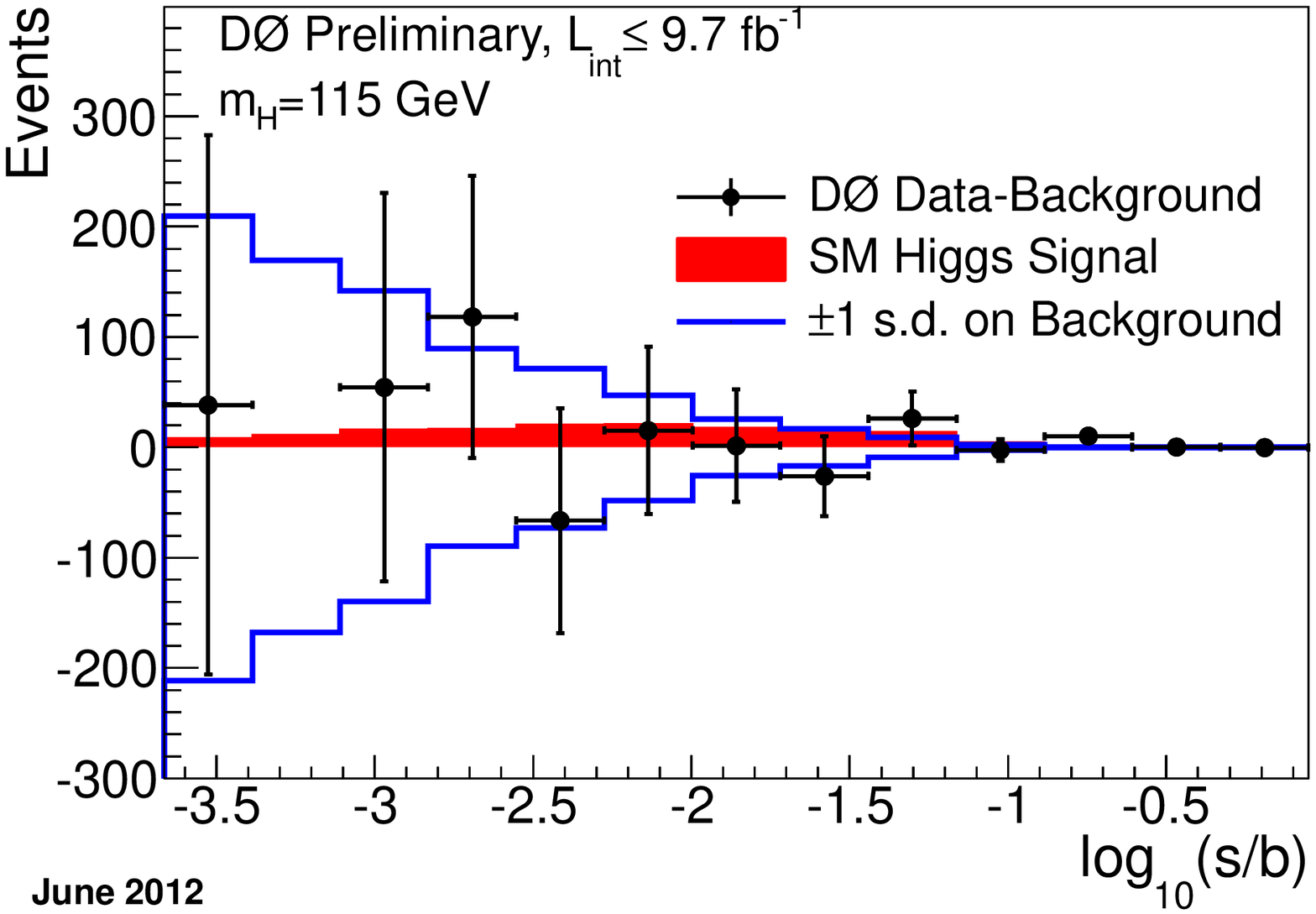}
\includegraphics[height=0.2\textheight]{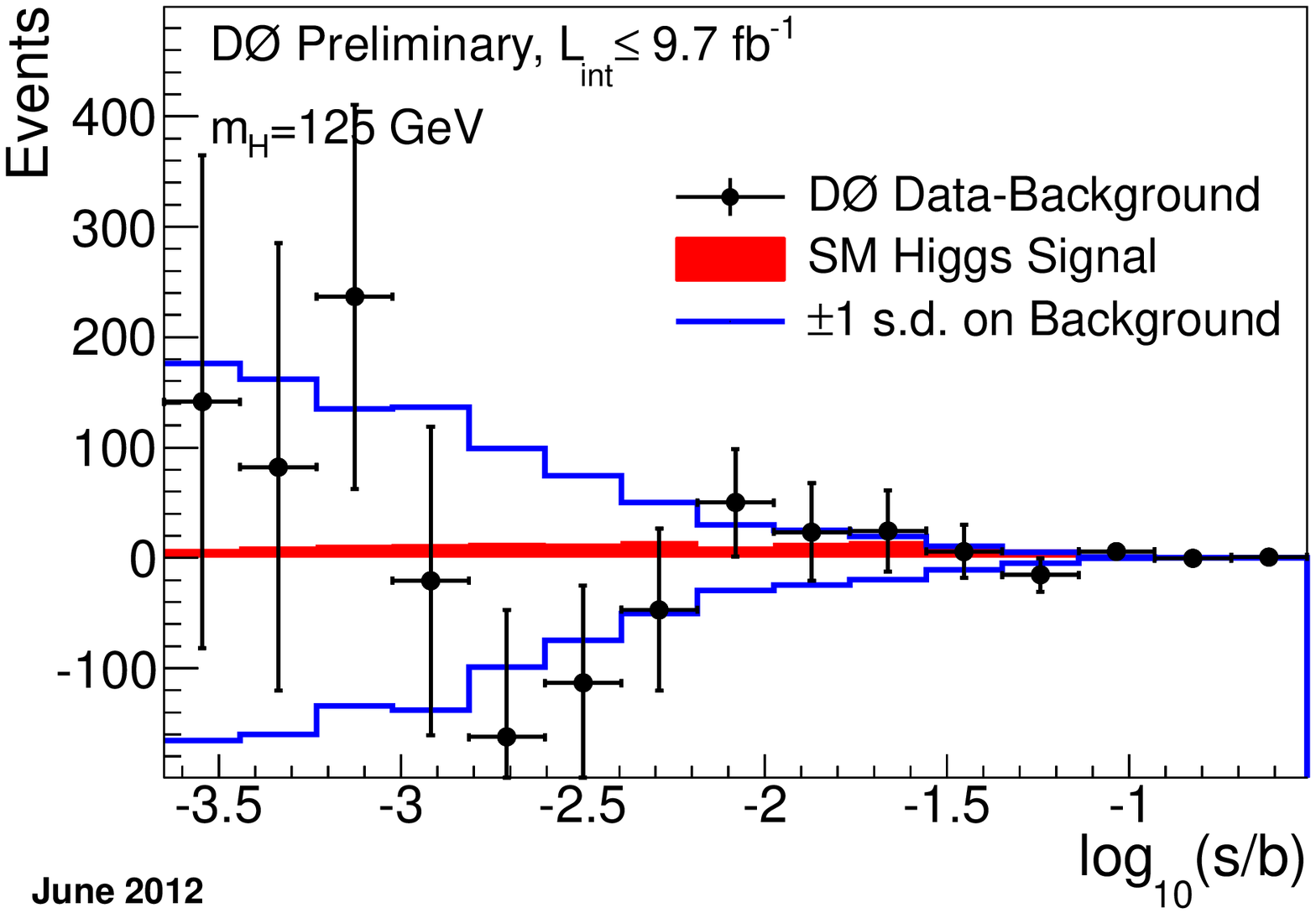} 
\includegraphics[height=0.2\textheight]{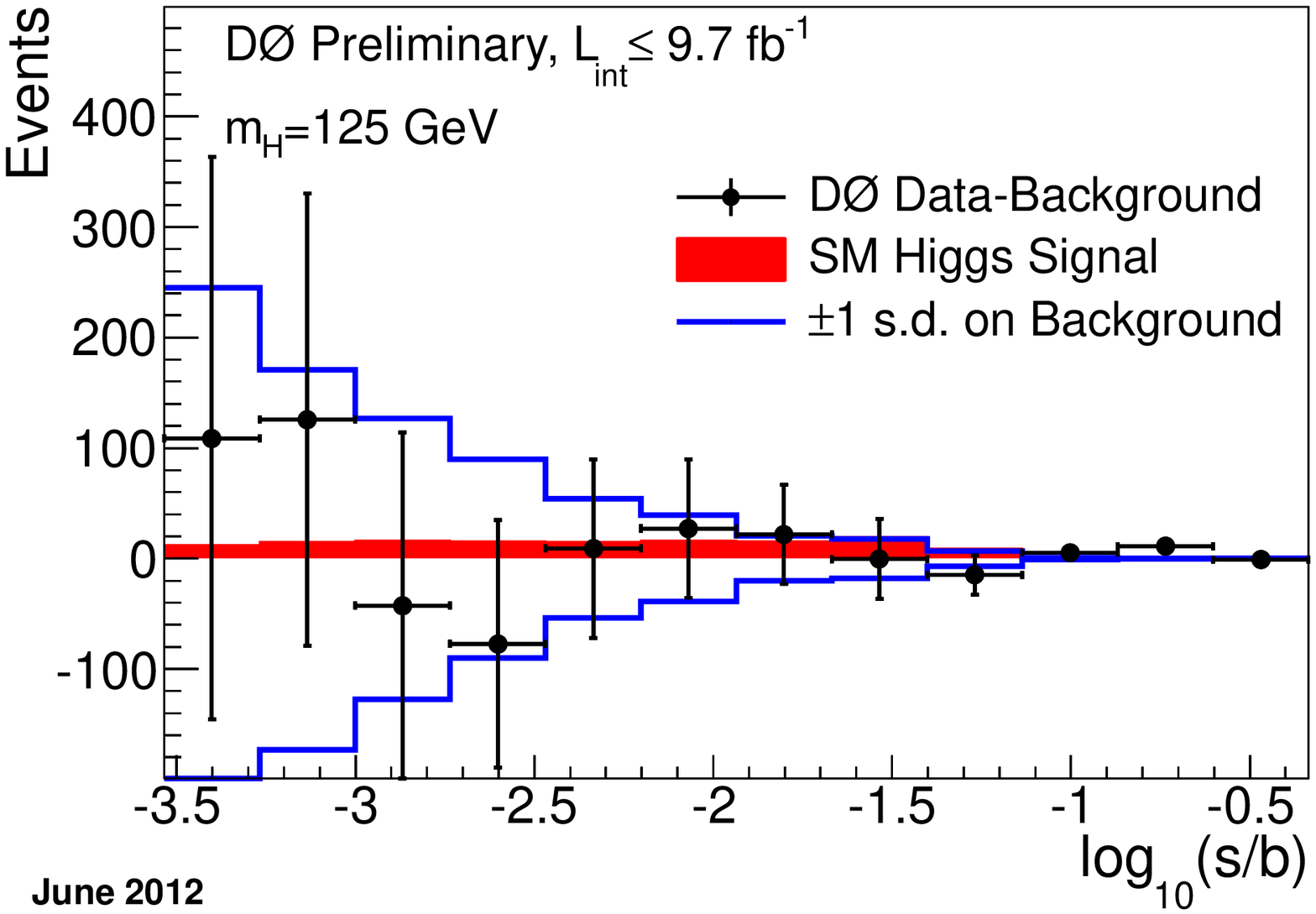}
\caption{\label{fig:sb_compare_subtract} Background-subtracted
  data distributions of $\log_{10}(s/b)$  for
  this result (right) and the result from Ref. \cite{M12dzcombo} (left) for
  assumed Higgs boson masses of 115\gev\ and 125\gev. 
  The background subtracted data are shown as
  points and the signal is shown as the red histograms.  The blue
  lines indicate the uncertainty on the background prediction.}
\end{figure}

\begin{figure}[tbp]
\includegraphics[height=0.2\textheight]{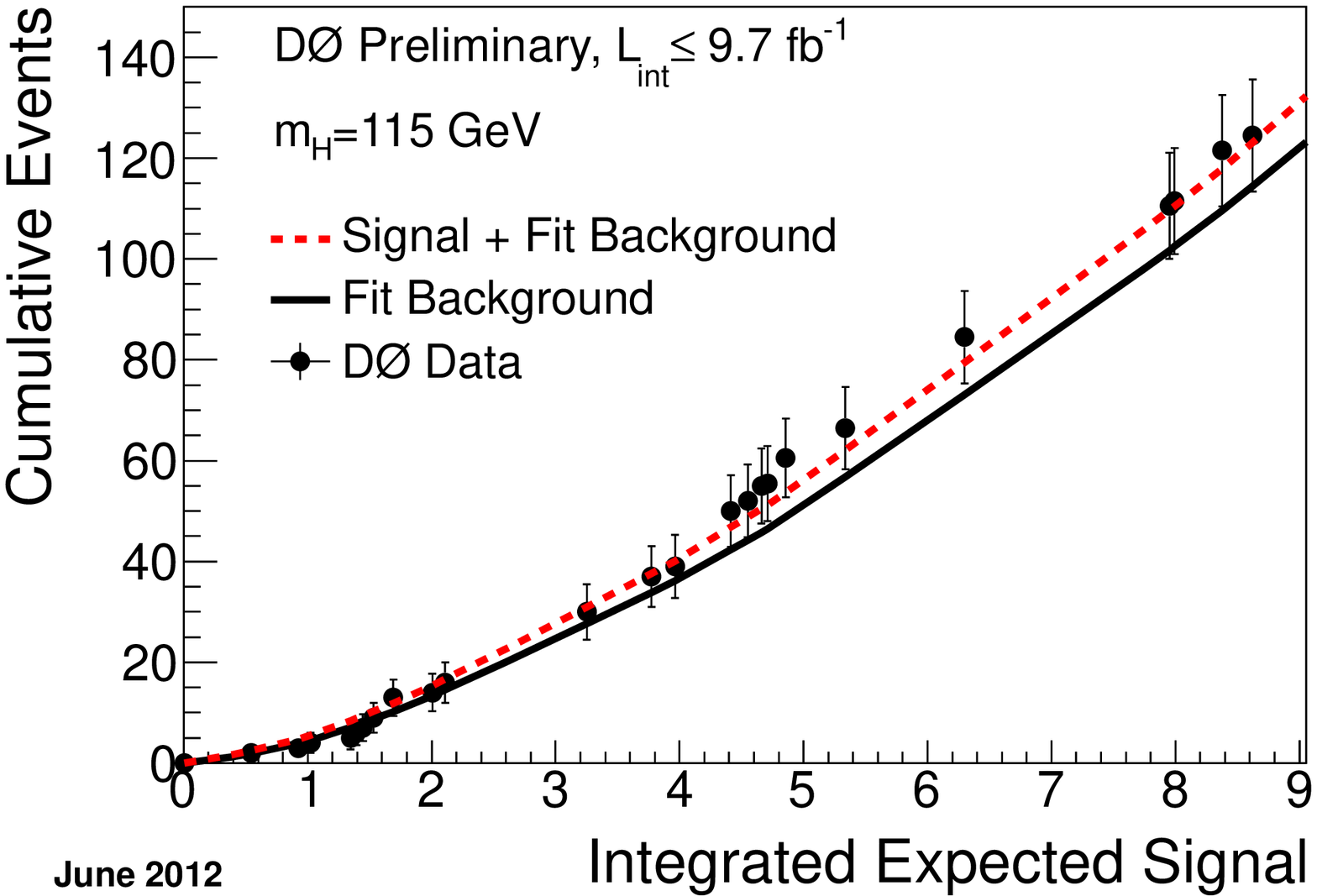} 
\includegraphics[height=0.2\textheight]{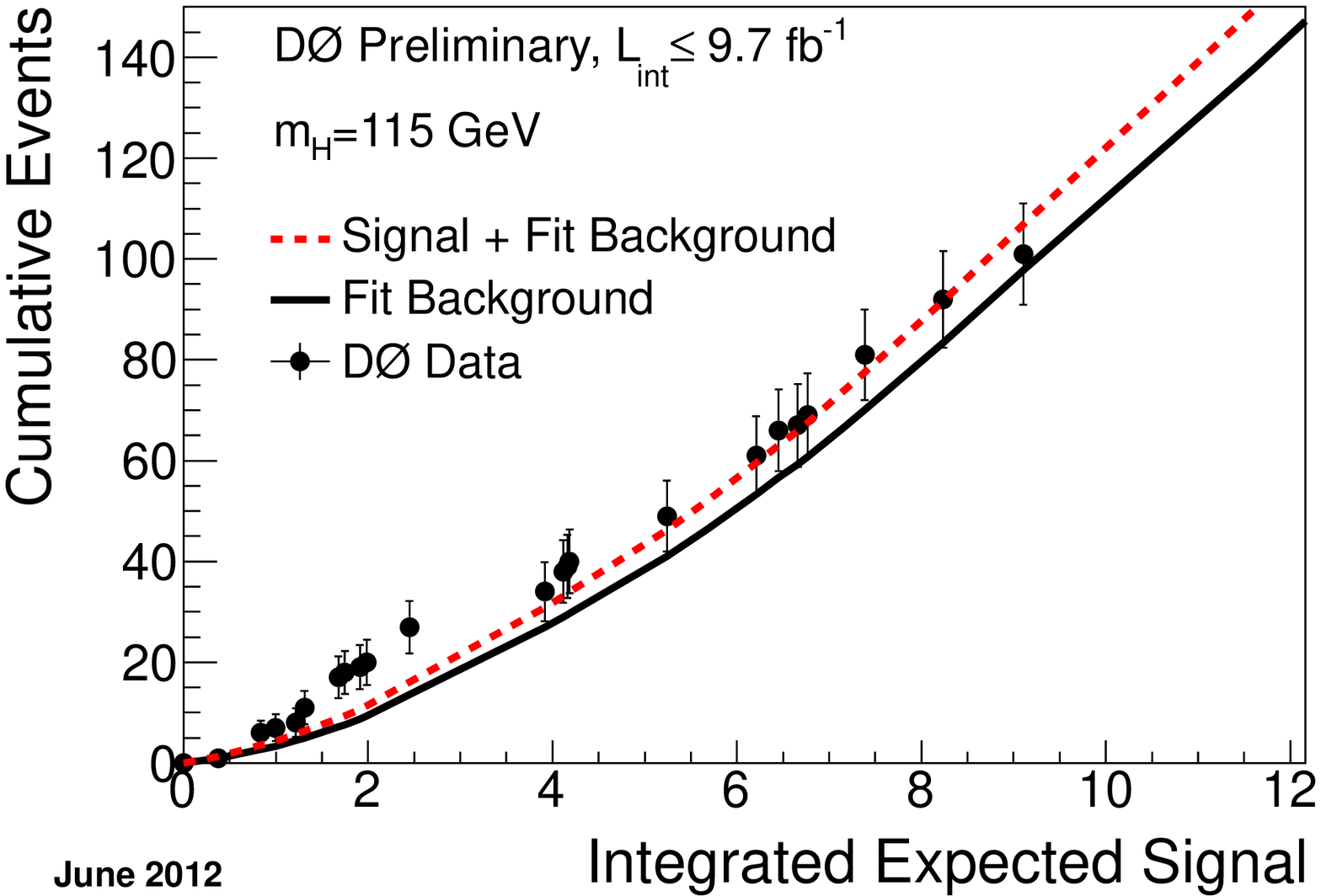}
\includegraphics[height=0.2\textheight]{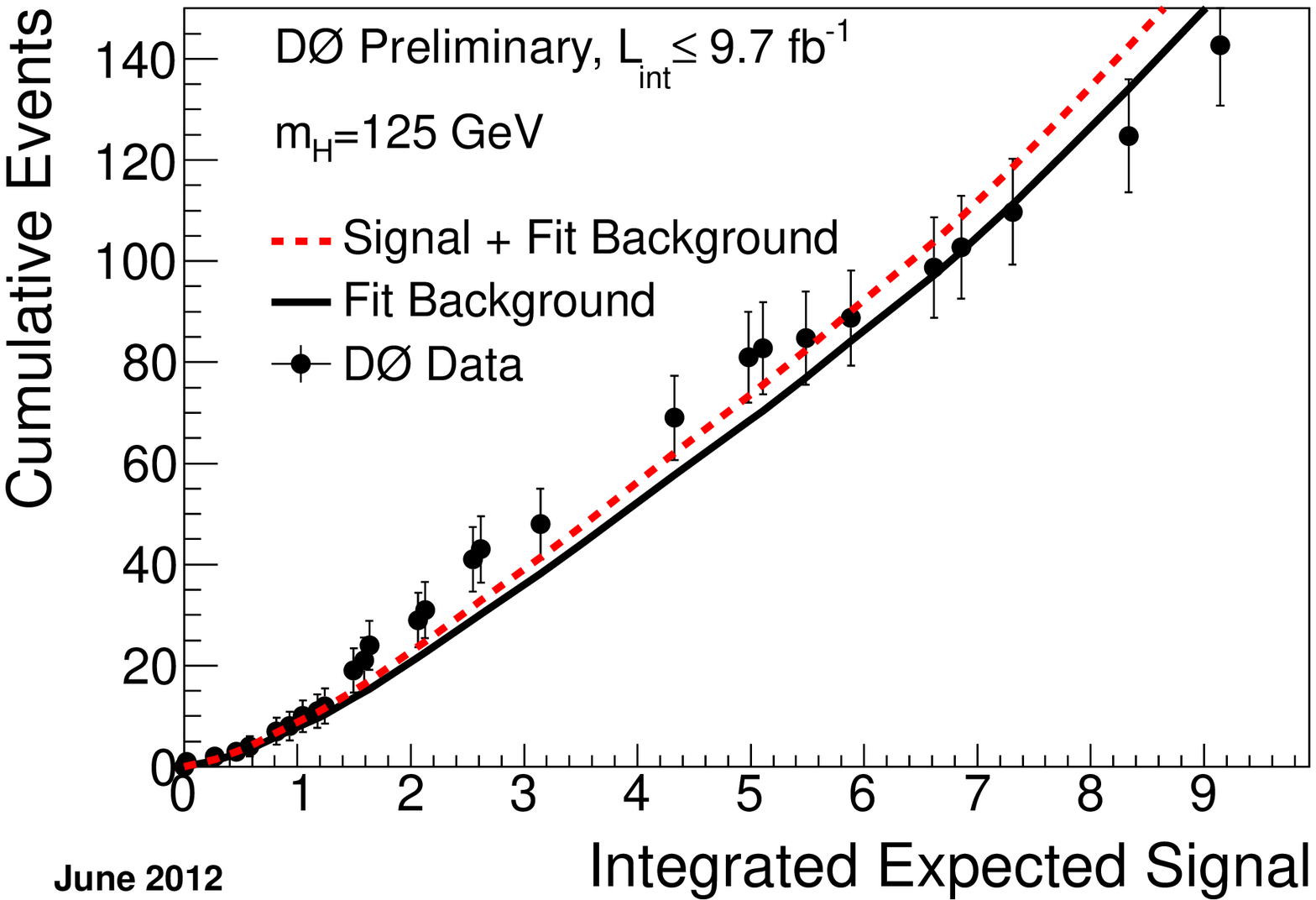} 
\includegraphics[height=0.2\textheight]{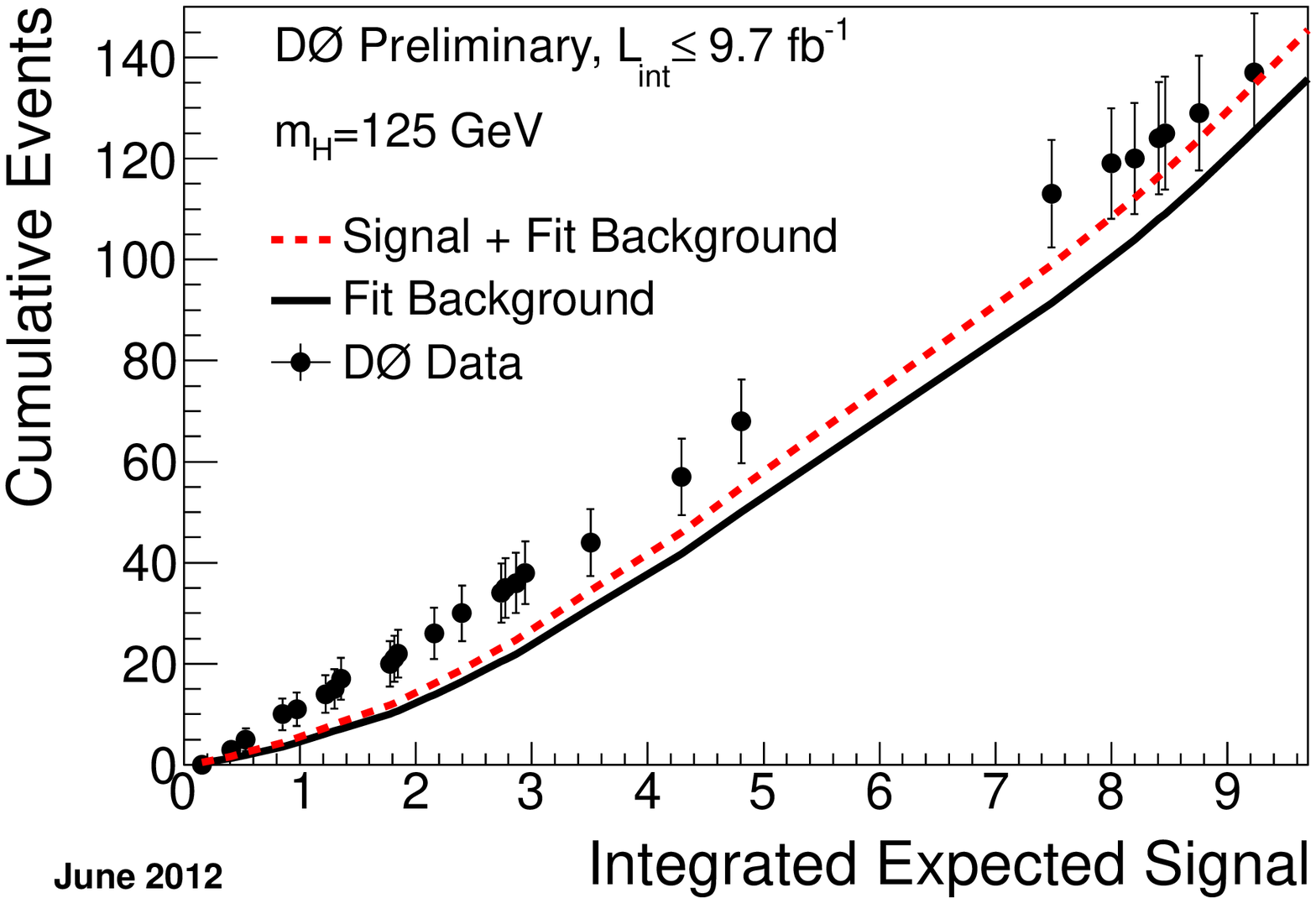}
\caption{\label{fig:compare_integral} 
Cumulative number of events for the highest $s/b$ bins for
this result (right) and the result from Ref. \cite{M12dzcombo} (left) for
assumed Higgs boson masses of 115\gev\ and 125\gev. 
The integrated background-only and signal+background predictions are shown as a
function of the accumulated number of signal events.  The points show the integrated 
number of observed events, including only the statistical uncertainty, which is correlated point-to-point.
Systematic uncertainties on the integrated background-only and signal+background
predictions are not displayed.}
\end{figure}

\begin{figure}[bp]
\centering
\includegraphics[height=0.2\textheight]{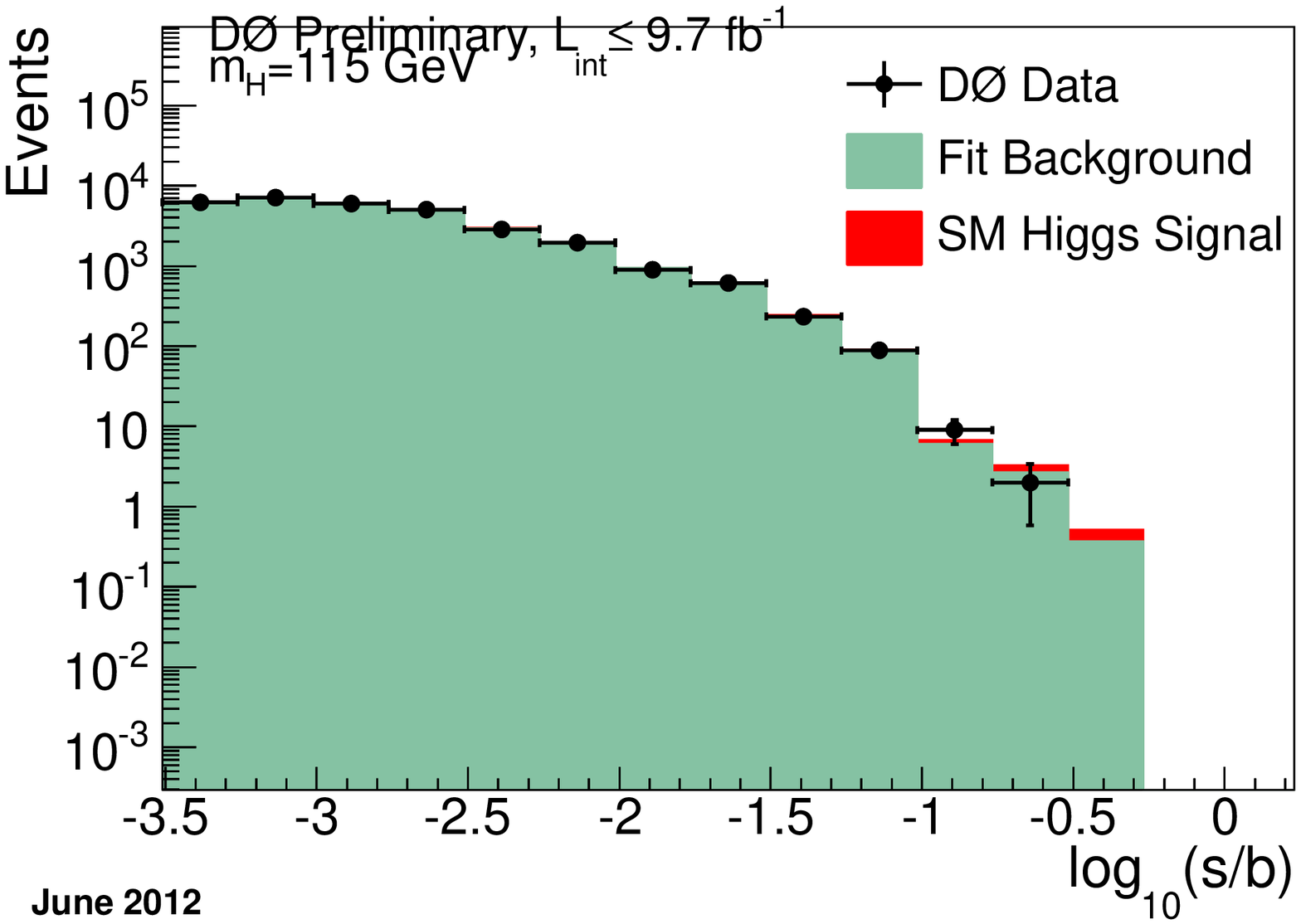} 
\includegraphics[height=0.2\textheight]{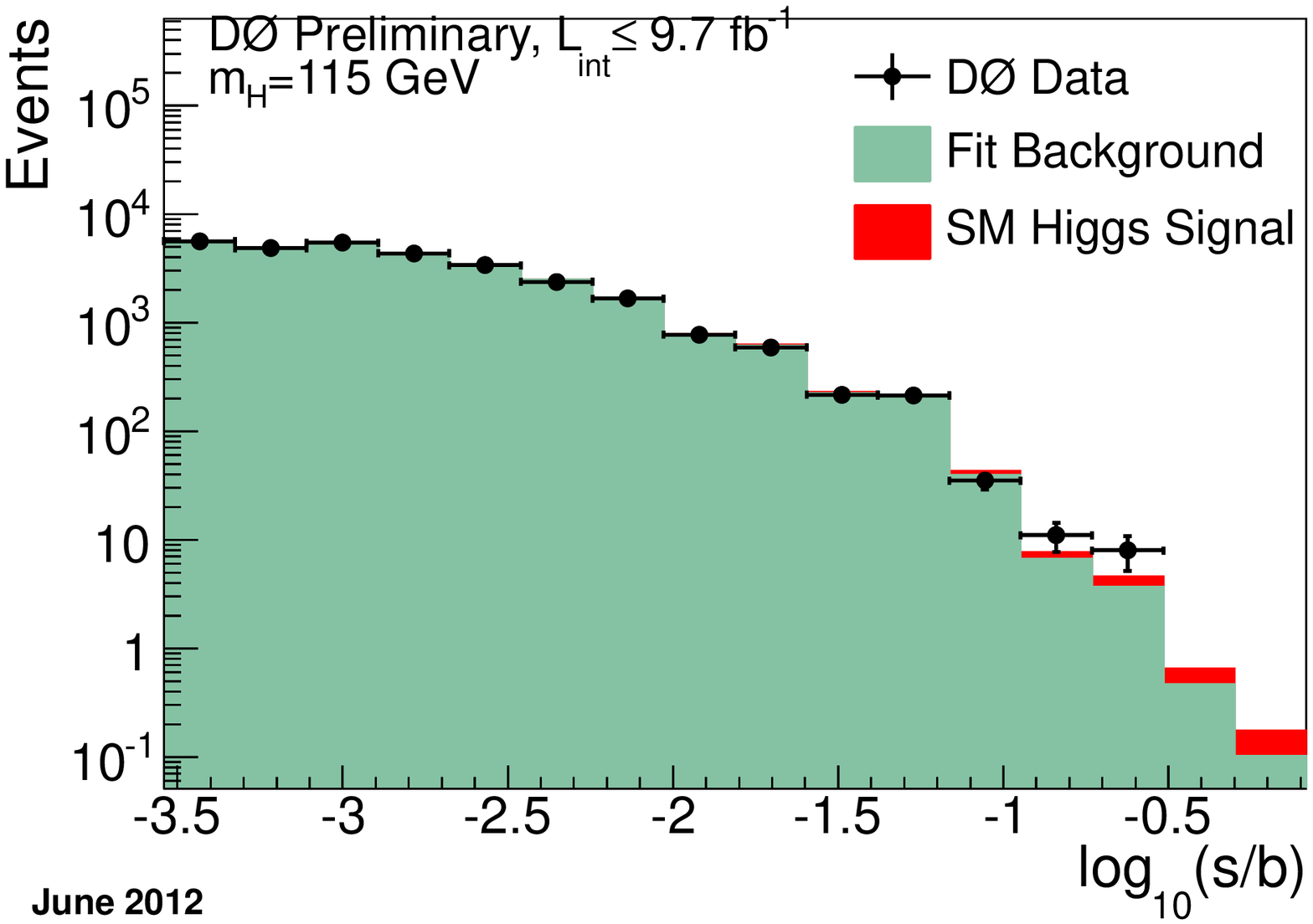}
\includegraphics[height=0.2\textheight]{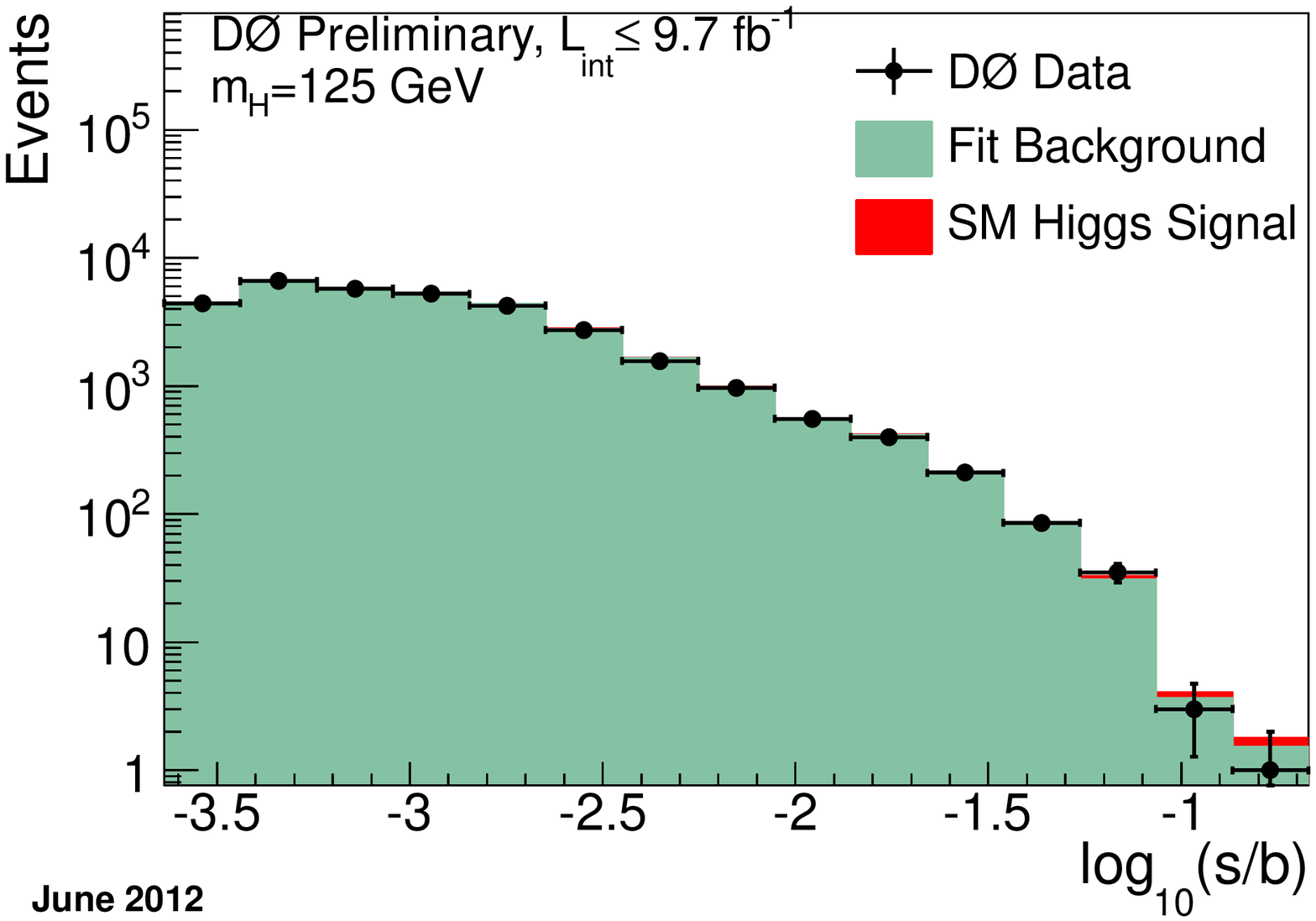} 
\includegraphics[height=0.2\textheight]{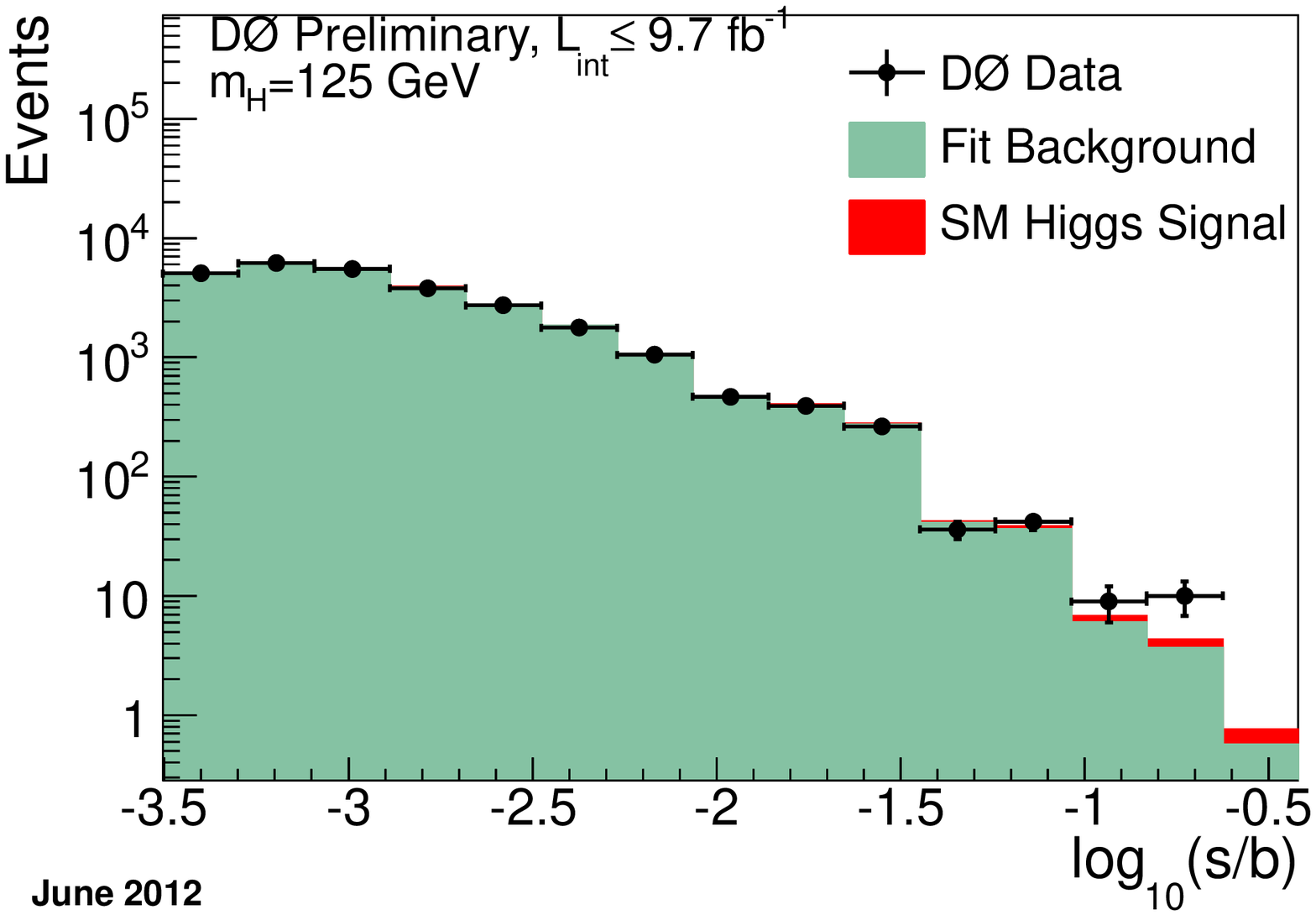}
\caption{\label{fig:HBBsb_compare} Distributions of $\log_{10}(s/b)$
  in the combined $WH/ZH, H$$\rightarrow$$ b\bar{b}$ analyses for
  this result (right) and the result from Ref. \cite{M12dzcombo} (left) for
  assumed Higgs boson masses of 115\gev\ and 125\gev.   The data are
  shown with points and the expected signal is stacked on top of the
  sum of backgrounds. Only statistical uncertainties
on the data points are shown. }
\end{figure}

\begin{figure}[tbp]
\centering
\includegraphics[height=0.2\textheight]{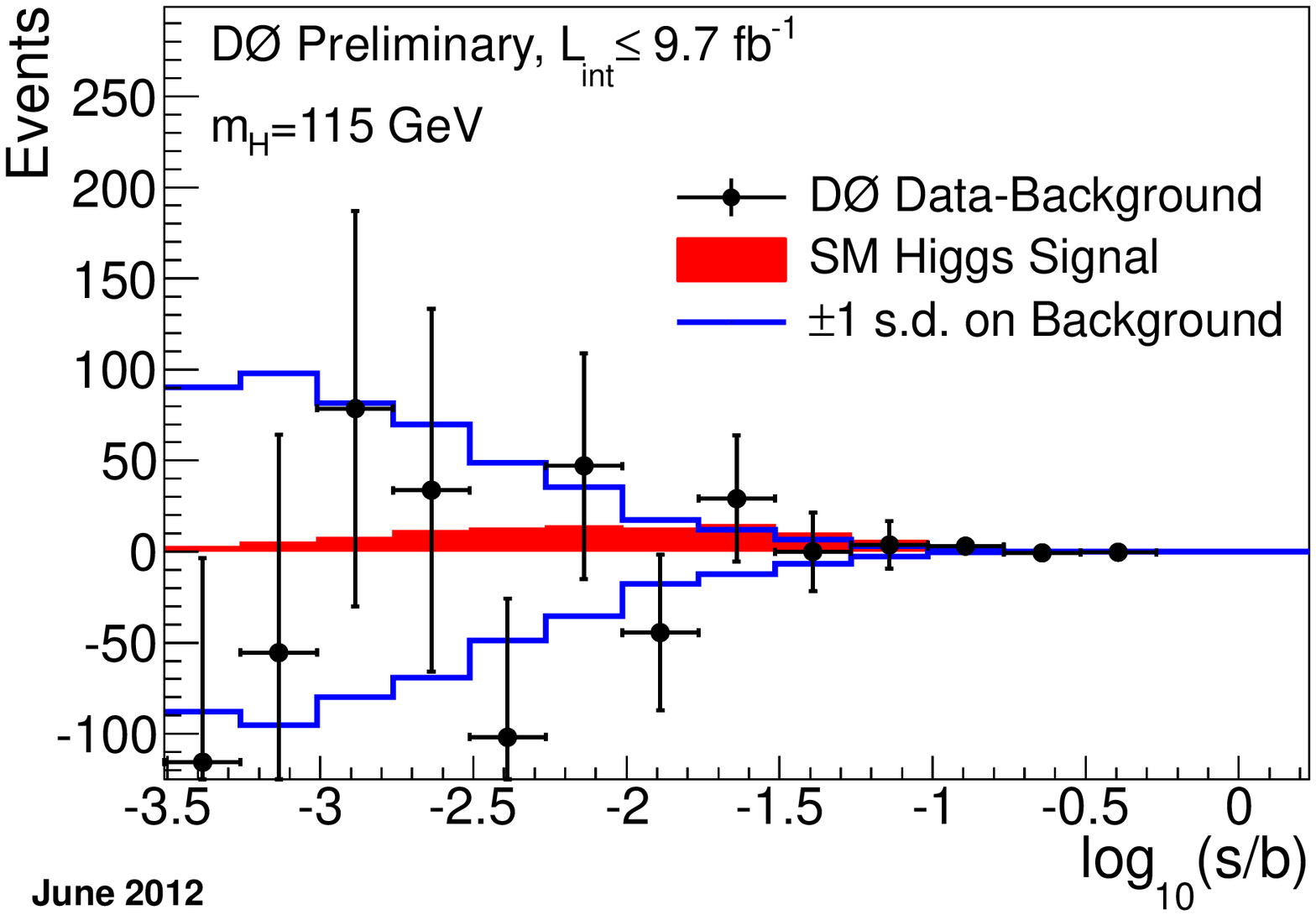} 
\includegraphics[height=0.2\textheight]{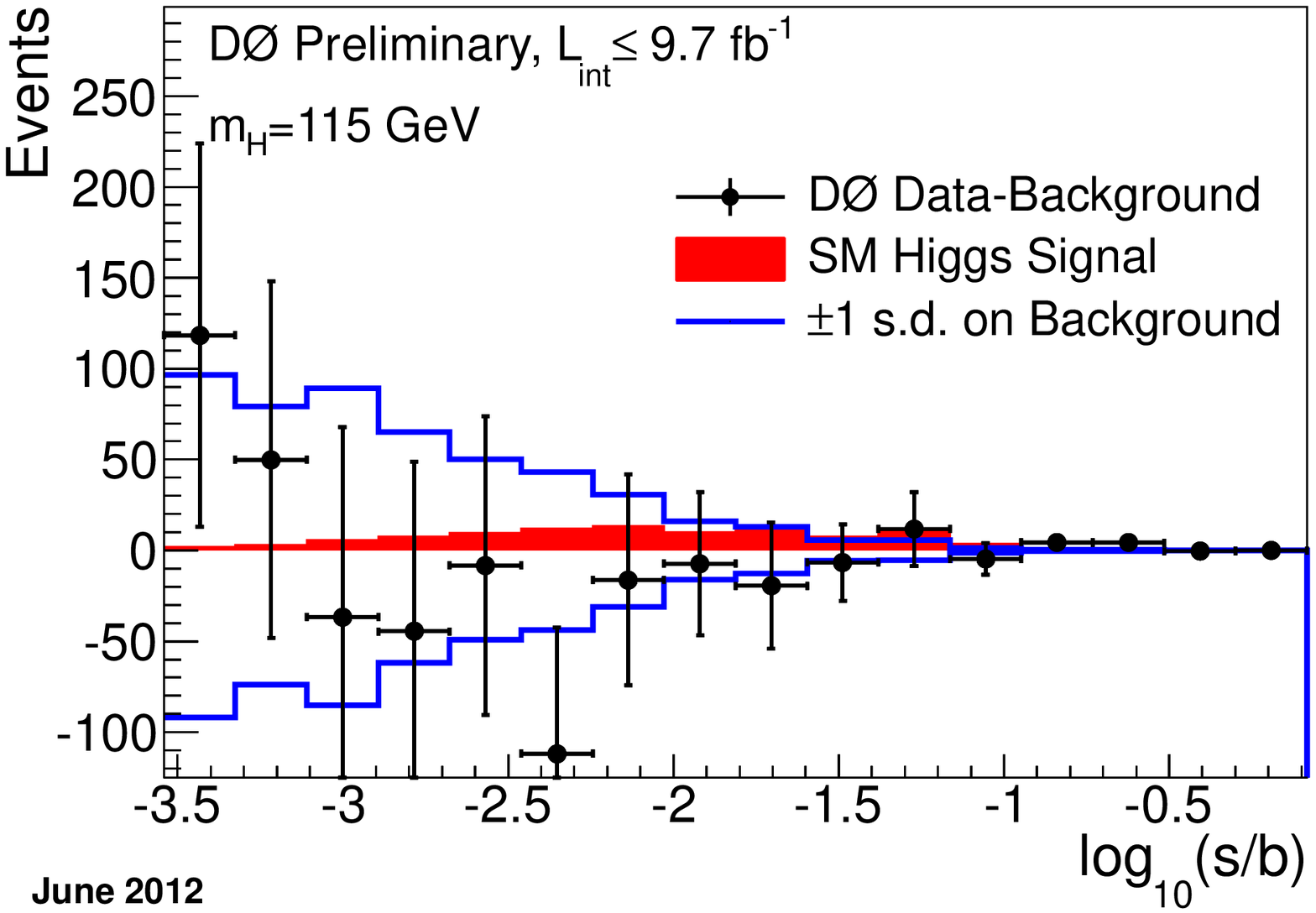}
\includegraphics[height=0.2\textheight]{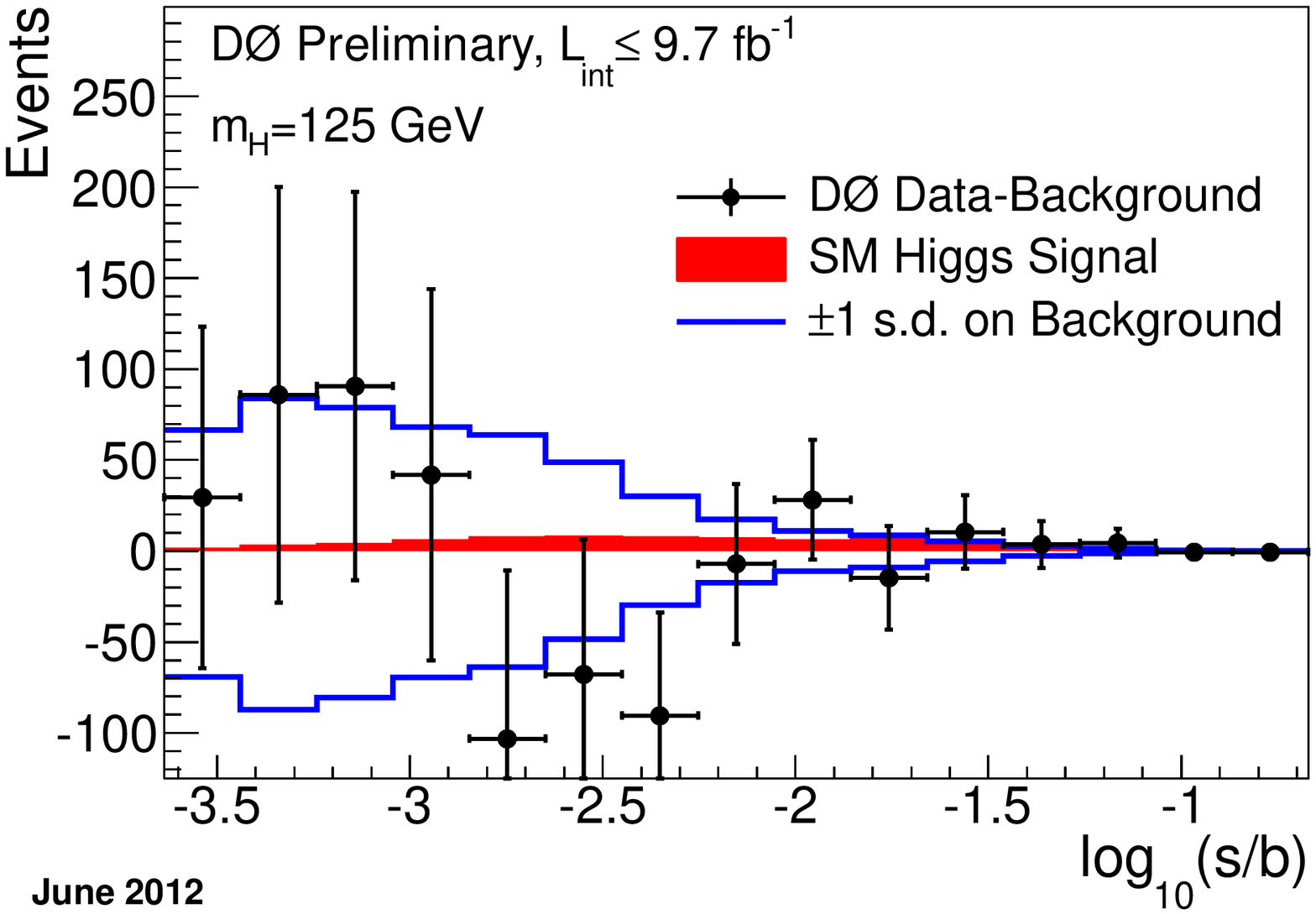} 
\includegraphics[height=0.2\textheight]{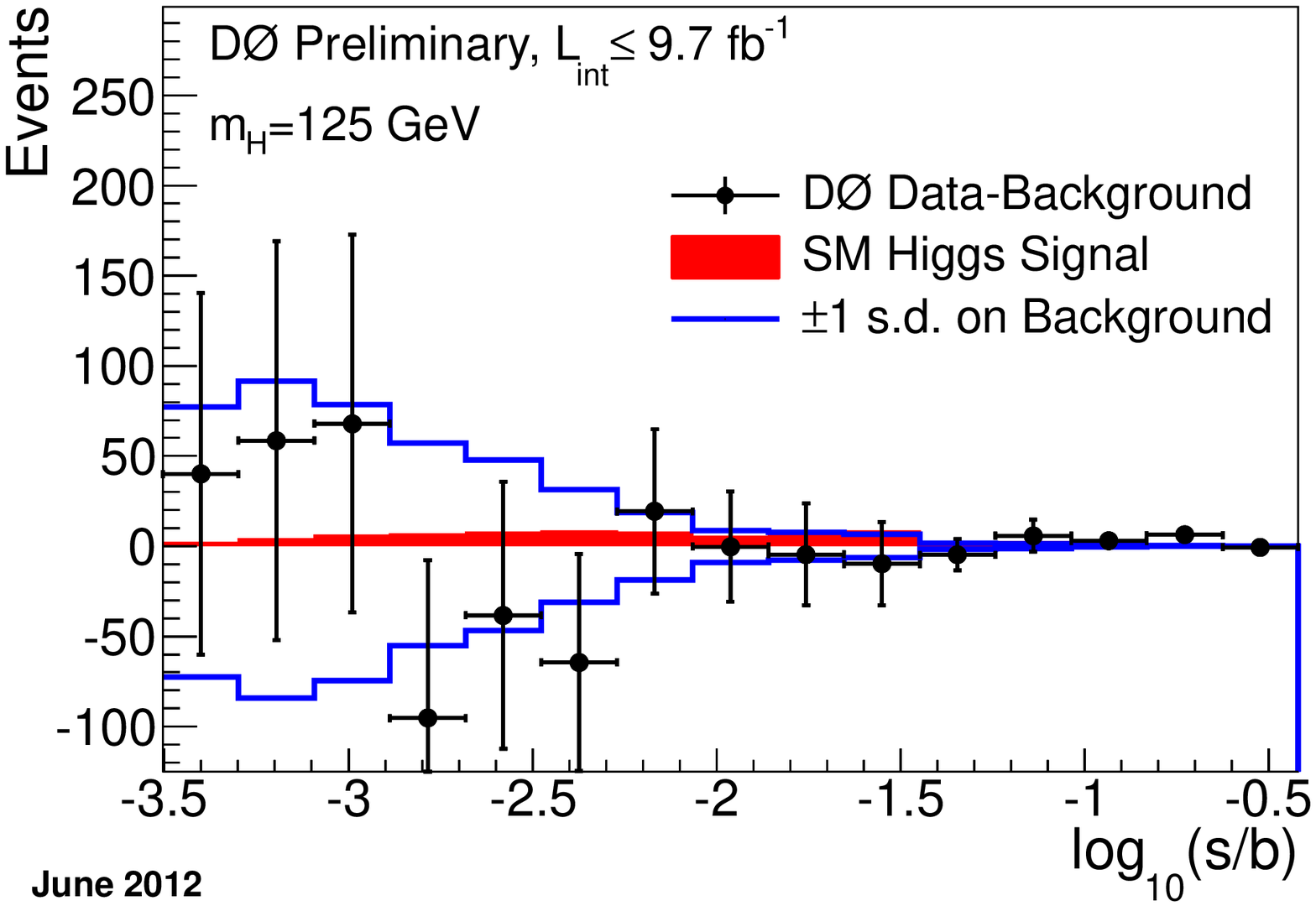}
\caption{\label{fig:HBBsb_compare_subtract} Background-subtracted
  data distributions of $\log_{10}(s/b)$ in the combined 
   $WH/ZH, H$$\rightarrow$$ b\bar{b}$ analyses for
  this result (right) and the result from Ref. \cite{M12dzcombo} (left) for
  assumed Higgs boson masses of 115\gev\ and 125\gev. 
  The background subtracted data are shown as
  points and the signal is shown as the red histograms.  The blue
  lines indicate the uncertainty on the background prediction.}
\end{figure}

\begin{figure}[tbp]
\includegraphics[height=0.2\textheight]{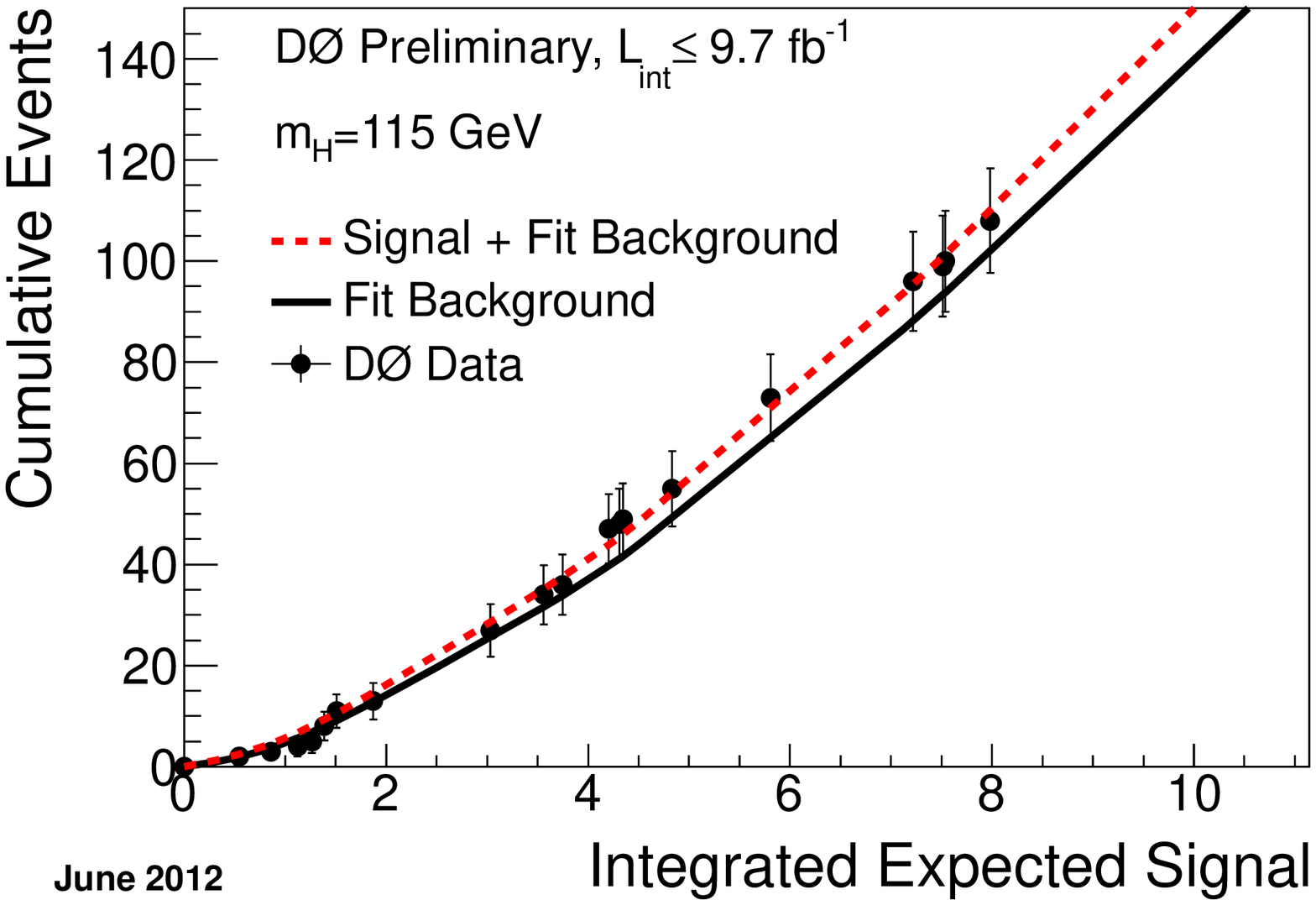} 
\includegraphics[height=0.2\textheight]{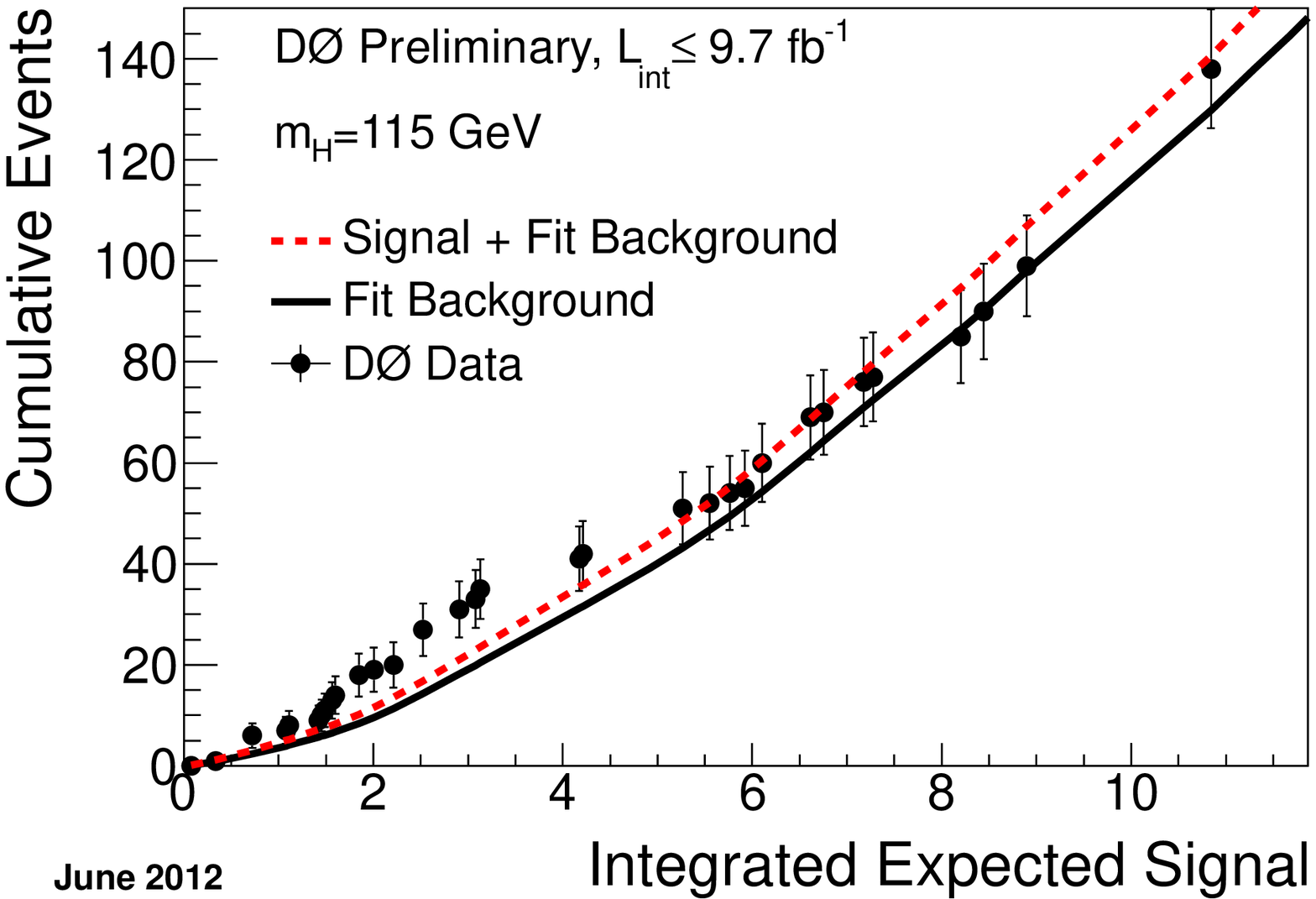}
\includegraphics[height=0.2\textheight]{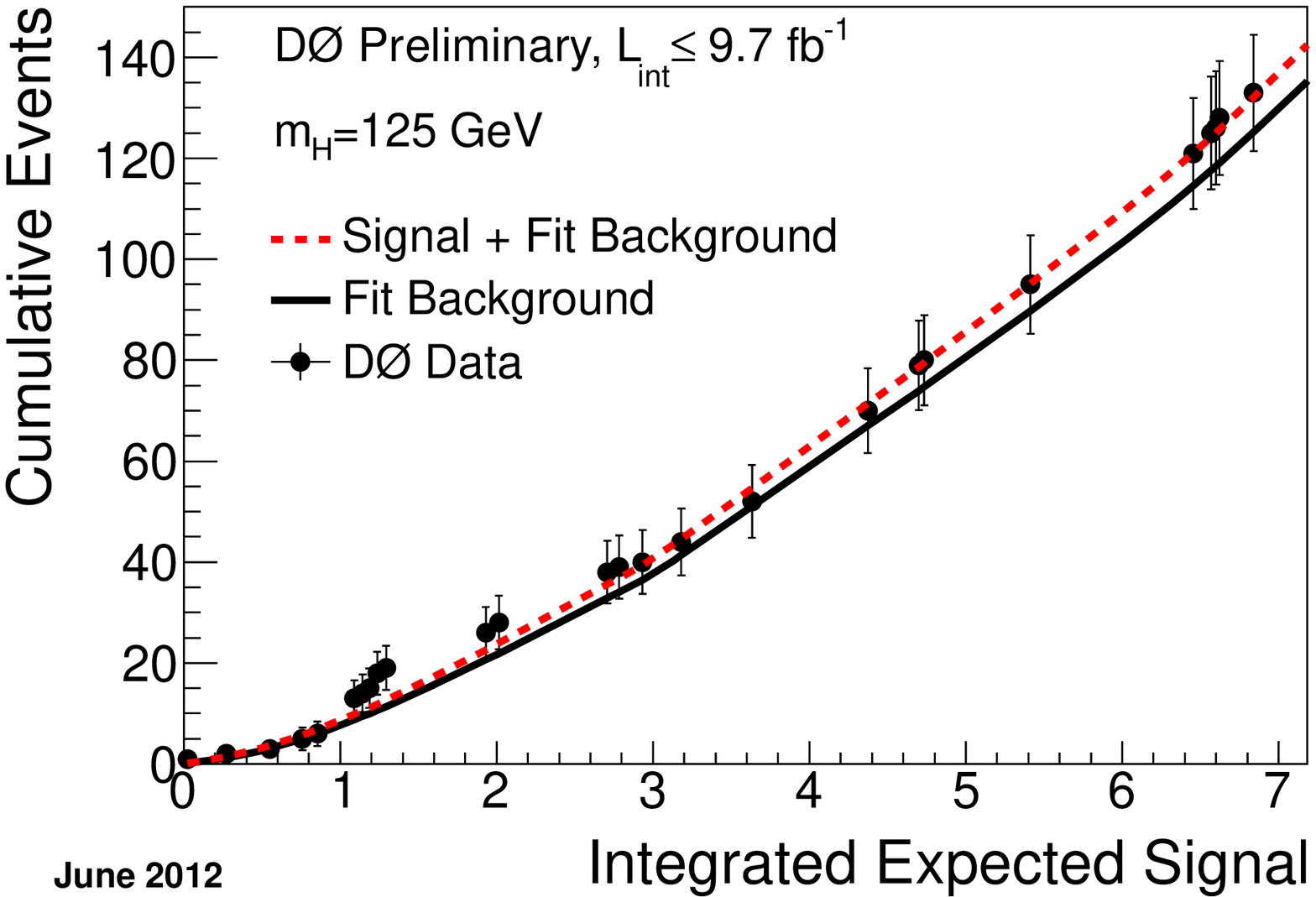} 
\includegraphics[height=0.2\textheight]{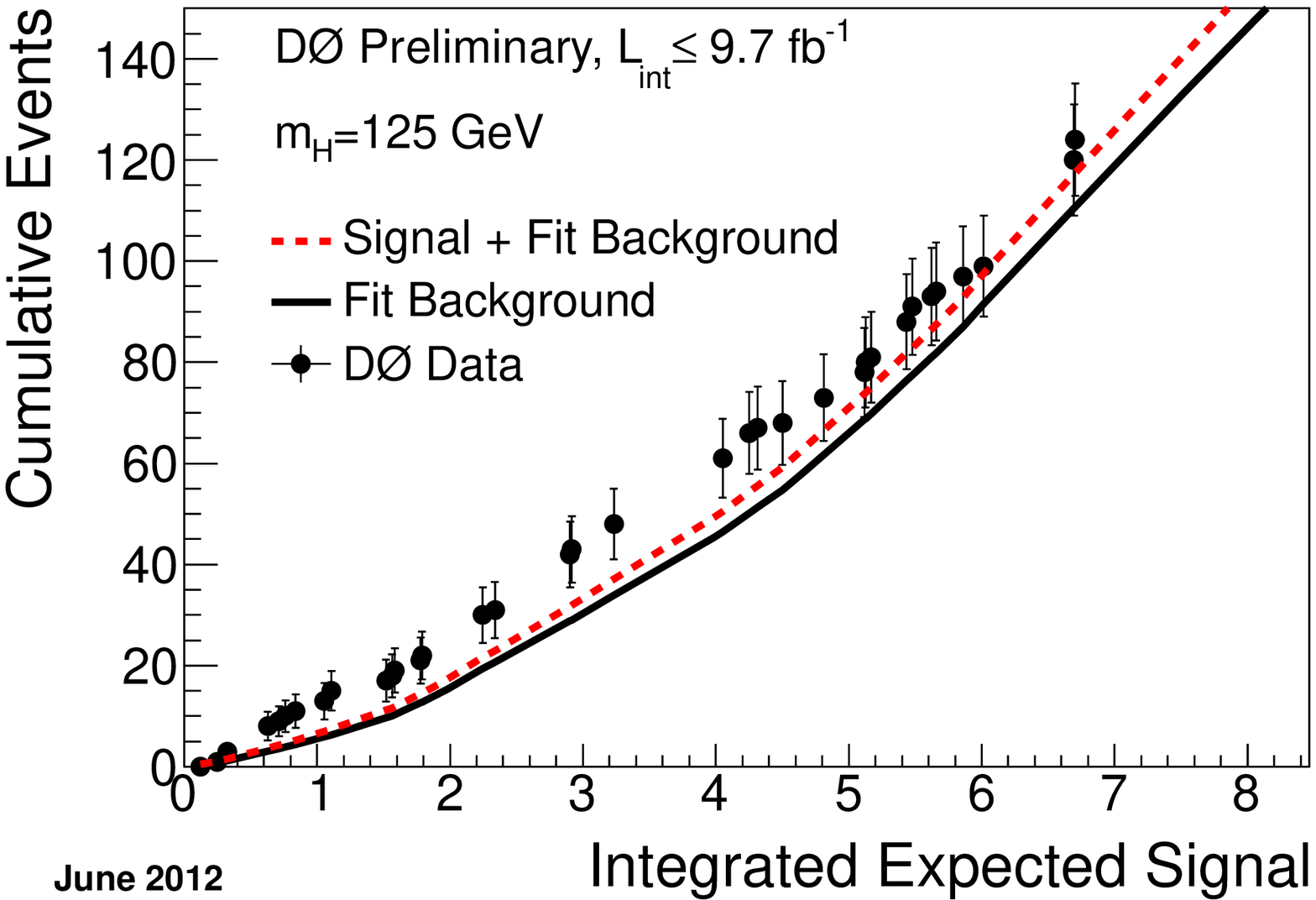}
\caption{\label{fig:HBBcompare_integral} 
Cumulative number of events for the highest $s/b$ bins 
in the combined $WH/ZH, H$$\rightarrow$$ b\bar{b}$ analyses for
  this result (right) and the result from Ref. \cite{M12dzcombo} (left) for
  assumed Higgs boson masses of 115\gev\ and 125\gev. .  The integrated background-only and
signal+background predictions are shown as a function of the
accumulated number of signal events.  The points show the integrated number of observed
events, including only the statistical uncertainty, which is correlated point-to-point.
Systematic uncertainties on the integrated background-only and signal+background
predictions are not displayed.}
\end{figure}

\begin{figure}[bp]
\centering
\includegraphics[height=0.2\textheight]{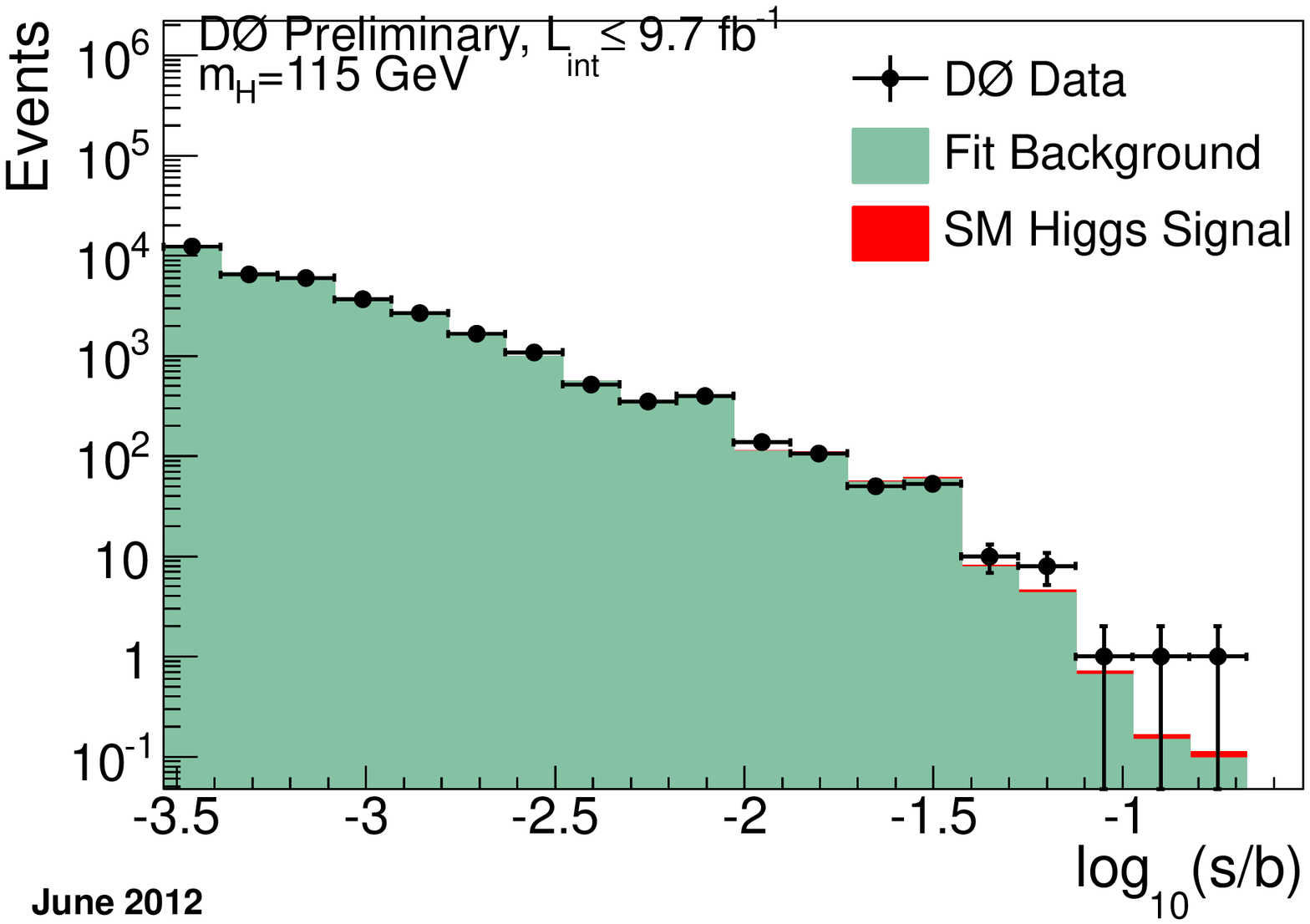} 
\includegraphics[height=0.2\textheight]{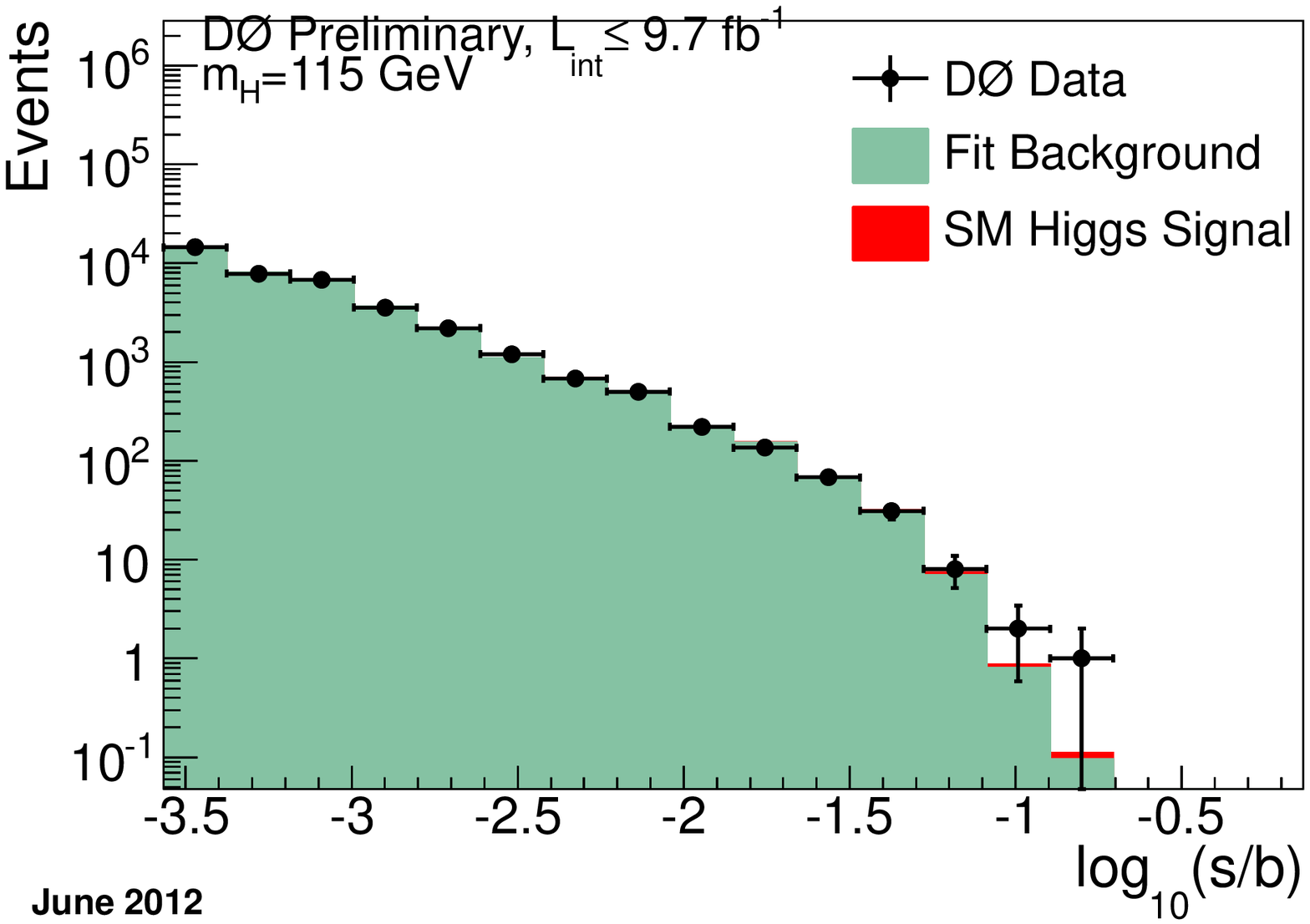}
\includegraphics[height=0.2\textheight]{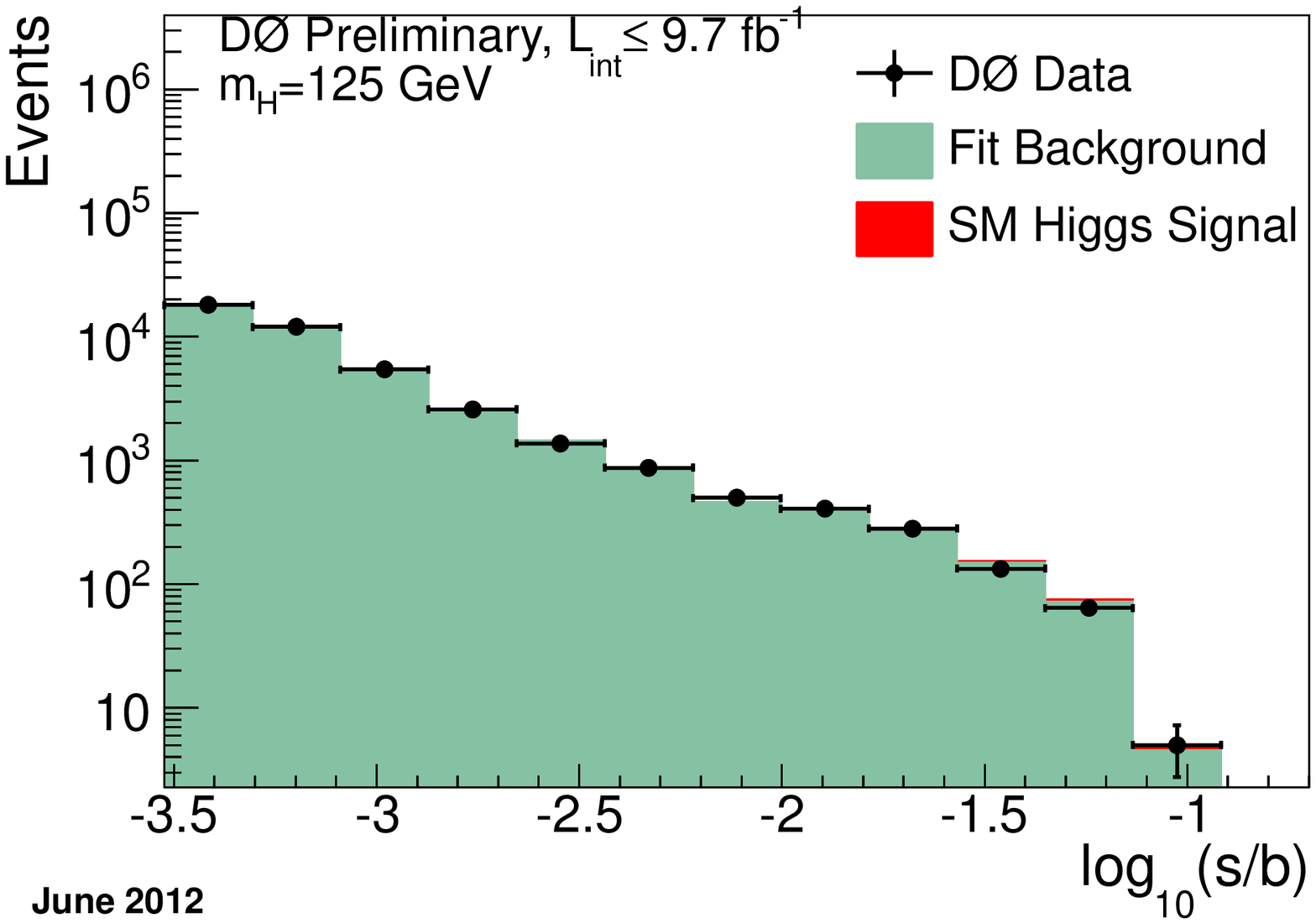} 
\includegraphics[height=0.2\textheight]{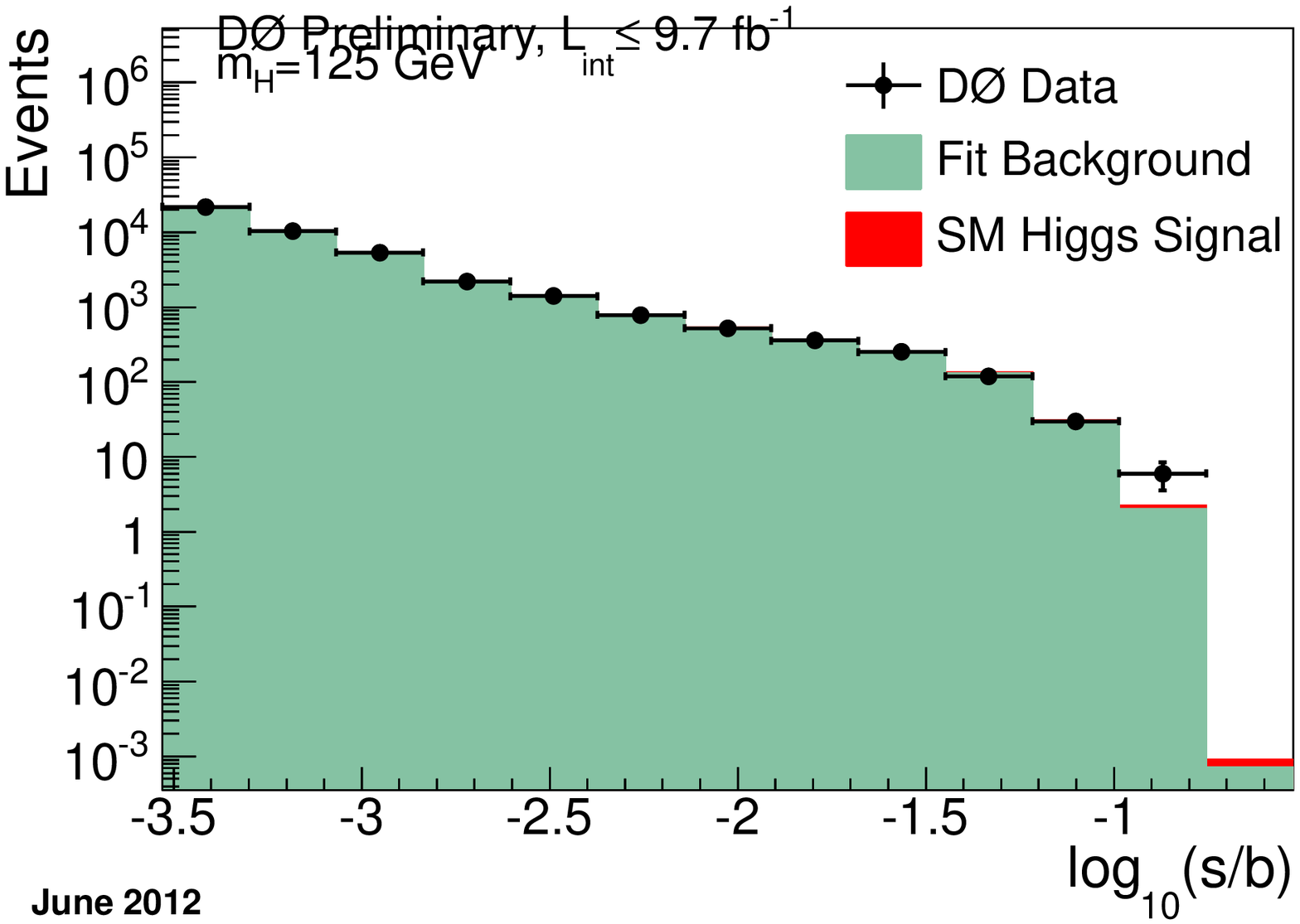}
\caption{\label{fig:HWWsb_compare} Distributions of $\log_{10}(s/b)$
  in the combined $WH/ZH/H, H$$\rightarrow$$ W^+W^-$ analyses
  for this result (right) and the result from Ref. \cite{M12dzcombo} (left) for
  assumed Higgs boson masses of 115\gev\ and 125\gev. .  The data are
  shown with points and the expected signal is stacked on top of the
  sum of backgrounds. Only statistical uncertainties
on the data points are shown. }
\end{figure}

\begin{figure}[tbp]
\centering
\includegraphics[height=0.2\textheight]{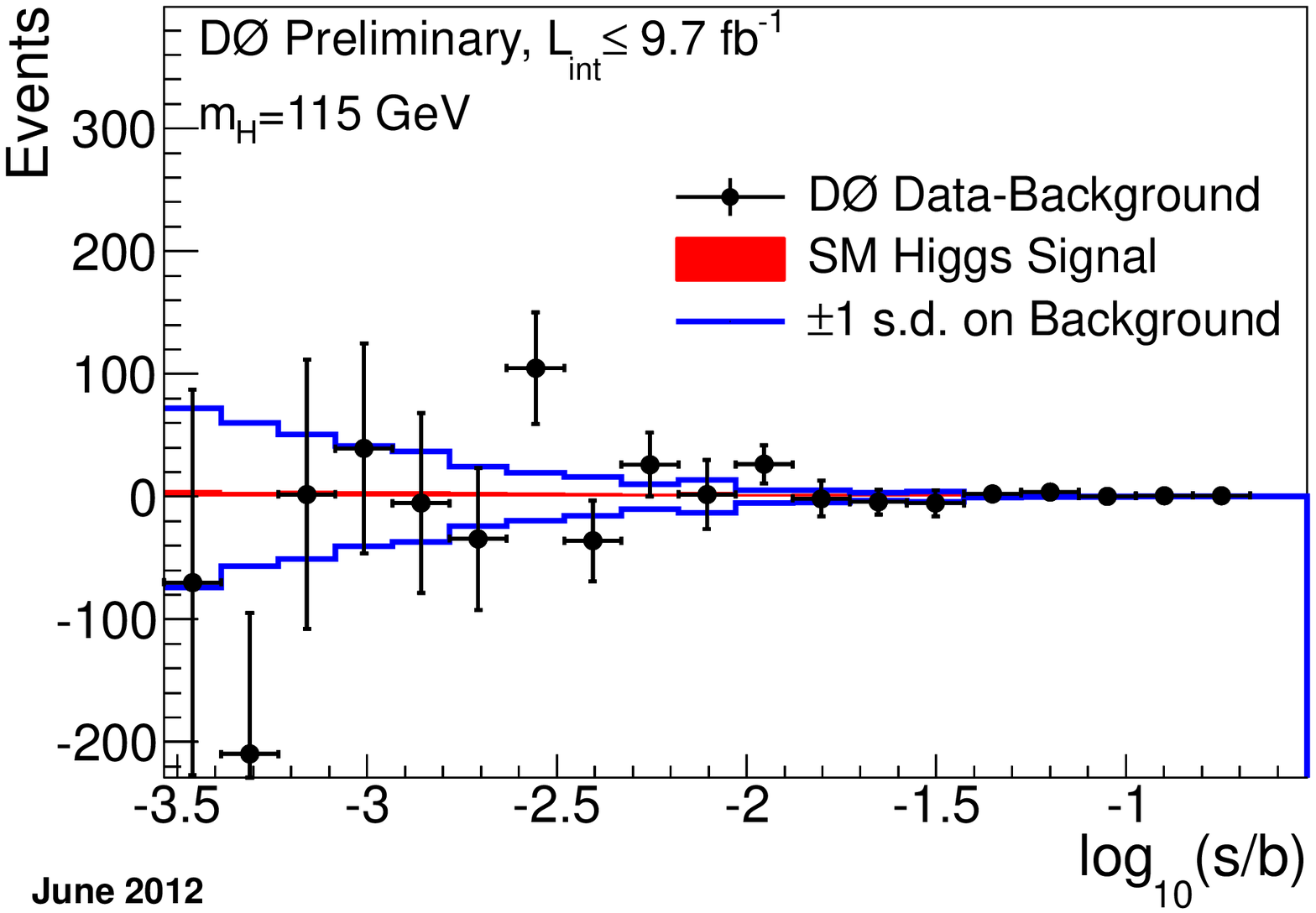} 
\includegraphics[height=0.2\textheight]{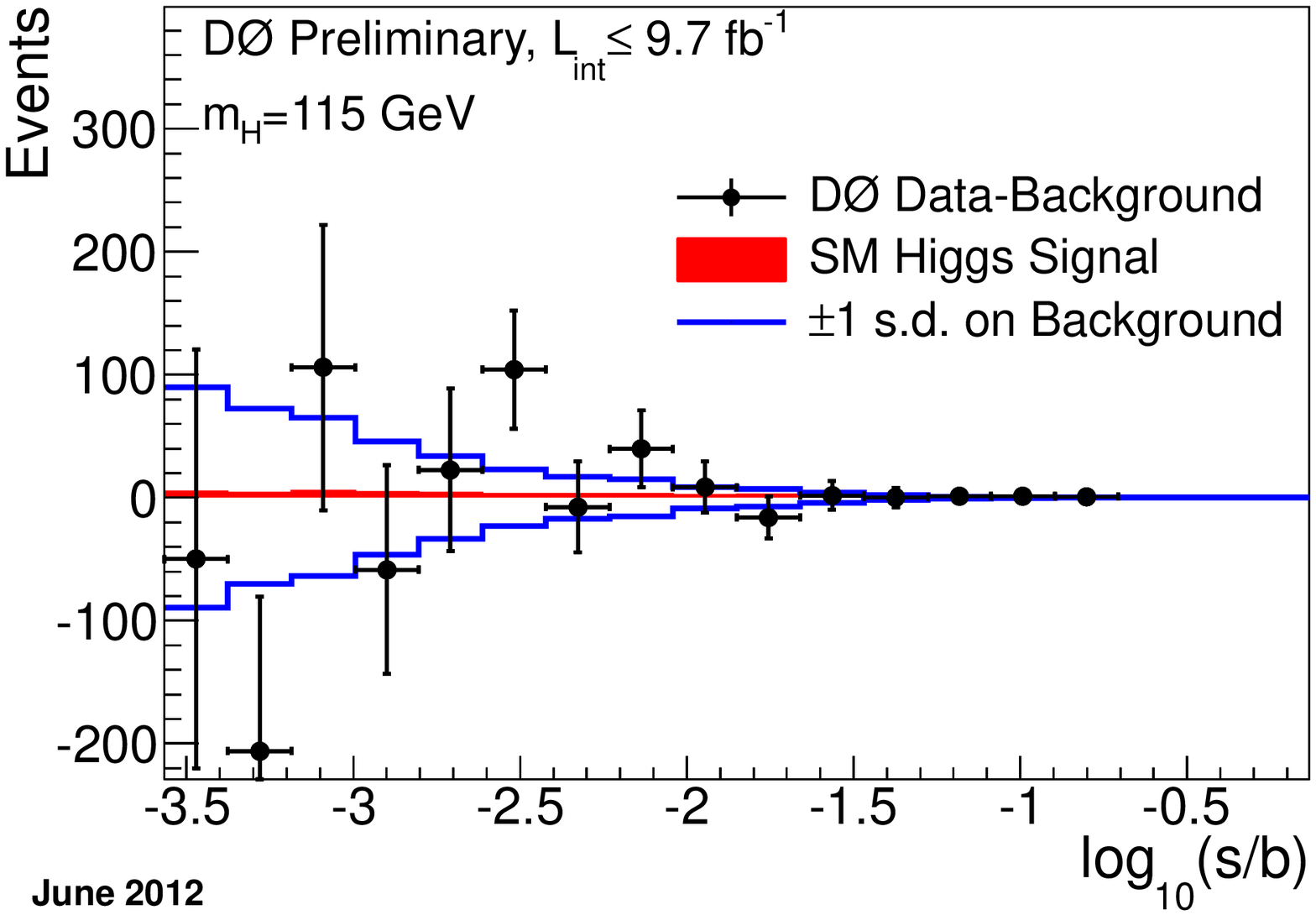}
\includegraphics[height=0.2\textheight]{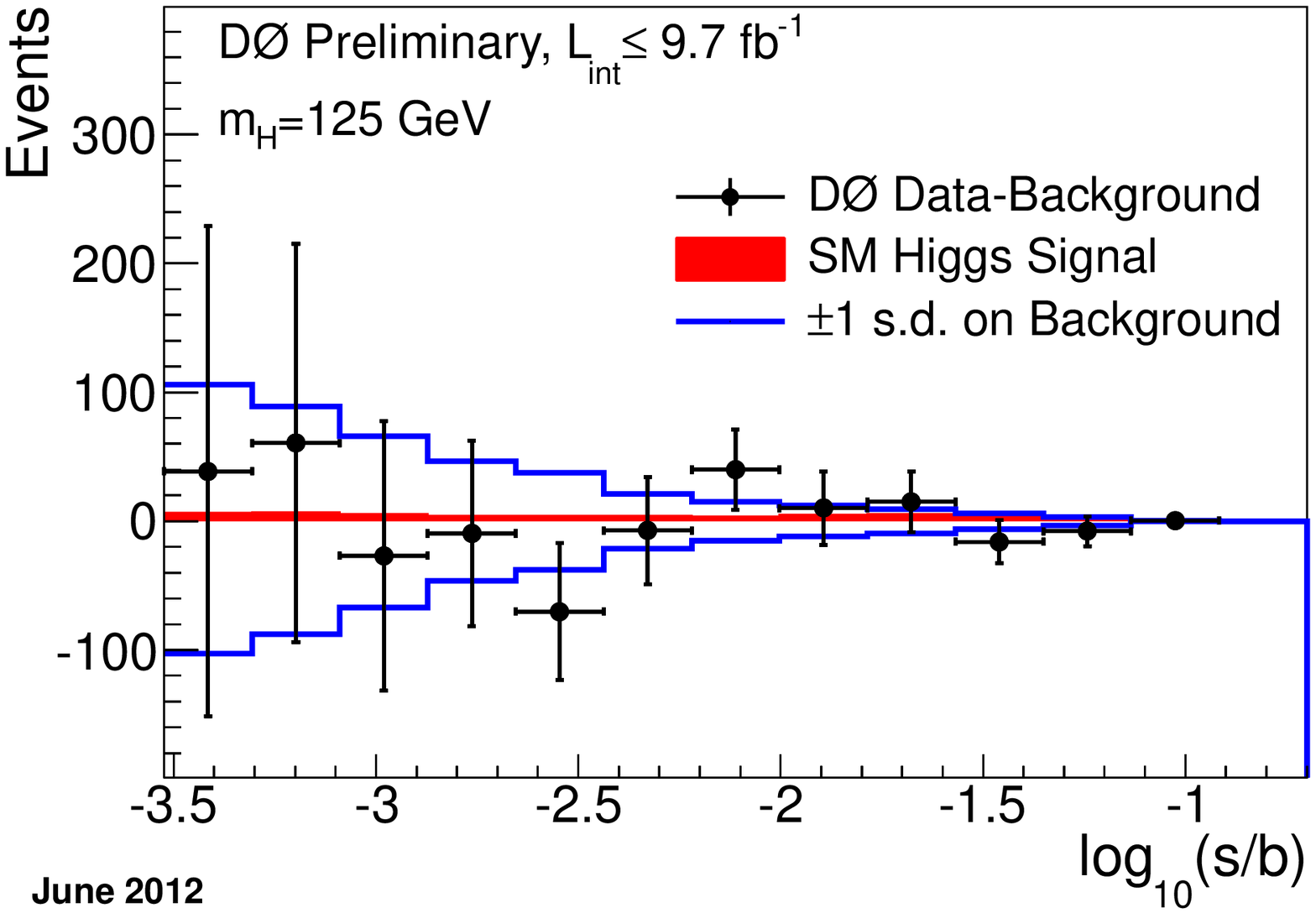} 
\includegraphics[height=0.2\textheight]{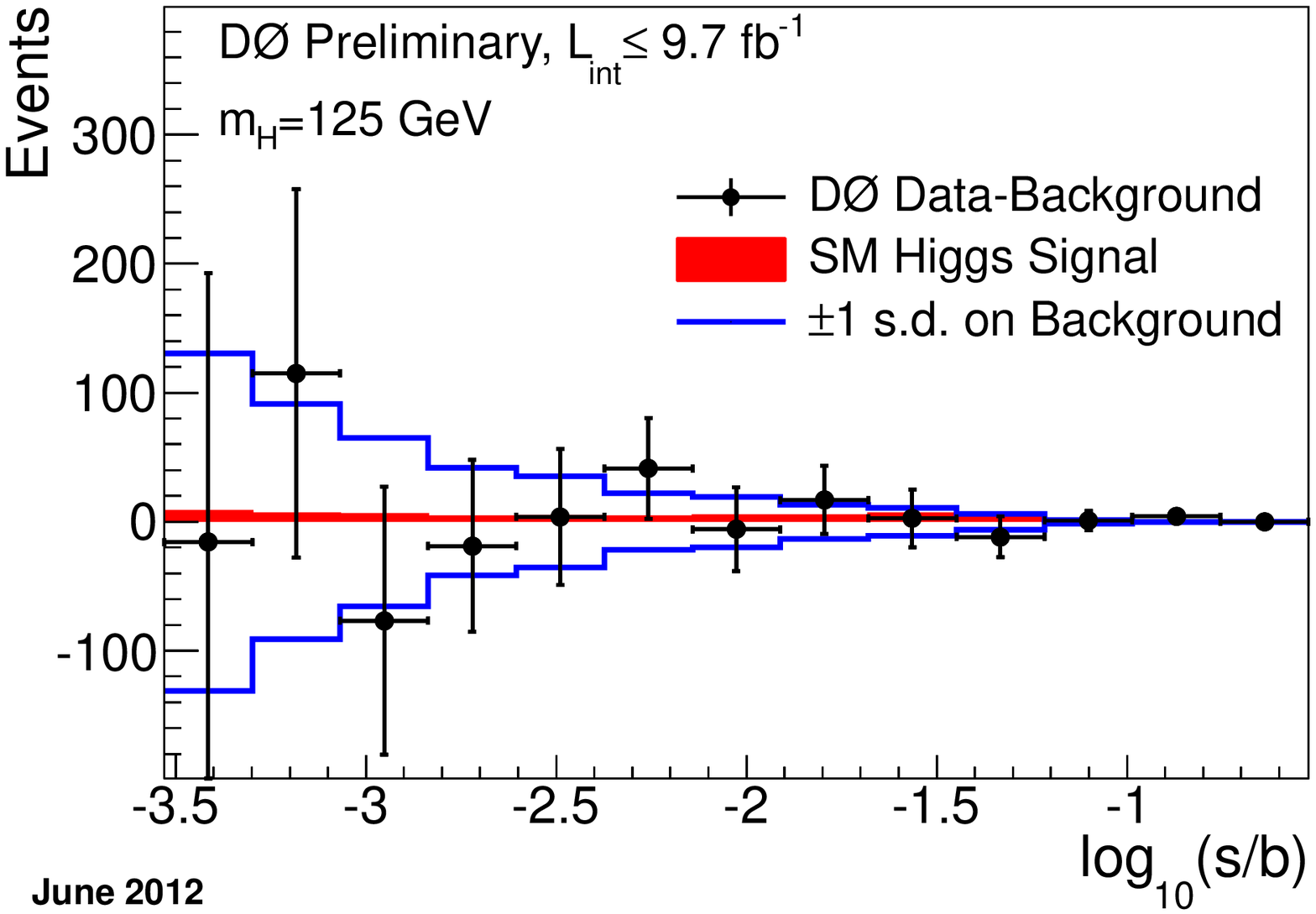}
\caption{\label{fig:HWWsb_compare_subtract} Background-subtracted
  data distributions of  $\log_{10}(s/b)$ in the combined 
   $WH/ZH/H, H$$\rightarrow$$ W^+W^-$ analyses for
  this result (right) and the result from Ref. \cite{M12dzcombo} (left) for
  assumed Higgs boson masses of 115\gev\ and 125\gev. 
   The background subtracted data are shown as
  points and the signal is shown as the red histograms.  The blue
  lines indicate the uncertainty on the background prediction.}
\end{figure}

\begin{figure}[tbp]
\includegraphics[height=0.2\textheight]{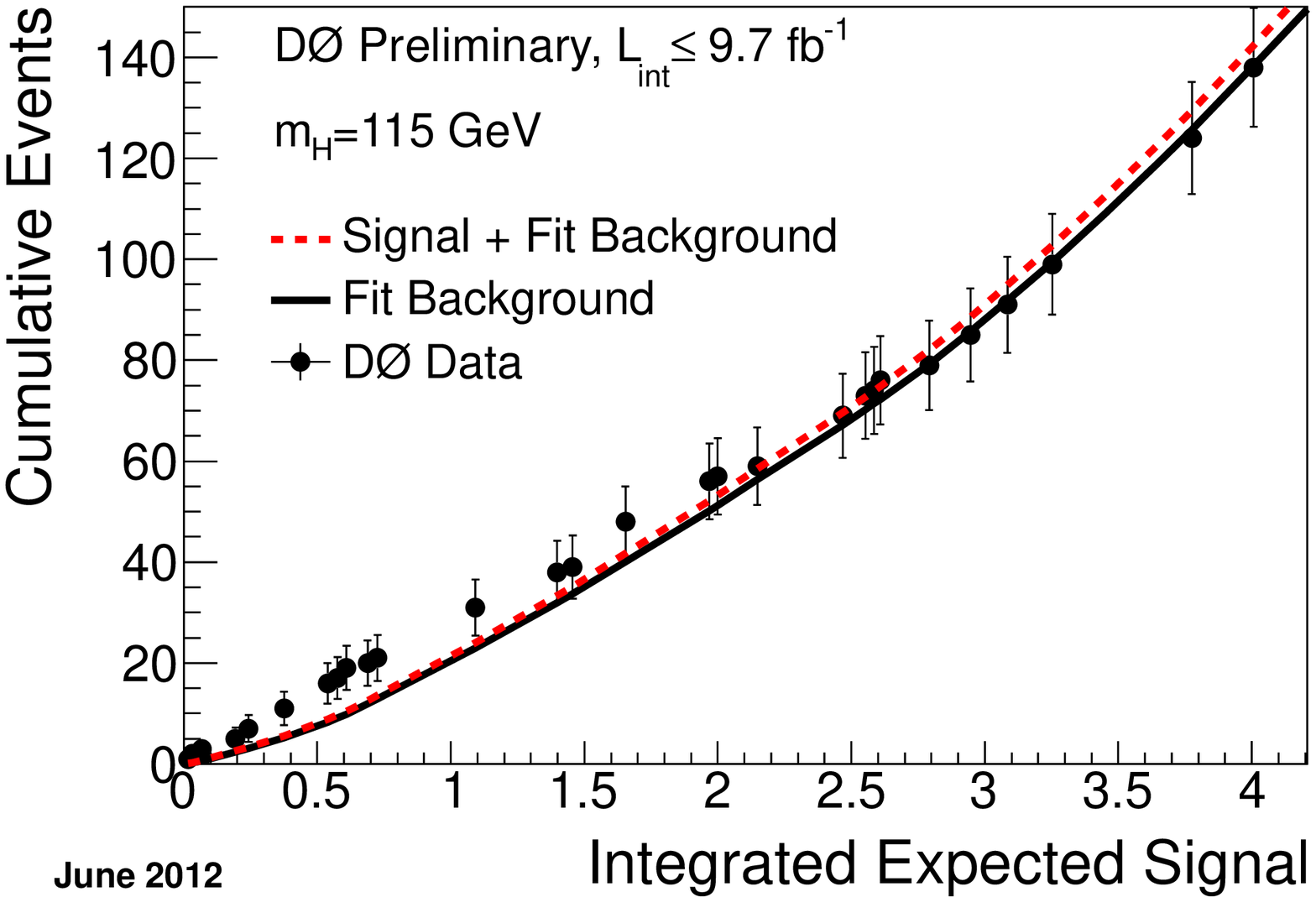} 
\includegraphics[height=0.2\textheight]{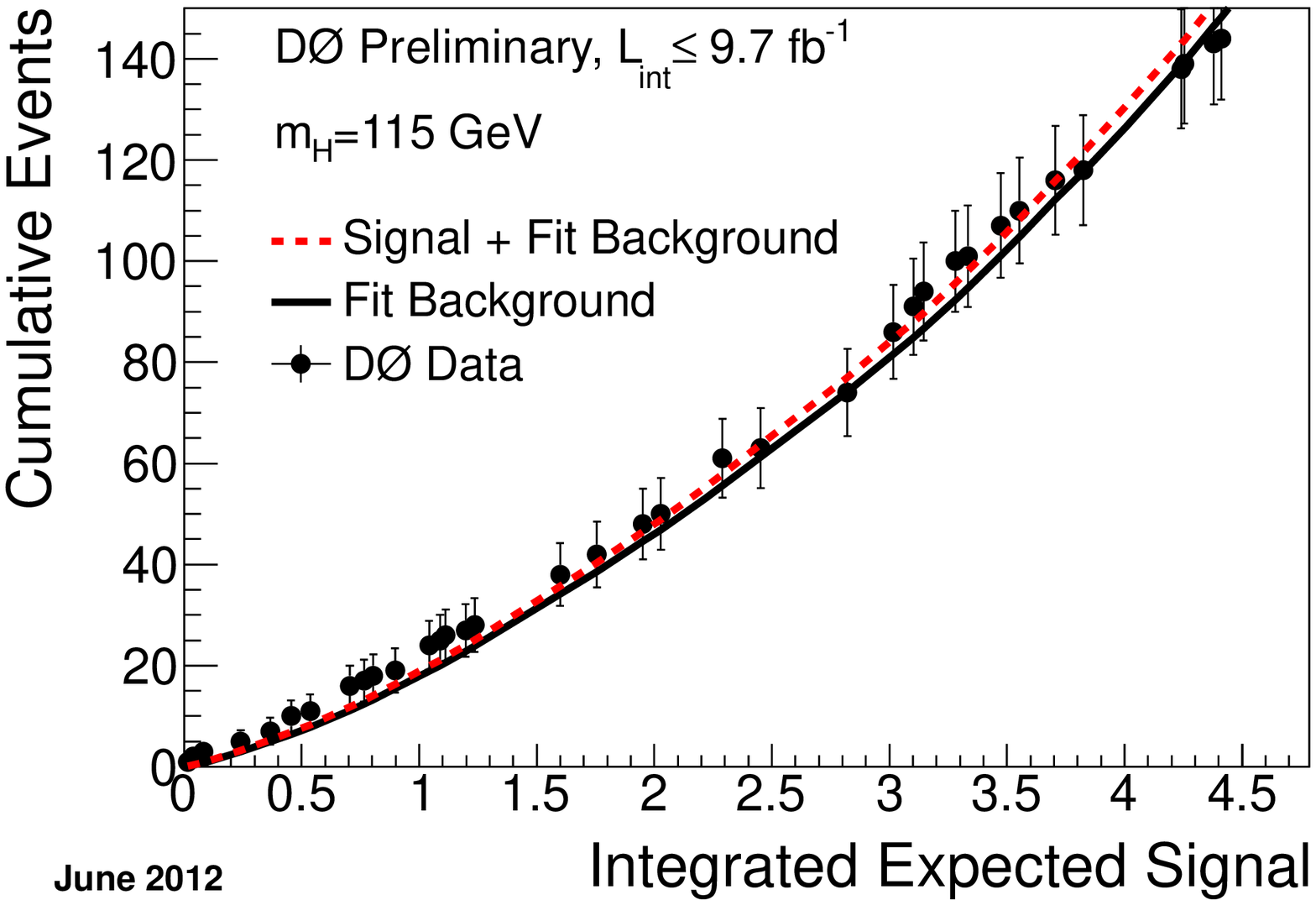}
\includegraphics[height=0.2\textheight]{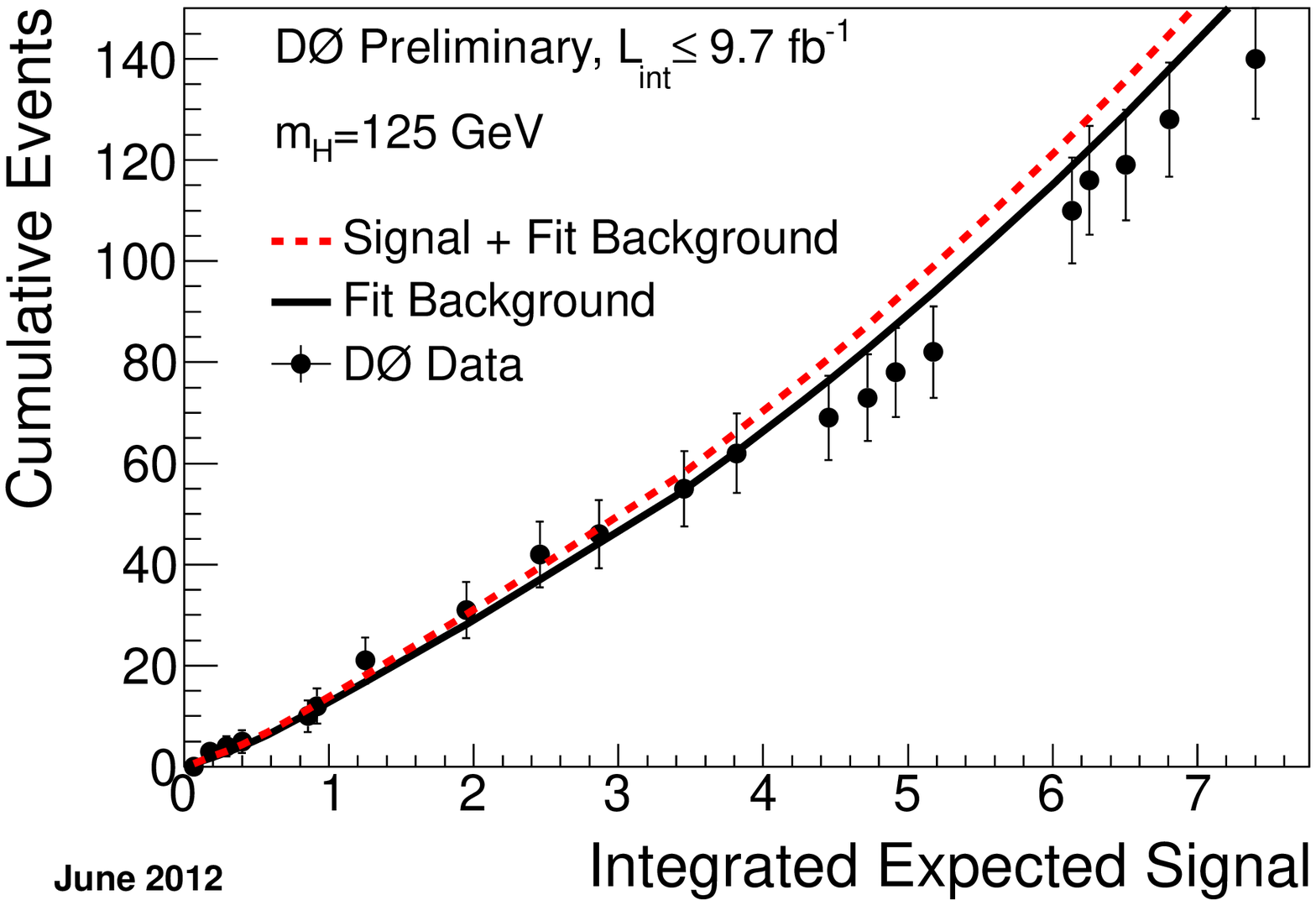} 
\includegraphics[height=0.2\textheight]{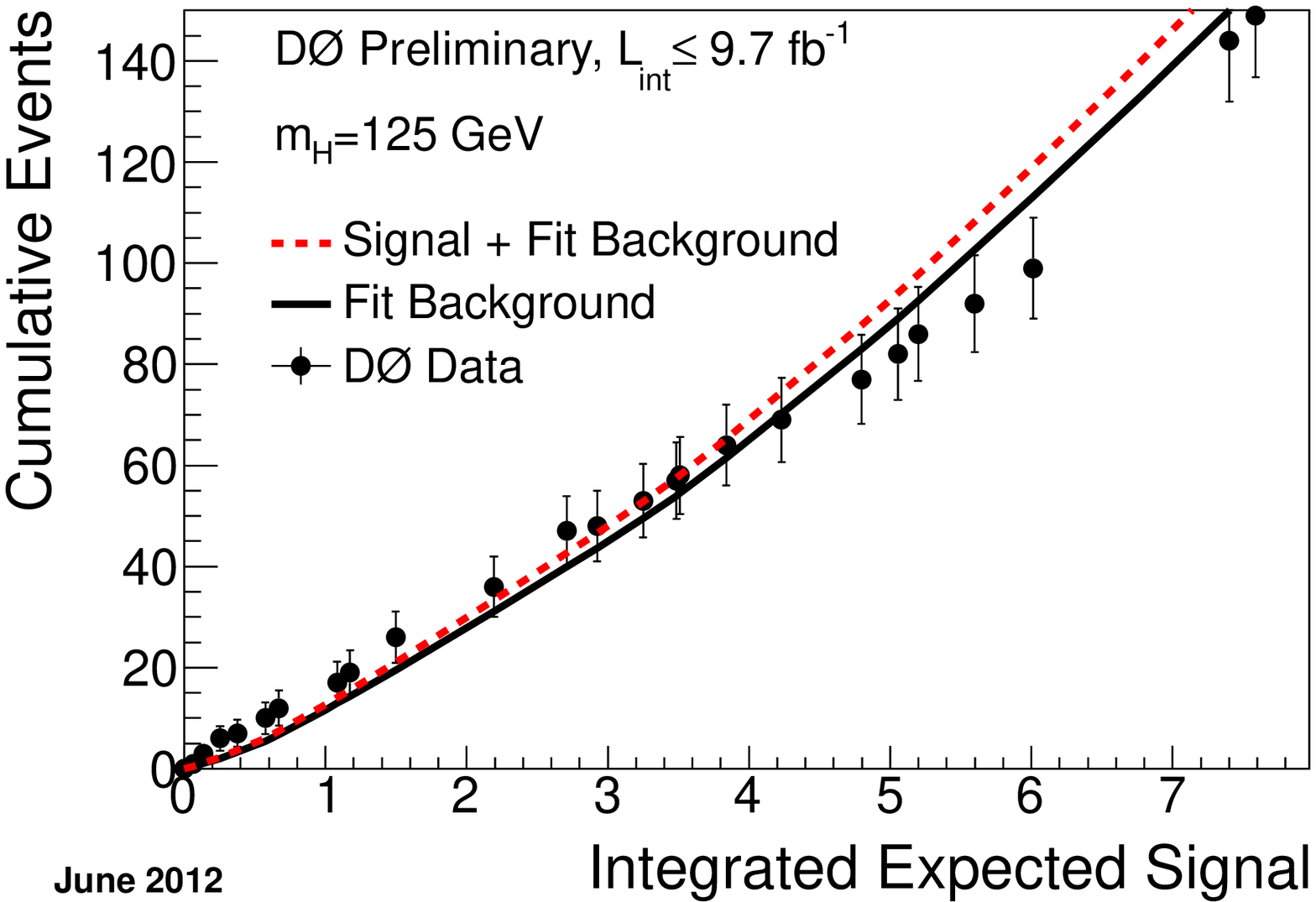}
\caption{\label{fig:HWWcompare_integral} 
Cumulative number of events for the highest $s/b$ bins in the combined $WH/ZH/H, H$$\rightarrow$$ W^+W^-$
analyses for this result (right) and the result from Ref. \cite{M12dzcombo} (left) for
assumed Higgs boson masses of 115\gev\ and 125\gev.   The integrated background-only and
signal+background predictions are shown as a function of the
accumulated number of signal events.  The points show the integrated number of observed
events, including only the statistical uncertainty, which is correlated point-to-point.
Systematic uncertainties on the integrated background-only and signal+background
predictions are not displayed.}
\end{figure}

 \end{document}